\documentclass[acmsmall]{acmart}

\AtBeginDocument{%
  }
    
\usepackage[a-2b]{pdfx}
\usepackage{amsmath}
\usepackage{bm}
\usepackage{graphicx}
\usepackage{apxproof}
\usepackage{etoolbox}
\usepackage{enumitem}
\usepackage{algorithmic}
\usepackage{multirow}
\usepackage{amsthm}
\usepackage{booktabs}
\usepackage{xspace}
\usepackage{amsmath,stmaryrd,pict2e,picture}
\usepackage{xcolor}
	
\usepackage{tabu}
\usepackage{bbold}
\usepackage{multirow}
\usepackage{diagbox}
\usepackage{balance}
\usepackage{float}
\usepackage[noend,lined,boxed,vlined,ruled,linesnumbered]
{algorithm2e}
\usepackage{makecell}

\usepackage{array}

\newcolumntype{C}[1]{>{\centering\arraybackslash}p{#1}}

\definecolor{myyellow}{HTML}{fcefd2}
\definecolor{mydarkyellow}{HTML}{FAD02E}

\definecolor{mygreen}{HTML}{d2fce6}
\definecolor{mypurple}{HTML}{f2d2fc}
\definecolor{peach}{HTML}{FFDAB9}
\definecolor{fpred}{RGB}{255,99,99}

\usepackage{caption, subcaption}

\usepackage[labelformat=simple]{subcaption}
\SetKwInOut{Input}{input}
\SetKwInOut{Output}{output}
\newcommand{\codeq}[1]{{\tt {\small ``#1''}}}

\newcommand{\code}[1]{\texttt{\small #1}}

\newcommand{\at}{\textsc{Auto-Test}\xspace}

\newcommand{\ignore}[1]{}
\newcommand{\sdc}{\text{Semantic-Domain Constraints}\xspace}
\newcommand{\sdca}{\text{SDC}\xspace}
\newcommand{\sdcas}{\text{SDCs}\xspace}

\newcommand{\css}{\text{Coarse-grained \sdca Selection}\xspace}
\newcommand{\fss}{\text{Fine-grained \sdca Selection}\xspace}

\newcommand{\cssa}{\text{CSS}\xspace}
\newcommand{\fssa}{\text{FSS}\xspace}

\newcommand{\sttest}{\textsc{Spreadsheet-Table-Bench}\xspace}
\newcommand{\sttrain}{\textsc{Spreadsheet-Tables}\xspace}

\newcommand{\sttesta}{\textsc{ST-Bench}\xspace}

\newcommand{\rttest}{\textsc{Relational-Table-Bench}\xspace}
\newcommand{\rttrain}{\textsc{Relational-Tables}\xspace}

\newcommand{\rttesta}{\textsc{RT-Bench}\xspace}

\newcommand{\tablib}{\textsc{Tablib}\xspace}

\newtheorem{df}{Definition}
\newtheorem{ex}{Example}
\newtheorem{pr}{Proposition}
\newtheorem{lemm}{Lemma}
\newtheorem{lem}{Theorem}

\newenvironment{example}{\begin{ex} \nopagebreak
\begin{rm}}{{\hfill$\Box$}\end{rm}\end{ex}} 

\newenvironment{definition}{\begin{df} \nopagebreak
\begin{rm}}{{}\end{rm}\end{df}}

\newtoggle{full}
\toggletrue{full}
\newcommand{\ar}{\textsc{All-Constraints}\xspace}
\newcommand{\fs}{\textsc{Fine-Select}\xspace}
\newcommand{\cs}{\textsc{Coarse-Select}\xspace}

\newtheorem{observation}{Observation}

\begin{document}
% \pagestyle{plain} % removed header with this

%\title{Auto-Test: Unsupervised  Error Detection in Tables using Automatically Designed Tests for Semantic Domains}
\title{Auto-Test: Learning Semantic-Domain Constraints for Unsupervised  Error Detection in Tables%in Tables using Automatically Designed Tests for Semantic Domains
}

%\title{Auto-Test: Unsupervised  Error Detection in Tables using Auto-Designed Hypothesis Tests on Table Corpora}  

\author{Qixu Chen}
\authornote{Part of work done while at Microsoft Research, email: qchenax@connect.ust.hk}
\affiliation{%
  \institution{Hong Kong University of Science and Technology}
  \country{China}
}
%\email{qchenax@connect.ust.hk}

\author{Yeye He}
\authornote{Correspondence: yeyehe@microsoft.com}
\affiliation{%
  \institution{Microsoft Research}
  %\streetaddress{1 Th{\o}rv{\"a}ld Circle}
  %\city{Hekla}
  \country{USA}
}
%\email{yeyehe@microsoft.com}

\author{Raymond Chi-Wing Wong}
\affiliation{%
  \institution{Hong Kong University of Science and Technology}
  \country{China}
}
%\email{coyan@microsoft.com}

\author{Weiwei Cui}
\affiliation{%
  \institution{Microsoft Research}
  \country{China}
} 
%\email{wanyue@microsoft.com}

% \linebreak   % Forces a line break after the fourth author

\author{Song Ge}
\affiliation{%
  \institution{Microsoft Research}
  \country{China}
}
%\email{surajitc@microsoft.com}

\author{Haidong Zhang}
\affiliation{%
  \institution{Microsoft Research}
  \country{China}
}
%\email{surajitc@microsoft.com}

\author{Dongmei Zhang}
\affiliation{%
  \institution{Microsoft Research}
  \country{China}
}
%\email{surajitc@microsoft.com}

\author{Surajit Chaudhuri}
\affiliation{%
  \institution{Microsoft Research}
  \country{USA}
}
%\email{surajitc@microsoft.com}

\renewcommand{\shortauthors}{Qixu Chen et al.}

% \acmConference
% \acmBooktitle
\setcopyright{cc}
\setcctype{by}
\acmJournal{PACMMOD}
\acmYear{2025} \acmVolume{3} \acmNumber{3 (SIGMOD)} \acmArticle{133} \acmMonth{6} \acmPrice{}\acmDOI{10.1145/3725396}

%%
%% The abstract is a short summary of the work to be presented in the
%% article.
\begin{abstract}
Data cleaning is a long-standing challenge in data management. While  powerful logic and statistical algorithms have been developed to detect and repair data errors in tables, existing algorithms  predominantly rely on domain-experts to first manually specify data-quality constraints specific to a given table, before data cleaning algorithms can be applied.

In this work, we propose a new class of data-quality constraints that we call \emph{Semantic-Domain Constraints}, which can be reliably inferred and automatically applied to \emph{any tables}, without requiring domain-experts to manually specify on a per-table basis. 
We develop a principled framework to systematically learn such constraints from table corpora using large-scale statistical tests, which can further be distilled into a core set of constraints using our optimization framework, with provable quality guarantees. Extensive evaluations show that this new class of constraints can be used to both (1)  directly detect errors on real tables in the wild, and (2) augment existing expert-driven data-cleaning techniques as a new class of complementary constraints.

Our extensively labeled benchmark dataset with 2400 real data columns, as well as our code are available at \url{https://github.com/qixuchen/AutoTest} to facilitate future research.
\end{abstract}

%%
%% The code below is generated by the tool at http://dl.acm.org/ccs.cfm.
%% Please copy and paste the code instead of the example below.
%%

% \ccsdesc{Do Not Use This Code~Generate the Correct Terms for Your Paper}
% \ccsdesc[100]{Do Not Use This Code~Generate the Correct Terms for Your Paper}

%%
%% Keywords. The author(s) should pick words that accurately describe
%% the work being presented. Separate the keywords with commas.

\received{October 2024}
\received[revised]{January 2025}
\received[accepted]{February 2025}

\maketitle

% %%% do not modify the following VLDB block %%
% %%% VLDB block start %%%
% \pagestyle{\vldbpagestyle}
% \begingroup\small\noindent\raggedright\textbf{PVLDB Reference Format:}\\
% \vldbauthors. \vldbtitle. PVLDB, \vldbvolume(\vldbissue): \vldbpages, \vldbyear.\\
% \href{https://doi.org/\vldbdoi}{doi:\vldbdoi}
% \endgroup
% \begingroup
% \renewcommand\thefootnote{}\footnote{\noindent
% This work is licensed under the Creative Commons BY-NC-ND 4.0 International License. Visit \url{https://creativecommons.org/licenses/by-nc-nd/4.0/} to view a copy of this license. For any use beyond those covered by this license, obtain permission by emailing \href{mailto:info@vldb.org}{info@vldb.org}. Copyright is held by the owner/author(s). Publication rights licensed to the VLDB Endowment. \\
% \raggedright Proceedings of the VLDB Endowment, Vol. \vldbvolume, No. \vldbissue\ %
% ISSN 2150-8097. \\
% \href{https://doi.org/\vldbdoi}{doi:\vldbdoi} \\
% }\addtocounter{footnote}{-1}\endgroup
% %%% VLDB block end %%%

% % %%% do not modify the following VLDB block %%
% %%% VLDB block start %%%
% \ifdefempty{\vldbavailabilityurl}{}{
% \vspace{.3cm}
% \begingroup\small\noindent\raggedright\textbf{Artifact Availability:}\\
% The data and/or other artifacts have been made available at \url{\vldbavailabilityurl}.
% \endgroup
% }
% % %%% VLDB block end %%%

\begin{sloppy}

%\yeye{add kb baselines?}

%\yeye{add google sheet baselines?}

\section{Introduction}
\label{sec:intro}

%\yeye{can you please provide some examples of good rules that we discover, that are intuitive and easy to explain (to use in the intro)? We can aim to have maybe 2 rules per type (pattern, embedding, cta). Can list these rules in a table, and write a running example around this example.}

%\yeye{add excel screenshot to describe the problem. Motivate using gpt4's lack of accuracy}

Data cleaning is a long-standing challenge in the data management community. While there is a long and fruitful line of research that developed powerful techniques using \emph{data-quality constraints} (e.g., FD, CFD, etc.) to detect and repair data errors in tables~\cite{dc-beskales2013relative, dc-chu2013holistic, dc-ge2020hybrid, dc-khayyat2015bigdansing, mahdavi2020baran, dc-rekatsinas2017holoclean, dc-yakout2013don, dc-khayyat2015bigdansing}, existing methods largely  depend  on domain-experts to first  specify  data-quality constraints that are specific to a given table, before data-cleaning algorithms can be performed (while constraint discovery methods also exist, they are mainly intended to discover \emph{candidate rules} that still require humans to  verify~\cite{discoverrule-chu2014ruleminer, discoverrule-fan2010discovering, discoverrule-huhtala1999tane, discoverrule-wyss2001fastfds, discoverrule-berti2018discovery}). We term this class of sophisticated and powerful approaches as ``\underline{\emph{expert-driven data cleaning}}''. 

%While rule discovery algorithms have been proposed that can run on individual datasets, rules so discovered are candidate rules, that still require on humans to verify on a per-dataset basis before they can be applied, in order to achieve a high level of accuracy. 

% \iftoggle{full}
% {
%     \begin{figure*}
%     \centering
%     \includegraphics[width=1 \linewidth]{figures/excel-demo-1.png}
%     \caption{Data cleaning feature for non-technical end-users in Excel. Each automatically detected data-quality issue in user table is shown as a ``suggestion cards'' on the side-pane (right), which users only need to review and accept. [\underline{{\href{https://drive.google.com/file/d/1kIVLVOZQfZn2Dqd2M-fblo7EwpP_O4tw/view?usp=drive_link}{This link}}}] gives an end-to-end demo of how users can stay in control while leveraging capabilities like \sdca to clean data automatically.  
%     }
%     \label{fig:excel-demo}
%     \end{figure*}
% }
% {
    \begin{figure}
    \centering
    \includegraphics[width=1 \linewidth]{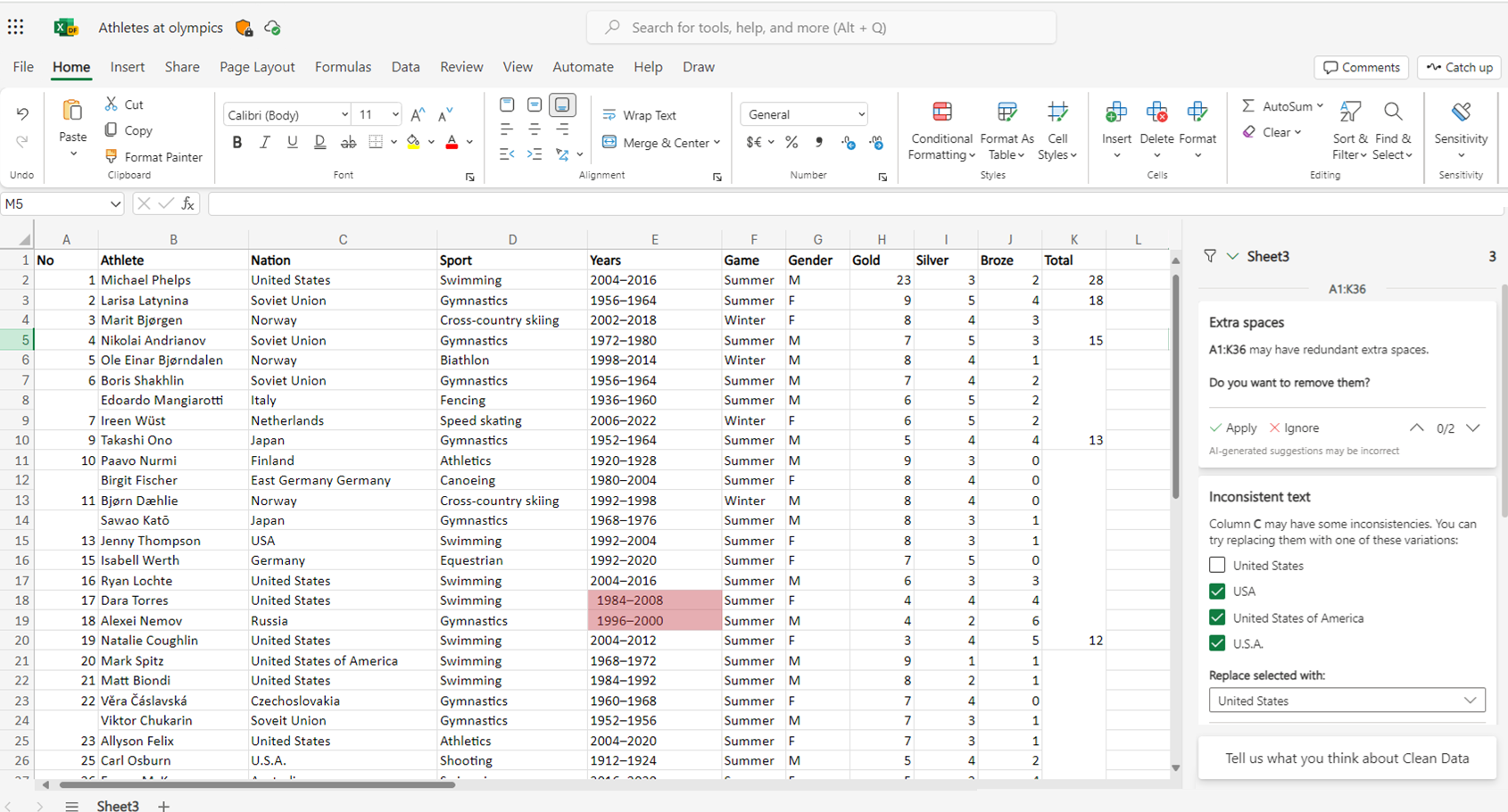}
    % \vspace{-7mm}
    \caption{Example data cleaning feature for end-users in Microsoft Excel. Data quality issues in user tables are automatically detected using techniques such as~\cite{HH18, wang2019uni, chakrabarti2016data, xing2024table}, and are presented as intuitive ``suggestion cards'' on the side-pane (right), for users to review and accept. [\underline{{\href{https://drive.google.com/file/d/1kIVLVOZQfZn2Dqd2M-fblo7EwpP_O4tw/view?usp=drive_link}{This link}}}]\cite{clean-data-demo} gives an end-to-end demo of how users can leverage such automated capabilities to easily clean data (without needing to define any constraints first), while staying in full control over any suggested changes that may be applied to their data.
    }
    \vspace{-3mm}
    \label{fig:excel-demo}
    \end{figure}
%}

While such expert-driven approaches to data cleaning are extremely powerful, when experts are available to inspect each table and define relevant constraints, we observe that there is an emerging class of ``\underline{\emph{end-user data-cleaning}}'' use cases that aim to democratize data-cleaning for the average non-technical users, by working out-of-the-box and without requiring experts to be involved.

For example, in end-user spreadsheet tools such as Microsoft Excel~\cite{Excel} and Google Sheets~\cite{Google-sheets} that are used by billions of non-technical users, 
there is a growing need to automatically detect and repair data errors in user tables out-of-the-box, \emph{without requiring users to define constraints or provide labeled data first}.

Figure~\ref{fig:excel-demo} shows 
a screenshot of such a feature in Excel, which uses techniques such as~\cite{HH18, wang2019uni, chakrabarti2016data, xing2024table} to automatically detected data-quality issues are presented as ``suggestion cards'' shown on the side-pane, that users can easily review and accept with the click of a button (without needing to define any constraints first). Google Sheets has a similar feature for error-detection~\cite{google-smart-cleanup}. %(Note that users always stay ``in control'' in these situations, as they choose to apply or ignore data cleaning suggestions that are presented to them).

%A key gap between existing data-cleaning techniques, and the capabilities 

In this work, we study a new class of data-quality constraints previously  overlooked  in the literature, which we call \emph{\sdc} \emph{(\sdca)}.  Importantly, such constraints can be reliably applied to in a generic fashion to relevant tables, without needing human experts, making them suitable for both ``end-user data cleaning'' (e.g., in spreadsheets), and ``expert-driven data cleaning'' (as they serve as a new class of constraints to complement existing constraints).

\begin{figure*}[t!]

    \centering
    \includegraphics[width=1 \linewidth]{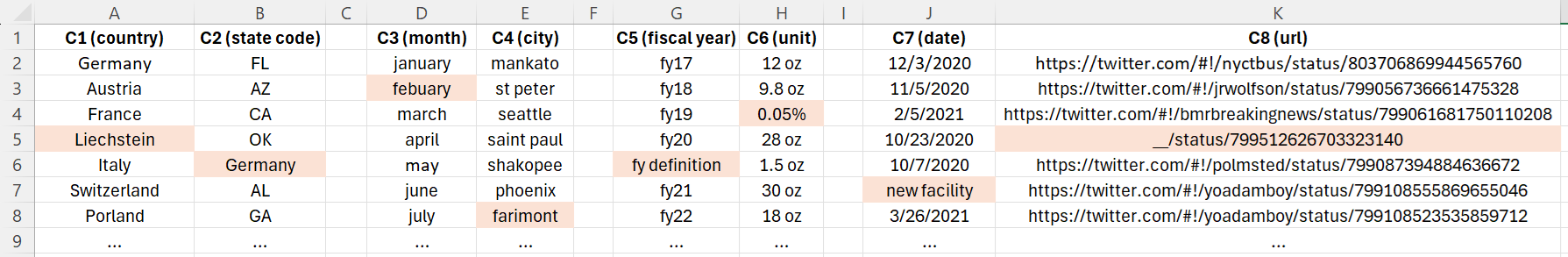}
    \vspace{-7mm}
    \caption{Real examples of table columns, each representing a distinct ``semantic domain'' (annotated in the column header). Each column $C_i$ has a real \colorbox{peach}{data error} (which may be a typo, or a semantically incompatible value), %(highlighted in \colorbox{peach}{color}), 
     that is detected by a corresponding ``semantic domain constraint'' $r_i$  in Table~\ref{tab:rule_example}, which are constraints automatically learned from running \at.}
    %\vspace{-1mm}
    \label{fig:example-table-columns}
\end{figure*}

\begin{small}
\begin{table*}[t]
    \caption{Example \sdc (\sdcas), instantiated using CTA-classifiers (in $r_1, r_2$), text-embedding ($r_3, r_4$), regex-patterns ($r_5, r_6$), and program functions ($r_7, r_8$), all sharing the same \sdca structure. Each \sdca has a \underline{(a) Pre-condition}: if \colorbox{mygreen}{matching-percentage\%} of column values in column $C$, evaluated using a  \colorbox{mypurple}{domain evaluation function} satisfy \colorbox{myyellow}{inner-distance threshold}, we recognize that constraint $r$ should apply to column $C$, and \underline{(b) Post-condition}: any value evaluated using the same \colorbox{mypurple}{domain evaluation function} satisfy \colorbox{mydarkyellow}{outer-distance threshold}, are predicted as data errors.   Each example constraint  $r_i$ in this table would  trigger the detection of a real data error shown in column  $C_i$ of Figure~\ref{fig:example-table-columns}. All color-coded components (matching-percentage, evaluation function, etc.) are parameters to \sdca that are learned using \at.}
     \vspace{-2mm}
    \label{tab:rule_example}
    \centering
    \scalebox{0.7}{
    \begin{tabular}{|p{0.2cm}|p{1.4cm}|p{8.2cm}|p{7.8cm}|C{0.5cm}|} \hline
       ID & Type & Pre-condition $P$: (on what columns should this constraint apply)  &  Post-condition $S$: (what values will be predicted as errors) & Conf. \\ \hline
       
       $r_1$ & CTA & \colorbox{mygreen}{85\%} col vals have their \colorbox{mypurple}{\textit{country-classifier}} scores > \colorbox{myyellow}{0.75}   & values whose \colorbox{mypurple}{\textit{country-classifier}} scores  < \colorbox{mydarkyellow}{0.01} & ... \\ \hline
       
       $r_2$ & CTA & \colorbox{mygreen}{90\%} col vals have their \colorbox{mypurple}{\textit{state-classifier}} scores > \colorbox{myyellow}{0.55} & values whose \colorbox{mypurple}{\textit{state-classifier}} scores < \colorbox{mydarkyellow}{0.05} & ... \\ \hline

       $r_3$ & Embedding & \colorbox{mygreen}{85\%} col vals have their \colorbox{mypurple}{\textit{Glove} distances to ``january''} < \colorbox{myyellow}{4.0} & values whose \colorbox{mypurple}{\textit{Glove}  distances to ``january''} > \colorbox{mydarkyellow}{5.5} & ... \\ \hline

       $r_4$ & Embedding & \colorbox{mygreen}{80\%} col vals have their \colorbox{mypurple}{\textit{S-BERT} distances to ``seattle''} < \colorbox{myyellow}{1.2} & values whose \colorbox{mypurple}{\textit{S-BERT} distances to ``seattle''} > \colorbox{mydarkyellow}{1.35} & 0.88  \\ \hline

       $r_4'$ & Embedding & \colorbox{mygreen}{90\%} col vals have their \colorbox{mypurple}{\textit{S-BERT} distances to ``seattle''} < \colorbox{myyellow}{1.1} & values whose \colorbox{mypurple}{\textit{S-BERT} distances to ``seattle''} > \colorbox{mydarkyellow}{1.4} & 0.93  \\ \hline

       $r_5$ & Pattern & \colorbox{mygreen}{95\%} col vals match pattern \colorbox{mypurple}{``$\backslash$[a-zA-Z]+$\backslash$d+''} (match = \colorbox{myyellow}{1}) & values not matching pattern \colorbox{mypurple}{``$\backslash$[a-zA-Z]+$\backslash$d+''} (match = \colorbox{mydarkyellow}{0}) & ... \\ \hline

       $r_6$ & Pattern & \colorbox{mygreen}{95\%} col vals match pattern \colorbox{mypurple}{``$\backslash$d+$\ \backslash$[a-zA-Z]+''} (match = \colorbox{myyellow}{1}) & values not matching pattern \colorbox{mypurple}{``$\backslash$d+$\ \backslash$[a-zA-Z]+''} (match = \colorbox{mydarkyellow}{0}) & ... \\ \hline

       $r_{7}$ & Function & \colorbox{mygreen}{98\%} col vals return true on function \colorbox{mypurple}{\textit{validate\_date()}} %\footnotemark[2] 
       (ret = \colorbox{myyellow}{1}) & values return false on function \colorbox{mypurple}{\textit{validate\_date()}} (ret = \colorbox{mydarkyellow}{0}) & ... \\ \hline

       $r_{8}$ & Function & \colorbox{mygreen}{99\%} col vals return true on function \colorbox{mypurple}{\textit{validate\_url()}} %\footnotemark[3] 
       (ret = \colorbox{myyellow}{1})  & values that return false on function \colorbox{mypurple}{\textit{validate\_url()}} (ret = \colorbox{mydarkyellow}{0}) & ... \\ \hline
       
    \end{tabular}
    }
    \vspace{-2mm}
\end{table*}
\end{small}
% \footnotetext[2]{Python function validate$\_$date(): \url{https://docs.dataprep.ai/api_reference/dataprep.clean.html#dataprep.clean.clean_date.validate_date}}
% \footnotetext[3]{Python function validate$\_$url(): \url{https://docs.dataprep.ai/api_reference/dataprep.clean.html#module-dataprep.clean.clean_url}}

\textbf{Intuition: leverage ``semantic domain'' for error detection.} The new class of constraints we study in this work are based on the intuitive notion of ``\emph{semantic domains}''. Specifically, given any relational table, values in the same column are expected to be \emph{homogeneous} and drawn from a ``domain'' of same semantics,  such as date, url, city-name, address, etc. Figure~\ref{fig:example-table-columns} shows an example table, where the semantics of each column  can then be inferred by humans from its values (annotated in column-headers to assist readability).

% \iftoggle{full}
% {
    The semantics of a column often implicitly define the ``domain'' of valid values that can appear in this column -- values falling outside of the ``domain'' can be picked up by humans as possible data errors, which we show in the example below.
    
    \begin{example}
    In Figure~\ref{fig:example-table-columns}, we as humans can see that column $C_1$ is likely a country column given its values, which makes \codeq{Liechstein} (a misspelling of \codeq{Liechtenstein}) an obvious error in the context of the ``semantic domain'' (country).
    
    Similarly, most values in column $C_2$ suggest this to be about state abbreviation codes, which makes \codeq{Germany} semantically incompatible in the column, and a likely error.
    
    As additional examples, \codeq{febuary} in $C_3$ and \codeq{farimont} in $C_4$ are clearly misspelled in the context of other values in the columns (which are month-names, and city-names, respectively).

    The values \codeq{fy definition} in   $C_5$ and \codeq{new facility} in $C_7$ are not compatible with other values in the columns and are likely errors (these are likely meta-data as opposed to actual data values). 
    
    Finally, the highlighted values in $C_6$ and $C_8$ are also inconsistent with the domains implied by other values in the columns (units and urls, respectively), which are therefore likely errors too.
    \end{example}
% }
% {
    % % revised{} for space in revision
    % The semantics of a column often implicitly define the ``domain'' of valid values that can appear in this column -- values falling outside of the ``domain'' can be picked up by humans as possible data errors, like the example in Figure~\ref{fig:example-table-columns} would show.
% }

Given that we as humans can reliably infer column semantics, and then use the underlying ``domain of valid values''  to identify likely  errors in Figure~\ref{fig:example-table-columns} (without needing to understand the domain-specific semantics of a table), the question we ask in this work is whether algorithms can mimic the human intuition, by codifying the intuition into precise and executable data-quality constraints to automatically detect data errors. 

% revised{} removing figure associated with an example for space
% \iftoggle{full}
% {
    \begin{figure*}[t]
    %\vspace{-18mm}
        \centering
        \includegraphics[width=0.85 \linewidth]{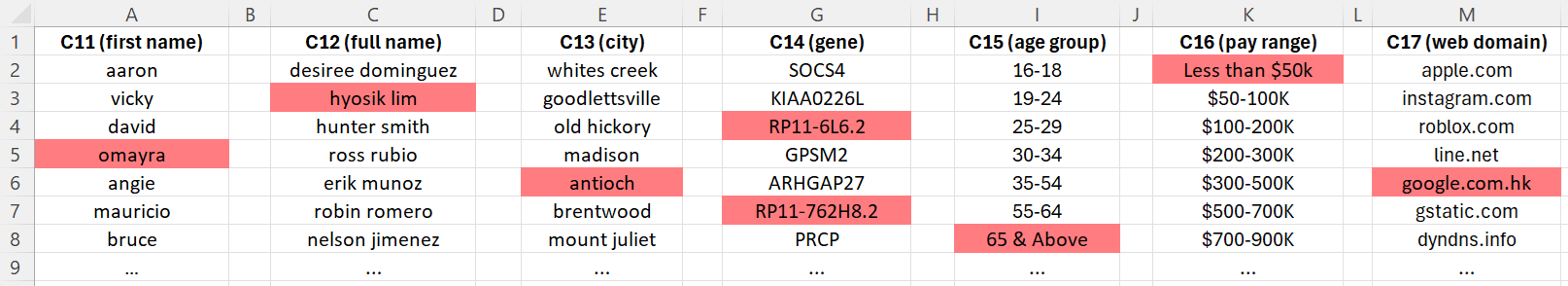}
        \vspace{-3mm}
        \caption{Real examples of table columns, where \colorbox{fpred}{false-positive} detection of errors are produced in highlighted cells, when existing column-type detection techniques are used directly to the task of error-detection. }
        \vspace{-3mm}
        \label{fig:example-table-columns-hard-fp}
    \end{figure*}
%}

\textbf{``Column-type detection'': insufficient for error detection.} 
There is a large literature on the related topic of ``column-type detection'', where the goal is to predict the semantic  type for a given column $C$, from a pre-defined list of types (e.g., people-names, locations, etc.). 
The problem has been studied in different settings, leading to techniques tailored to different types of tabular data. 

For example, for \underline{natural-language-oriented data columns} (e.g., people-names, company-names, address, etc.),  this is typically studied as a multi-class classification problem, also known as ``column type annotation'' (CTA)~\cite{cta-hulsebos2023adatyper, doduo, sherlock, auto-em-with-type, sato, cta-1, cta-2, cta-3, cta-4}, where techniques based on machine learning (ML)-classifiers and text embedding (e.g., Glove~\cite{PSM14} and SentenceBERT~\cite{RG19}) are developed. 

On the other hand, for \underline{machine-generated data columns} that are often number-heavy, and with strong regularity (e.g., ip-address, upc-code, time-stamps, etc.), both synthesized and curated regex-patterns / program-functions are used to detect types for such columns~\cite{auto-type, Dataprep-clean, PWL21, validator}.

While ``column-type detection'' is closely related to our goal of using ``semantic domain'' to detect errors, we observe that applying  column-type detection  directly to the task of error detection is  insufficient, because column-type detection focuses on the \emph{macro-level prediction} of whether a column $C$ belongs to a type $T$, without being calibrated to make \emph{fine-grained, micro-level predictions} of whether a value-instance $v \in C$ must be an error or not in the context of $C$ and type $T$. As a result, directly applying column-type detection to error detection  can  lead to lots of false-positive, like shown below.

% revised{} remove example for space
% \iftoggle{full}
% {
    
    \begin{example}
    Figure~\ref{fig:example-table-columns-hard-fp} shows an example table, where column semantics can again be inferred from data values, and are annotated in column headers to assist readability. 
    
    Column-type detection methods, such as CTA classifiers and text embedding, can reliably classify column $C_{11}$, $C_{12}$ and $C_{13}$ as \codeq{first-names}, \codeq{full-names} and \codeq{city}, respectively. However, when using such classifiers on \emph{individual values} $v \in C$ that happen to be uncommon names, such as \codeq{omayra}, \codeq{hyosik lim} and \codeq{antioch}, these classifiers will produce low scores, suggesting that the relevant value $v$ may not belong to the type $T$. 
    
    For instance, \codeq{omayra} (an uncommon name) in $C_{11}$ is not in the vocabulary of Glove embedding~\cite{PSM14}, and therefore  predicted by both embedding and CTA-classifiers to have low-scores for the type \codeq{first-names}.  When using these classifiers directly for error-detection, we arrive at the incorrect conclusion that values like \codeq{omayra} are data errors, which are, in fact, valid (but uncommon) names. The same is true for other uncommon names, such as \codeq{hyosik lim} in $C_{12}$  and \codeq{antioch} in $C_{13}$. 
    
    Similarly, in columns $C_{14}$, $C_{15}$, $C_{16}$ and $C_{17}$, a combination of syntactic patterns and NLP-classifiers can be used to recognize the semantics of these columns as  \codeq{gene}, \codeq{age-group}, etc. However, just because a value $v$ is in the minority and falls outside of the dominating pattern of a column, does not necessarily make $v$ an error, like the highlighted values in column $C_{14}$ -- $C_{17}$ would show, all of which are \emph{not} data errors. Note that these are in contrast to previous examples in $C_5$ -- $C_8$ from Figure~\ref{fig:example-table-columns}, where dominating patterns can help to identify both column-types and data errors.
    %It is therefore important to distinguish between cases where patterns are suitable for error detection (e.g., in Figure~\ref{fig:example-table-columns}), and cases where they are not reliable (e.g., in Figure~\ref{fig:example-table-columns-hard-fp}). % (which is something that our proposed optimization method  can learn-to-differentiate using a large table corpus, as we will demonstrate).
    \end{example}
    
    We can see from the example that column-type detection  techniques such as CTA classifiers and patterns cannot be used to detect errors directly, because they are designed to detect semantic types at the column-level (a macro-level prediction), which however are not suited to make fine-grained instance-level predictions for whether a value is erroneous or not (a micro-level prediction).
%}

\textbf{\at: Learn reliable ``semantic domain constraints''.}
In this work, we propose a new class of data-quality constraints called ``\sdc (\sdca)'', that \emph{builds upon and unifies diverse prior techniques for column-type detection}, for our new task of error detection.

%, while seamlessly unifying diverse types of column-type detection techniques (CTA-classifiers, text embedding, patterns, and functions), all using the same \sdca construct for the purpose of detecting errors in tables.

Table~\ref{tab:rule_example} shows examples of \sdca constraints. Briefly, each \sdca $r$ consists of a ``pre-condition'' that tests whether a given column $C$ is in the relevant semantic domain for $r$ to apply, and if so, a ``post-condition'' would calibrate the confidence of predicting $v \in C$ as an error, mirroring our fuzzy intuition of using ``semantic domains'' for error detection, but codified in precise constraints.  The exact parameters (color-coded components in the table), can be automatically learned from large table corpora  using statistical tests  in our proposed \at, which is the focus of this work.

Note that \sdca unifies prior techniques for column-type detection, using an abstraction we call ``domain evaluation functions'' (colored in purple in Table~\ref{tab:rule_example}), as they can be instantiated using diverse column-type detection, such as CTA-classifiers, text embedding, patterns, and functions, like shown in Table~\ref{tab:rule_example}.

%Each \sdca is further parameterized with suitable parameters as color-coded in the table, which are hard to set manually, that we show can be learned reliably over a large table corpora, using rigorous statistical hypothesis tests. 

Extensive evaluations using  benchmarks with real spreadsheet and relational tables in the wild, suggest that \at can reliably learn high-quality \sdca constraints for accurate error-detection. 

Overall, our proposed \at has the following key features: 

(1) \underline{Consistently more accurate} than alternative methods, including language models like GPT-4; %(we give a detailed analysis of why language models may not be a good fit for the error-detection task in our experiments); 

(2) \underline{Generalizable} to new and unseen tables, making it possible to apply \sdca with human experts; 

(3) \underline{Extensible} to new column-type detection techniques, all unified in the same framework; 

(4) \underline{Highly efficient} even on large tables, with negligible runtime and memory overhead; 

(5) \underline{Explainable} to humans, as our constraints leverage the natural notion of ``semantic domains'', which are not black-box models. 

\vspace{-2mm}
\section{Related work}
\label{sec:related}

% \iftoggle{full}
% {
    \textbf{Column-type detection.} Identifying the semantic types of columns in a table is an important  problem, where different techniques are developed that tailor to different types of columns. 
    
    For example, for natural-language heavy data content (name, address, etc.), ML-classifiers utilizing embedding features are shown to be effective~\cite{cta-4, cta-hulsebos2023adatyper, doduo, sherlock, auto-em-with-type, sato}. 
    For machine-generated data and data with strong regularity (e.g., ip-address, emails, time-stamps, etc.), regex-patterns~\cite{auto-tag, auto-validate, Dataprep-clean, systemx, pattern-profiling} and program-functions~\cite{auto-type, Dataprep-clean, PWL21, validator} are often used instead.
    We defer a detailed treatment of existing column-type detection methods to Section~\ref{sec:preliminary}.
% }
% {}

\textbf{Data Cleaning.}
Data cleaning is a long-standing  challenge in the data management community, with an influential line of research developing constraint-based techniques, including functional dependency (FD), conditional functional dependency (CFD), denial constraints (DC), etc., to detect and repair data errors in tables~\cite{dc-beskales2013relative, dc-chu2013holistic, dc-ge2020hybrid, dc-khayyat2015bigdansing, mahdavi2020baran, dc-rekatsinas2017holoclean, dc-yakout2013don}. 
\iftoggle{full}
{
Differential dependencies~\cite{song2011differential} and Metric FD~\cite{koudas2009metric} are additional powerful examples that can effectively relax equality constraints using distance functions. 
}

While existing data cleaning techniques are both flexible and powerful, they generally rely on complex data-quality constraints (e.g., in first-order logic) to be first defined by experts. Constraint discovery methods have also been studied, though they are generally designed to discover \emph{candidate rules} that still require human experts to verify, in order to ensure accuracy~\cite{discoverrule-chu2014ruleminer, discoverrule-fan2010discovering, discoverrule-huhtala1999tane, discoverrule-wyss2001fastfds, discoverrule-berti2018discovery}.
% \iftoggle{full}
% {
    In contrast, there is an emerging class of ``end-user data-cleaning'' use cases, especially in spreadsheet software (e.g., Excel and Google Sheets)~\cite{google-smart-cleanup, excel-clean-data},  that aim to empower the masses of non-technical spreadsheet users to clean their data, without the help of experts. % to program constraints or provide training data.
    
    We aim to show that the new class of \sdca constraints we introduce, when learned over large table corpora, can apply reliably to new and unseen tables without humans experts, making it applicable to ``end-user data-cleaning'', while also augmenting existing constraint-based data cleaning by serving as a new and complementary class of constraints.
%}

\textbf{Outlier detection}. There is a large literature of outlier detection methods in the machine learning and data mining literature, as reviewed in~\cite{chandola2009anomaly, markou2003novelty-1, markou2003novelty-2}, which are conceptually related to the problem we study. However, classical outlier detection methods predominantly operate only on \emph{local statistical features} (e.g., value distribution \emph{within a single target column} in our context) to determine outliers, without considering more \emph{global corpus-level information} (e.g., inferred semantic types and global data distributions) that our proposed method  specifically leverages  for error detection on tabular data. We will show in experiments that this gives our method a unique edge, which substantially outperforms SOTA outlier detection methods from the literature~\cite{domingues2018comparative}. %(selected based on a benchmark  study~\cite{domingues2018comparative}).  

% \iftoggle{full}
% {
    \textbf{Language models.}  Recent advances in NLP show that language models are applicable in a range of table tasks, including data cleaning~\cite{table-gpt, narayan2022can}.  Since language models would also understand column semantics, we will empirically compare with state-of-the-art language models like GPT-4. %, as well as fine-tuned versions of GPT-4 specifically targeting data-cleaning use cases (using the same training data as \at).
% }
% {

% }
% }

\section{Preliminary:  semantic column types}
\label{sec:preliminary}

Since our \sdca constraints are based on ``\emph{semantic types}'', we start with an overview of semantic types, and existing techniques to detect them.

\textbf{Semantic column-type detection methods.}
%\label{subsec:type-detection-methods}
As humans, we read tables columns not as string vs. numbers (primitive types), instead, we interpret the semantics of columns (column types), as \code{date}, \code{url}, \code{people-name}, \code{address},  etc., as shown in Figure~\ref{fig:example-table-columns}.  

``\emph{Column-type detection}'' refers  techniques to identify the ``semantic types'' for a given column $C$. 
Diverse techniques have been developed, including ML-classifiers and NL-embedding that are effective for natural-language data (e.g., \code{people name}, \code{address}, etc.), and regex-like patterns or program-functions that are suitable for machine-generated data with syntactic structures (e.g., \code{ip address}, \code{time-stamps}, etc.).

We survey existing column-type techniques and summarize them into four categories below:

%Given a table column, it is expected that normal values in the column should fall into the same \textit{semantic domain}, or \textit{domain} for short, which refers to the space of values that share certain homogeneous characteristics. 
%Although prior works approach the column-type detection from different angles, we in this work show that these methods can be put into a unifies framework to support various types of domains.
%Mathematically, a domain $D$ can be described as $\{v \in \mathcal{V} \, | \, f_D(v) \in A_{in}\}$ where $\mathcal{V}$ is the value space (i.e., the space of all possible values). $f_D$ is a \textit{evaluator} function based on a specific column-type detection function, which maps a value $v$ to its \textit{domain score} $f_D(v)$. $A_{in}$ is a range of domain scores.

%We survey the field of column-type detection and broadly classify existing works into four major types, namely CTA-based, embedding-based, pattern-based and function-based. Next, we introduce how to leverage each type of methods to construct corresponding evaluators. Readers may refer to Table~\ref{tab:rule_example} for examples of evaluators, which we marked in purple.

\noindent
\underline{(1) CTA-based methods~\cite{doduo, sherlock, sato}.} In 
Column Type Annotation (CTA), column-type detection is treated as an ML problem of multi-class classification, that predicts a semantic type from a fixed set of options.  %, including \textit{Sherlock} \cite{sherlock}, \textit{Doduo} \cite{doduo}, \textit{Sato} \cite{sato}, etc.
Various ML-classifiers have been developed for this problem, such as \textit{Sherlock}~\cite{sherlock} classifiers that can detect 78 semantic types (\codeq{type-city}, \codeq{type-country}, etc., from DBPedia), and \textit{Doduo} \cite{doduo} can further detect 121 semantic types (based on Freebase).  

At a conceptual level, a CTA classifier for a semantic type $t_i$ (say \codeq{type-country})\footnote{While some CTA-classifiers such as Sherlock are framed as multi-class classification, they can be equivalently interpreted as  multiple binary-classifications (one for each type), to simplify our discussions.}, 
can be viewed as a function $f_{\text{cta}}$, that given a value $v$ (say \codeq{Germany}) as input\footnote{Note that while some CTA-classifiers take an entire column $C$ as input, they also produce valid scores for individual values $v \in C$ (since CTA-classifiers need to make  predictions for single-value columns such as $C' = \{v\}$ too). %, which is necessary for error-detection, as predictions have to be made for individual values $v$.
},  can produce a classifier score $\text{CTA-classifier}(t_i, v)$ in the range of $[0, 1]$, to indicate the likelihood of $v$ in type $t_i$, written as: 
\begin{equation*}
%\label{eqn:cta}
f_{\text{cta}}(t_i, v) = \text{CTA-classifier}(t_i, v)
\end{equation*}
For example, we may get $f_{\text{cta}}(\text{\codeq{type-country}}, \text{\codeq{Germany}}) = 0.8$, and 
$f_{\text{cta}}(\text{\codeq{type-city}}, \text{\codeq{Germany}}) = 0.1$, from CTA-classifiers. 

Observe that  $f_{\text{cta}}(t_i, v)$ measures ``similarity'' between type $t_i$ and value $v$. To unify CTA with other column-type detection methods, we standardize $f_{\text{cta}}$ into a ``distance function'', written as $f^d_{\text{cta}}$:
\begin{equation}
\label{eqn:cta-d}
f^d_{\text{cta}}(t_i, v) = 1-  f_{\text{cta}}
\end{equation}
With this distance function, we can equivalently write $f^d_{\text{cta}}(\text{\codeq{type-country}}, \text{\codeq{Germany}})$ $ = 0.2$, and 
$f^d_{\text{cta}}(\text{\codeq{type-city}},$ $ \text{\codeq{Germany}})$ $ = 0.9$, etc., where a smaller distance indicates a closer association.% between value $v$ and type $t_i$.
%The classifier takes a column/value as input and computes its prediction score on each of the predetermined semantic types, and assigned it to the type of which it has the highest score. 
%Therefore, a evaluator in CTA domain is simply a function that computes the score of $v$ generated by a classifier (e.g., \textit{Sherlock}) on a specific type (e.g., country).

%$F_{\text{CTA}-t_1}(C)$

%$f_{\text{CTA-t1}}(C)$

%$f_{\text{cta}}(t_i, v)$

% $f_{\text{emb}}(c_i, v)$

% $f_{\text{pat}}(p_i, v)$

% $f_{\text{fun}}(f_i, v)$

% $f_{D_i}(v)$

%f^{\text{cta}}_{\text{t1}}(C)$

% Therefore, a CTA domain can be considered as the collection of values whose prediction scores exceed a predetermined threshold on a particular class.
% For example, a possible CTA domain could be the set of values whose prediction scores > 0.8 on class ``country''. 

\noindent
\underline{(2) Embedding-based methods~\cite{PSM14, RG19}.}
% CTA models uses embedding as features but requires a separate training process for each class, which does not scale. Using embedding directly can be scalable to many potential classes without separate training.
Text embedding, such as \textit{Glove}~\cite{PSM14} and \textit{SentenceBERT}~\cite{RG19}, are popular vector-based representations of text in NLP. In the embedding space, texts with similar semantic meanings (e.g., \code{month-names} like \codeq{january}, \codeq{feburary}, etc.) tend to cluster closely together, while those with unrelated meanings (e.g., \codeq{january} and  \code{color-names} like  \codeq{yellow}) are positioned further apart~\cite{PSM14, RG19, mikolov2013efficient}. 

Such embedding provides an effective method to detect semantic types. Specifically, it is natural to select a  ``\emph{centroid}'', say \codeq{january}, to represent the semantic-type we want to detect (in this case \code{month-name}), and for a given a column $C$, if most or all values $v \in C$ fall within a small radius of \codeq{january}, we may predict column $C$ as type \code{month-name} (implied by the centroid \codeq{january}).

%For example, in the embedding vector space. For example, , will all be clustered together and close to each other in the embedding space, while \code{color-names} like  \codeq{red}, \codeq{yellow}, etc., will be close together in a different region.

We view text-embedding as providing another function, 
$f^d_{\text{emb}}(c_i, v)$, that calculates the ``distance'' between a given value $v$, and a centroid $c_i$ (representing a semantic-type):
\begin{equation}
\label{eqn:emb-d}
f^d_{\text{emb}}(c_i, v) = \text{dist}(\text{emb}(c_i), \text{emb}(v))
\end{equation}
For example, let  $c$ = \codeq{january} be a centroid,  we may have 
$f^d_{\text{emb}}(c, \text{\codeq{february}}) = 0.1$, indicating the close proximity of the two values. Alternatively, let $c'$ = \codeq{yellow} be another centroid (for \code{color-name}), and we may have 
$f^d_{\text{emb}}(c', \text{\codeq{february}}) = 0.7$, showing that \codeq{february} is likely not in the same type as \codeq{yellow}. 
Note that $f^d_{\text{emb}}$ is already a distance-function, like $f^d_{\text{cta}}$ (Equation~\ref{eqn:cta-d}).

%are used to generate embedding vector for textual input, which are subsequently utilized for semantic type detection. In the embedding space, texts with similar semantic meanings tend to cluster closely together, while those with unrelated meanings are positioned further apart. The evaluator function in embedding domain computes the distance of $v$ to a specific center (e.g., ``january'') in a certain embedding space (e.g., \textit{Glove}).

\noindent
\underline{(3) Pattern-based methods~\cite{auto-validate, auto-tag, systemx, pattern-profiling}.} 
For machine-generated data with clear syntactic structures (e.g., \code{date}, \code{email}, \code{timestamp}, etc.), regex-like patterns can often detect semantic types~\cite{auto-validate, auto-tag, systemx, pattern-profiling}. 
For example, if most values in $C_7$ of Figure~\ref{fig:example-table-columns} follow the pattern \codeq{$\backslash$d\{1,2\}$/\backslash$d\{1,2\}$/\backslash$d\{4\}}, we may predict the column as type \code{date}. 

%Existing works in this line of research use regex patterns to detect column types, which has been proved to be effective for types with clear syntatical structures, such as email and timestamps. For pattern domains, a evaluator function simply checks if $v$ matches a specific pattern (e.g., ``$\backslash$[a-zA-Z]+$\backslash$d+'').

Similar to CTA and embedding, given a semantic type implied by pattern $p_i$ (e.g., \codeq{$\backslash$d\{1,2\}$/\backslash$d\{1,2\}$/\backslash$d\{4\}} for \code{date}), and  a value $v$, we can view the pattern-based detection as a different ``similarity'' function $f_{\text{pat}}(p_i, v)$ between value $v$, and a type represented by $p_i$:
\begin{equation*}
%\label{eqn:pat}
f_{\text{pat}}(p_i, v) = 
\begin{cases} 
1, & \text{if } v \text{ matches } p_i\\
0, & \text{if } v \text{ does not match } p_i
\end{cases}
\end{equation*}
We also normalize $f_{\text{pat}}(p_i, v)$ into a distance-function, $f^d_{\text{pat}}$:
\begin{equation}
\label{eqn:pat-d}
f^d_{\text{pat}}(p_i, v) = 1 - f_{\text{pat}}(p_i, v)
\end{equation}
For example, let $p =$ \codeq{$\backslash$d\{1,2\}$/\backslash$d\{1,2\}$/\backslash$d\{4\}}, $v_1 = $ \codeq{12/3/2020} and $v_2 =$ \codeq{new facility} in Figure~\ref{fig:example-table-columns}.  We have $f^d_{\text{pat}}(p, v_1) = 0$, indicating ``distance= 0''  between type $p$ and a compatible value $v_1$; and $f^d_{\text{pat}}(p, v_2) = 1$, indicating ``distance = 1'' between $p$ and an incompatible value $v_2$.

\noindent
\underline{(4) Function-based methods~\cite{auto-type, PWL21, validator, Dataprep-clean}.} Finally, various 
``validation-functions'' (in python and other languages) have been developed, to validate rich semantic types.   For example,  \code{credit-card-number} and \code{UPC-code} are not just random-numbers, but have internal check-sums and can be validated using special validation-functions\footnote{For example,  \url{https://yozachar.github.io/pyvalidators/stable/api/card/} for ``credit-card-number'', and \url{https://pypi.org/project/barcodenumber/} for type ``UPC-code''} (e.g., Luhn's checksum~\cite{validator-luhn}). Similarly, \code{date} and \code{timestamps} can also be validated  precisely with functions (in place of simple patterns)\footnote{For example, \url{https://docs.dataprep.ai/user_guide/clean/clean_date.html} and \url{https://gurkin33.github.io/respect_validation/rules/DateTime/} for ``date'' and ``timestamp''.}.
Such ``validation functions'' 
 are curated in popular open-source repositories like \textit{DataPrep} \cite{PWL21} and \textit{Validators} \cite{validator},  to reliably detect semantic column types.

%includes customized code that validates the correctness of various data types, such as URLs, dates, and SSN numbers. Examples of these functions includes libraries such as \textit{DataPrep} \cite{PWL21} and \textit{Validators} \cite{validator}.  Works in this direction harness these functions for column-type detection. Therefore, a evaluator in this domain checks if $v$ can be validated by a specific library function (e.g., \textit{validate\_date()}).
% is the set of values that can be validated by some specified function. 
% A possible function domain could be the set of values that can be validated by \textit{validate\_date}, a function in \textit{DataPrep} that verifies date strings.

For each validation-function $f_i$ (to validate a semantic-type), we similarly view it as a function $f_{\text{fun}}(f_i, v)$, that measures the ``similarity'' between value $v$ and a type represented by $f_i$:
\begin{equation*}
%\label{eqn:fun}
f_{\text{fun}}(f_i, v) = 
\begin{cases} 
1, & \text{if } f_i(v) \text{ returns true  }  \\
0, & \text{if } f_i(v) \text{ returns false  }  
\end{cases}
\end{equation*}
which can again be standardized into a distance-function, $f^d_{\text{fun}}$:
\begin{equation}
\label{eqn:fun-d}
f^d_{\text{fun}}(f_i, v) = 1 - f_{\text{fun}}(f_i, v)
\end{equation}
where a distance $f^d_{\text{fun}}(f_i, v) = 0$  indicates that a value $v$ is validated true by function $f_i$. %, and may belong to the type represented by $f_i$.
%where $f_{\text{fun}}$ measures whether $v$ can be validated by program-function $f_i$, and the result $s$ is either true or false.
For example, let $f_i$ be the \textit{validate\_date()} function. Then for $C_7$  in Figure~\ref{fig:example-table-columns}, we  have $f^d_{\text{fun}}(f_i, \text{\codeq{12/3/2020}}) = 0$,   and $f^d_{\text{fun}}(f_i, \text{\codeq{new facility}}) = 1$.

%\textbf{Discussions}.
Observe that different column-type detection methods  can  have \emph{overlapping coverage} in the types they detect -- for example, different CTA-classifiers (e.g., from Sherlock, Doduo, Sato, etc.) all have their own implementations to detect the same semantic type (e.g., \code{type-city}). 
Similarly, both pattern-based and function-based methods can detect similar types (e.g., \code{timestamps}).
We do not attempt to manually determine which method  is the best for a type $t$ -- we simply ingest all type-detection methods into our framework, which can be reasoned consistently to automatically select suitable \sdca constraints, which is a salient feature of \at.

\textbf{Domain-evaluation function}. Note that we intentionally characterize all column-type detection methods as distance functions between value $v$ and a type $t$ (e.g., $f^d_{\text{cta}}$, $f^d_{\text{emb}}$, $f^d_{\text{pat}}$, and $f^d_{\text{fun}}$ in Equation~\eqref{eqn:cta-d}-\eqref{eqn:fun-d}), %, that gives a measure of ``distance'' between a value $v$ and a type $t$,
%(in Equation~\eqref{eqn:cta-d},~\eqref{eqn:emb-d},~\eqref{eqn:pat-d}, and~\eqref{eqn:fun-d}), 
so that they can be reasoned consistently. 
Specifically, to quantify whether a value $v$ may be ``in'' vs. ``out of'' type $t_i$ for error-detection, we use a notion of ``\emph{domain-evaluation functions}'' that naturally generalizes these distance-functions.

% \subsection{Domain and domain-evaluation function}
% \label{subsec:domain-eval-func}
% Suppose that we know column $C$ is of type $t_i$. For error detection, we want to test whether a value $v \in C$ (say \codeq{germany} in $C_2$ of Figure~\ref{fig:example-table-columns}) may be an error in the context of type $t_i$ (say  \codeq{type-city}). To do that, we want to quantify whether a value $v$ may be ``in'' vs. ``out of'' type $t_i$, based on a notion of ``\emph{semantic domains}''.

% Intuitively, for a given type $t_i$,  its ``semantic domain'' defines the possible space of all valid values in $t_i$.  Depending on the type, its domain may be: (1) \underline{precisely defined}, e.g., for types like \emph{ip-address} or \emph{url} that have exact specifications of possible values; or (2)  \underline{fuzzily defined}, for in most cases (e.g., types involving natural-language), the domain of a type is usually \emph{imprecise}, with a fuzzy boundary separating what values fall in vs. out of the type. % (likely to be of type $t_i$), while what values fall outside of the domain boundary (likely to be  errors, if most other values in $C$ are of type $t_i$).  

% To characterize the ``domain boundary'' of a type $t$ for error-detection, we use a notion of ``\emph{domain-evaluation functions}'' that naturally generalizes the distance-functions for different column-type detection methods  in Equation~\eqref{eqn:cta-d}-\eqref{eqn:fun-d}.

\begin{definition}\label{def:func}
[\textbf{Domain-evaluation function}].
    Given a semantic type $t_i$ defined by an underlying column-type detection method (CTA, embedding, etc.), a \emph{domain-evaluation function} $f(t_i, v)$ measures the ``distance'' between  type $t_i$ and value $v$, where $f$ can be instantiated  as $f^d_{\text{cta}}$, $f^d_{\text{emb}}$, $f^d_{\text{pat}}$, and $f^d_{\text{fun}}$ in Equation~\eqref{eqn:cta-d}-\eqref{eqn:fun-d}. % also simply written as $f_{t_i}(v)$
\end{definition}

As a distance function, a smaller $f(t_i, v)$ naturally  indicates that $v$ is likely ``in'' the domain of type $t_i$, while a larger $f(t_i, v)$  indicates $v$ to be likely ``out'' of type $t_i$.

\section{\sdc}

%\subsection{\sdc}
\label{sec:sdc_struct}
Given the domain-evaluation functions $f(t_i, v)$ in Section~\ref{sec:preliminary}, which we will henceforth write as $f_{t_i}(v)$ for simplicity, we now describe a new class of data-quality constraints called \sdc (\sdca) we propose in this work.

\begin{definition}\label{def:SDC}
[\textbf{\sdc}]
    A Semantic-Domain Constraint  (SDC), denoted as $r_t = (P, S, c)$ for semantic type $t$, is a 3-tuple that consists of a \emph{pre-condition} $P$, a \emph{post-condition} $S$, and a \emph{confidence-score} $c$, where: 
    \begin{itemize}[leftmargin=*]
        \item The \underline{pre-condition} $P$: it determines whether the \sdca described in $r$ should apply to an input column $C$, defined as:
        \[
        P(C, f_t, d_{in}, m) = 
        \begin{cases} 
        \text{true} & \text{if } \frac{|\{v | v\in C,  f_t(v) \leq d_{in} \}|}{|\{v | v\in C\}|} \geq m, \\
        \text{false} & \text{otherwise}.
        \end{cases}
        \]
        When the fraction of values $v \in C$ with domain-evaluation function $f_t(v)$ no greater than an \emph{inner-distance threshold} $d_{in}$, denoted as $\frac{|\{v | v\in C,  f_t(v) \leq d_{in} \}|}{|\{v | v\in C\}|}$, is over a \emph{matching-percentage} $m$, the pre-condition $P$ evaluates true (in which case $r_t$ applies to $C$), .
        \item The \underline{post-condition} $S$: if the pre-condition $P$ evaluates true, it will be used to detect  values $v \in C$ as errors, whose domain-evaluation function $f_t(v)$ evaluates to be greater than an \emph{outer-distance threshold}, $d_{out}$, written as:
        \[
        S(C, f_t, d_{out}) = \{v |v \in C, f_t(v) > d_{out}\}
        \]
        \item The \underline{confidence} $c \in [0, 1]$: indicates the confidence of the  errors detected by the post-condition $S$.
    \end{itemize}
\end{definition}

%For a SDC $r$, the construction of its pre-condition and post-conditions are based on its underlying domain, while its confidence is set based on the training corpus, which we defer to Section~\ref{subsec:rule_quality_eval}. Specifically, given an underlying domain $D = \{v \in \mathcal{V} \, | \, f_D(v) \in A_{in}\}$, $r.pre$ is set to $\{C \,|\, f_D(v) \in A_{in}$ for at least $matching\mbox{-}percentage$  of  $v \in C \}$ where $matching\mbox{-}percentage \in [0,1]$ is a real number, and $r.post$ is set to $\{v \,|\, f_D(v) \in A_{out}\}$, where $A_{out}$ is a set of domain scores that is disjoint from $A_{in}$. 

% revised, removed for space
% \iftoggle{full}
% {
    \begin{figure}[t]
        \vspace{-4mm}
        \centering
        \includegraphics[width=0.27 \linewidth]{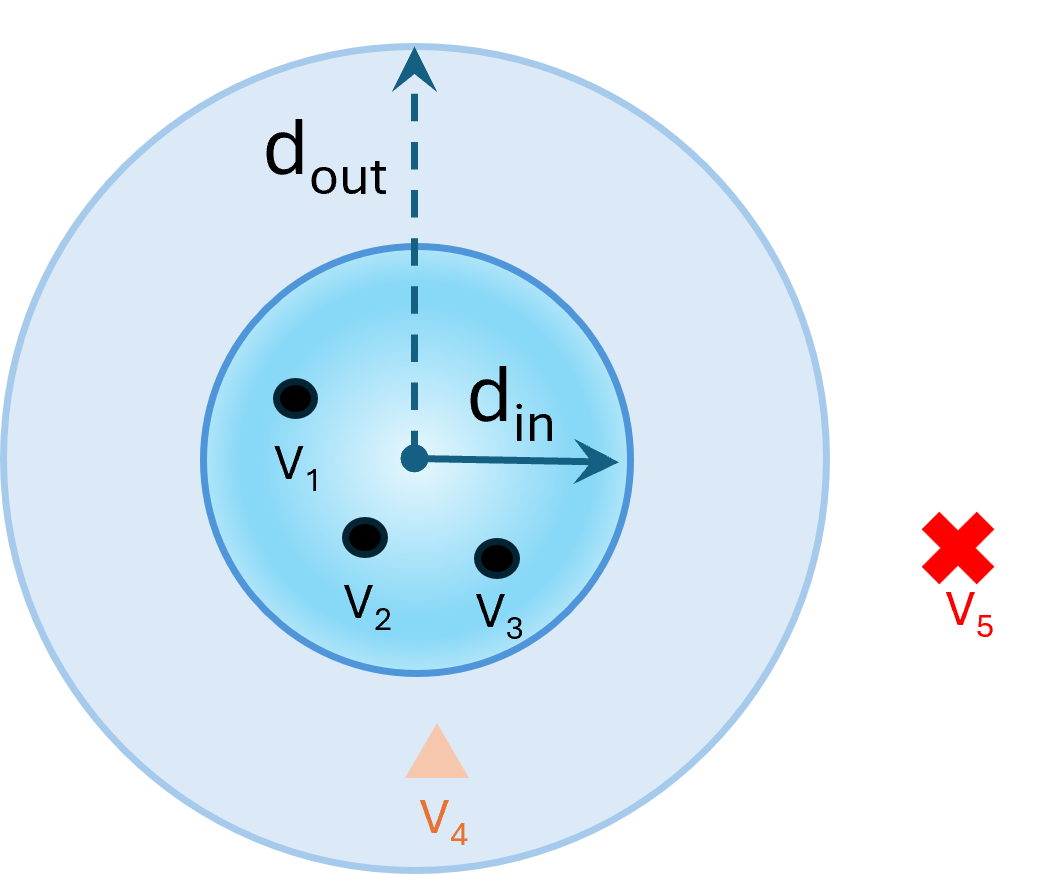}
        % \vspace{-3mm}
        \caption{Visual illustration of a constraint $r_t = (P, S, c)$, where the inner-ball with radius $d_{in}$ corresponds to the pre-condition $P$, the outer-ball with radius $d_{out}$ corresponds to the post-condition $S$. For a column $C = \{v_1, v_2, v_3, v_4, v_5\}$, $v_1$, $v_2$ and $v_3$ fall inside the inner-ball (indicating that these values are are likely in the domain of type $t$), while $v_5$ falls outside of the outer-ball (and likely not in the type $t$).}
        % \vspace{-3mm}
        \label{fig:example-balls}
    \end{figure}
    
    Intuitively, we can visualize any constraint $r_t = (P, S, c)$  for a type $t$, as two ``concentric balls'' depicted in Figure~\ref{fig:example-balls}, where the pre-condition $P$ corresponds to an inner ball of radius $d_{in}$, and the post-condition $S$ corresponds an outer ball of radius $d_{out}$, respectively. 
%}

The \underline{pre-condition $P(C, f_t, d_{in}, m)$} checks whether a given column $C$ is in the semantic domain of the type $t$ (before $r_t$ can apply). Specifically, it uses the domain-evaluation function $f_t(v)$ for type $t$, to calculate the fraction of values $v \in C$ that, when evaluated using $f_t(v)$, fall within the inner ball of radius $d_{in}$ (indicating that they belong to type $t$), written as $\frac{|\{v | v\in C,  f_t(v) \leq d_{in} \}|}{|\{v | v\in C\}|}$. 

% revised, removed for space
% \iftoggle{full}
% {
    In Figure~\ref{fig:example-balls} for instance, for an example column $C = \{v_1, v_2, v_3, v_4, v_5\}$, values $v_1$, $v_2$ and $v_3$ marked in circles fall inside the inner-ball, making this ratio $\frac{3}{5}=0.6$. If the matching-percentage $m=0.5$, then $C$ is determined to be in the domain of type $t$, so that the post-condition $S$ in $r_t$ will apply to $C$.
%}

The \underline{post-condition $S(C, f_t, d_{out})$} would then check whether there are any values $v \in C$ that fall substantially away from the inner ball, to be outside of the outer ball, written as $S = \{v |v \in C, f_t(v) > d_{out}\}$. If such values exist in $C$, values in $S$ will be predicted as \emph{errors}, with a confidence score $c$.

% revised, removed for space
% \iftoggle{full}
% {
    In Figure~\ref{fig:example-balls} for example, value $v_5$ marked by a red cross is outside of the outer-ball, and can be predicted as an error. Note that value $v_4$ marked as a triangle in the figure, falls outside of the inner-ball but inside of the outer-ball, will not be predicted as errors (because intuitively they are not sufficiently farther away from the inner-ball, to be reliably predicted as errors).
%}

\underline{Parameters.} Note that three parameters, $d_{in}$, $d_{out}$, $m$, are used in each constraint $r_t$. These are clearly hard to set manually, especially when there are many semantic-types, where each type $t$ has its own optimal parameters. A key technical challenge in this work is to automatically learn these parameters from real tables, both efficiently and with quality-guarantees (Section~\ref{sec:autotest}).

%As an analogy, a SDC can be considered as two concentric circles in the value space. The pre-condition corresponds to the inner sphere, which circles out the normal values that naturally belongs to the underlying domain. The post-condition, on the other hand, is the outer circle, beyond which all values are considered as errors. 

% revision to remove example for space
% \iftoggle{full}
% {
    We re-visit the table in  Figure~\ref{fig:example-table-columns} to illustrate \sdca below.
    
    \begin{example}
    \label{ex:inference}
    Consider $r_4$ in Table~\ref{tab:rule_example}. Its domain-evaluation function $f_t$ is based on \textit{Sentence-BERT} embedding distance, and its type $t$ is implied by the centroid \codeq{seattle} (of type \code{city}). The pre-condition $P$ has an inner-ball radius of $d_{in} = 1.2$, meaning values within distance $1.2$ to \codeq{seattle} are believed to be in the same type. Evaluating this $r_4$ against $C_4$ in Figure~\ref{fig:example-table-columns}, we find 95\% of values in $C_4$ to be within the inner-ball, greater than the required matching-percentage $m=80\%$, ensuring that $r_4$ applies to $C_4$. Checking two values outside of the inner-ball, \codeq{shakopee} (an uncommon name) and \codeq{farimont} (a typo), we find the latter to have a distance greater than $d_{out} = 1.35$, therefore falling outside of the outer-ball specified in its post-condition $S$, suggesting \codeq{farimont} may be an error with confidence $r_4.c = 0.88$. (\codeq{Shakopee} falls inside the outer-ball, and is not predicted as an error).
    
    Note that it can be checked that $r_4$ does not apply to any other columns in Figure~\ref{fig:example-table-columns}, as not enough fraction of values in these columns fall in the inner-ball to meet the matching-percentage $m$ requirement (intuitively they are not of type of \code{city}).
    
    Also note that there exists another constraint $r_4'$ that has the same centroid \codeq{seattle}, but with a smaller inner-ball radius (1.1), a larger outer-ball radius (1.4), and a larger $m$ (0.9), which  intuitively is a stricter and more confident version of $r_4$, with confidence $0.93$. It also triggers on \codeq{farimont} in $C_4$, and since $r_4'$ has a higher confidence than $r_4$ (0.88), it assigns  \codeq{farimont} a even higher confidence-score. There can be many such variants for the same semantic-type (e.g., $r_4$ and $r_4'$) in Table~\ref{tab:rule_example}, with different parameter configurations, and their corresponding calibrated confidence scores. We assign the confidence of a prediction based on its most confident \sdca (e.g., 0.93 for \codeq{farimont}), which is natural.
    
    Similarly, another constraint $r_2$ based on CTA-classifiers can detect the incompatible value \codeq{germany} from $C_2$. Recall that CTA-classifiers produce similarity-scores, which we standardize into distances (Equation~\eqref{eqn:cta-d}). The pre-condition requires a score greater than 0.55, which would translate to an inner-ball radius of $d_{in} = 0.45$. We find over $m=90\%$ of $C_2$ to be in the inner-ball, making $r_2$ applicable on $C_2$. Since \codeq{germany} lays outside of the outer-ball $d_{out} = 1-0.05=0.95$, making it an error predicted by $r_2$.
    
    As a final example, $r_6$ is based on patterns, where match=1 and 0 are transformed into distance of 0 and 1, respectively, using distance functions (Equation~\eqref{eqn:pat-d}), so that we have $d_{in} = 0$ and $d_{out} = 1$, respectively. This $r_6$  will  trigger on column $C_6$, since over $m=95\%$ of values match the pattern (distance=0), which fall inside the inner-ball. The only non-matching value \codeq{0.05\%} in $C_6$ has distance=1, which is outside of the outer-ball and detected as error.
    \end{example}
%}

% \begin{example}
% Consider $r_3$ in Table~\ref{tab:rule_example}. It's pre-condition and post-condition are shown in the table, with the evaluator $f_D(v)$ measures the \textit{Glove} embedding distance to ``january'', $matching\mbox{-}percentage = 80\%$, $A_{in} = [0, 4.0)$, and $A_{out} = (5.5, +\infty)$. 
% Using our analogy, the pre-condition can be considered as a inner circle centered at ``january'' with a radius of 4.0, and the post-condition is the outer circle with a radius of 5.5.
% In Figure~\ref{fig:example-table-columns}, column $C_3 \in r_3.pre$ since most of its values are months, whose \textit{Glove} embedding distances to ``january'' are within 4.0 and falls inside the inner circle. On the other hand, the typo \codeq{febuary} falls out of the outer circle and is thus considered as an error by $r_3$. Note that $r_3$ will not be applied on the remaining columns in Figure~\ref{fig:example-table-columns} since they do not fall in the pre-condition of $r_3$. 
% \end{example}

\underline{Why this design of \sdca.}
We design  \sdca with this structure for the following reasons. First, it mimics the human intuition of identifying data errors --  given a table in Figure~\ref{fig:example-table-columns}, humans would read values in a column, to first identify its semantic type (e.g., \code{city} vs. \code{date}, which is a ``pre-condition''), before using a fuzzy notion of ``domain'' of each type to identify  errors (post-conditions). 
Our design of \sdca  mimics the reasoning process -- by imposing a inner-ball/outer-ball ``structure'', we constrain  the search space of constraints using the strong ``prior'', effectively reducing the problem to a more tractable form that focuses solely on learning a suitable set of parameters ($d_{in}$, $d_{out}$, $m$).

Because \sdca is based on semantic-domains, the resulting constraints are ``explainable'' as they are often associated with types (e.g., the prediction of \codeq{germany} in $C_2$ can be explained using the CTA-classifier for \code{state}), making predictions interpretable. 

Finally,  the \sdca framework is extensible to different column-type detection methods, and it is easy to add/remove constraints in a white-box fashion, making it easy to deploy and operationalize. % in settings like in Figure~\ref{fig:excel-demo}.

%utilize existing domain detection for error detection, consistent with human intuition (or the way human reasons)

%constrain the search space, using the prior

%explainable

%unifying and extensible.

% Specifically, for a domain $D$, We generate preconditions in the form $\{c \,|\, f_D(v) \in A_{in}$ for at least $ratio$\%  of  $v \in c \}$, and constraints in the form $\{v \,|\, f_D(v) \in A_{out}\}$, where $ratio$ is a real number and $A_{in}$, $A_{out}$ are two disjoint set of domain scores. 
% For each domain type, we generate a large set of rules candidates by varying parameters related to that type. 

% \yeye{ after preliminary (domains), may need a section  to introduce the SDC construct we design, using the example in Table 1.}

% \yeye{move some of section 5.1 here}

%\section{Problem Statement: Learning \sdca}
%\label{sec:problem_definition}
\textbf{Problem Statement: Learning \sdca}.
In this work, we want to ``learn'' high-quality \sdcas  with appropriate parameters from a large table corpus, so that they can cover diverse semantic-types (e.g., in Table~\ref{tab:rule_example}), and be readily applicable to new and unseen tables.

We leverage a large corpus of tables $\mathcal{C}$ (e.g., millions of tables crawled from the web and enterprises), and model them as a collection of individual columns $\mathcal{C} = \{C\}$. %Note that the corpus $\mathcal{C}$ is \emph{unlabelled}, but assumed to be clean
%
%The main problem we study in the paper is how to learn a set of high-quality \sdca from large, unlabeled table corpora.  
%Since \sdca are applied on individual columns for error detection, we preprocess the training corpus $\mathcal{C}$ consisting of single columns extracted from the table corpus with some simple pruning criteria.
Ideally, we want to leverage $\mathcal{C}$ to learn a set of high-quality \sdcas, denoted by $R$, such that: 
\begin{itemize}[leftmargin=*]
    \item[] (1) \underline{\emph{recall}} of $R$ is maximized, or $R$ should detect as many errors as possible on unseen test set $\mathcal{C}_{test}$;
    \item[] (2) \underline{\emph{false positive rate (FPR)}} of $R$ is minimized, for $R$ should trigger few false-positives on $\mathcal{C}_{test}$;
    \item[] (3) \underline{\emph{size}} of $R$ is not exceedingly large for latency/efficiency reasons (the size of Table~\ref{tab:rule_example} is limited).
\end{itemize}

We give a high-level sketch of our problem below, which will instantiate into concrete problem variants in later sections.

\begin{definition}\label{def:prob_def}
[\textbf{Learning \sdc}]. Given a corpus $\mathcal{C}$, a size constraint $B_{size}$, and a FPR threshold $B_{FPR}$,  find a set of SDCs  $R$ that maximizes $\text{Recall}(R)$, while satisfying $\left | R \right | \leq B_{size}$ and $FPR(R) \leq B_{FPR}$, written as:
    \begin{align}
    \small
         \max_{R} ~~~~ & \text{Recall}(R)         \label{eqn:general_obj} \\
        \text{s.t. }        & \left | R \right | \leq B_{size}  \label{eqn:general_size} \\
                            & FPR(R) \leq B_{FPR} \label{eqn:general_fpr}
    \end{align}
\end{definition}

Note that in balancing the three requirements, we want to bound FPR (e.g., false-positive rate should not exceed $B_{FPR} = 1\%$, for scenarios like Figure~\ref{fig:excel-demo} has strict precision requirements), and the size of $R$ (e.g., $|R|$ should not exceed $B_{size} = 10000$) to limit its memory footprint and make inference efficient, while  maximizing recall as much as possible.

\section{\at: Learn \sdca using tables}
\label{sec:autotest}
We now describe our proposed \at that learns high-quality \sdcas from a large table corpus $\mathcal{C}$ in an unsupervised manner.
%For the ease of illustration, in the rest of this paper, we refer to the pre-condition, post-condition and confidence of the \sdca $r$ as $r.P$, $r.S$ and $r.c$, respectively. Besides, we use $r.S(C)$ to denote the values reported by \sdca $r$ on column $C$ (using post-condition $r.S$). 

% revised, removed for space
% \iftoggle{full}
% {
    \begin{figure}[t]
     \vspace{-2mm}
        \centering
        \includegraphics[width=0.7 \linewidth]{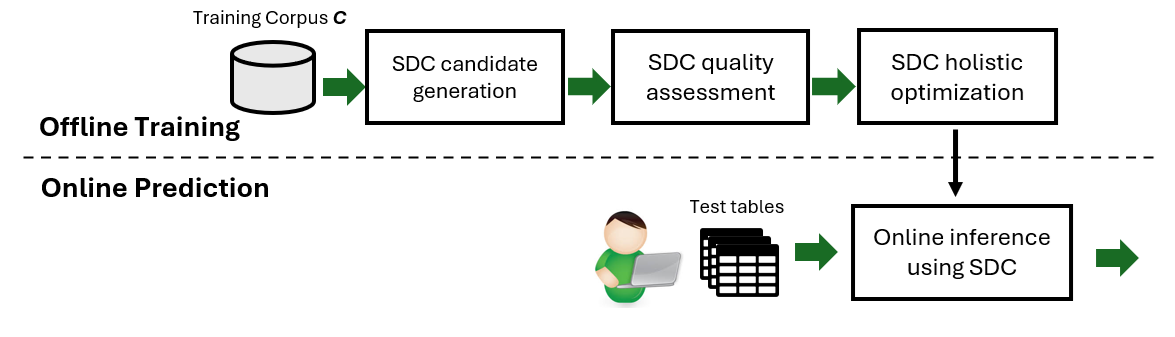}
        % \vspace{-8mm}
        \caption{Architecture diagram of \at}
         \vspace{-3mm}
        \label{fig:architecture_diagram}
    \end{figure}
%}

% revised, removed for space
% \iftoggle{full}
% {
    Figure~\ref{fig:architecture_diagram} shows the architecture of \at, which has an \underline{\textit{offline training stage}}, where \sdcas are learned from given a corpus $\mathcal{C}$, and an \underline{\textit{online prediction stage}}, where the learned \sdcas are applied on new test tables to make prediction. 
    Since online prediction is relatively straightforward and already explained in Example~\ref{ex:inference}, we focus on the offline training part in this section.

The offline training has three steps, which we will describe in turn below. At a high level,  we will first generate a set of \sdca candidates (Section~\ref{subsec:rule_cand_generation}), and assess their quality using principled statistical tests (Section~\ref{subsec:rule_quality_eval}), before we perform holistic optimization of the problem stated in Definition~\ref{def:prob_def}, to select an optimal set of \sdcas with quality guarantees (Section~\ref{subsec:selectionWithFPRGuarantees}). %We will describe these steps in offline training in turn below.

\subsection{\sdca Candidate Generation}
\label{subsec:rule_cand_generation}
Our first  step in offline training is a preprocessing step that generates a large set of \sdcas candidates. 

Recall that in Definition~\ref{def:SDC}, each \sdca has 4 parameters: domain-evaluation function $f_t$, inner-distance $d_{in}$, outer-distance $d_{out}$, and matching-percentage $m$ (color-coded in Table~\ref{tab:rule_example} for readability). 

Note that  $f_t$ may be instantiated using different ``domain-evaluation functions''  (Definition~\ref{def:func}) for different column-type detection methods, namely $f^d_{cta}$, $f^d_{emb}$, $f^d_{pat}$ and $f^d_{fun}$ in Equation~\eqref{eqn:cta-d}-\eqref{eqn:fun-d}, for CTA, embedding, patterns, and functions, respectively.  
Specifically, we instantiate $f_t$ as follows:
% Table~\ref{tab:sdc_generation_summary} provides a comprehensive summary of this process.
% Each \sdca type uses a domain-evaluation function instantiated from a corresponding distance function (i.e., $f^d_{cta}$, $f^d_{emb}$, $f^d_{pat}$ and $f^d_{fun}$) described in Section~\ref{sec:preliminary}, which in turn can be instantiated using different semantic-detection methods. 
% The methods included in this work for each distance function are:
\begin{itemize}[leftmargin=*]
    \item \underline{CTA}. We use the 78 classifiers in \textit{Sherlock}~\cite{sherlock} (designed to semantic-types in DPBpedia), and the 121 classifiers in \textit{Doduo}~\cite{doduo} (for semantic-types in Freebase), for a total of 199 $f^d_{cta}$ functions.  %are used as the classifier function CTA-classifier($\cdot$) in $f^d_{cta}$.
    \item \underline{Embedding}. We use the \textit{Glove}~\cite{PSM14} and \textit{SentenceBERT}~\cite{RG19} embedding, and randomly sample 1000 values as centroids (which may be values like \codeq{seattle} and \codeq{january} shown in Table~\ref{tab:rule_example}),  to create a total of 2000  $f^d_{emb}$ functions.
    \item \underline{Pattern}. We generate common patterns observed in our corpus $\mathcal{C}$, for a total of 45 $f^d_{pat}$ functions.
    \item \underline{Function}. We use validation-functions in \textit{DataPrep}~\cite{PWL21} and \textit{Validators}~\cite{validator}, as 8 $f^d_{fun}$ functions.
\end{itemize}

For parameters $d_{in}$, $d_{out}$, and $m$, we perform grid-search and enumerate parameters using fixed step-size (e.g., the matching-percentage $m$ is enumerated with a step-size of 0.05, or $m \in \{1.0, 0.95, 0.9, \ldots \}$, and $d_{in}$/$d_{out}$ are enumerated similarly). This generates a total of over 100,000 candidate \sdcas. %parameters and domain-specific parameters, and generate a \sdca candidate for each combination. 

Since these \sdca candidates are enumerated in an exhaustive manner, only a small fraction of appropriately parameterized \sdcas are suitable for error-detection, which we will identify using (1) statistical  tests and (2) principled optimizations, explained below.

% \begin{example}
%     Take CTA-based \sdca candidates (which use $f^d_{cta}$ as the domain-evaluation function) as an example. Assume that common parameters $m$, $d_{in}$ and $d_{out}$ each takes 5 possible values. For classifier \textit{Sherlock}, it has 78 semantic types, so the domain-specific parameter $t_i$ in $f^d_{cta}$ has 78 different values. We can generate 9,750 combinations for these parameters. %resulting in the same number of \sdca candidates. 
%     Similarly, \textit{Doduo} has 121 possible semantic types, which corresponds to 15,125 candidates.
% \end{example}

% It is worth mentioning that while generating and assessing a large number of candidates can be time-consuming, this process only needs to be performed once offline, making the runtime acceptable. 
% Furthermore, in \iftoggle{full}{Appendix~\ref{apx:rule_gen_optim}}{\cite{full}}, we present several optimization techniques that significantly speed up this process by pruning low-quality candidates early on.

\subsection{\sdca Quality Assessment by Statistical Tests}
\label{subsec:rule_quality_eval}
In this section, we take all \sdca candidates, and use statistical hypothesis tests to assess the quality of each candidate $r$.

% As can be expected, most candidates in $R_{can}$ are ineffective for error detection. Therefore, we develop several metrics to assess the quality of a candidate. Candidates that do not meet the standard set on these metrics are considered low-quality and pruned from $R_{can}$.

Given an \sdca $r = (P, S, c)$, where $r.P$ is its pre-condition, $r.S$ its post-condition, and $r.c$ its confidence, from Definition~\ref{def:SDC}.  
We say a column $C$ is ``\underline{\textit{covered by}}'' $r$, if $r.P(C) = $ true (i.e., more than $m$ fraction of values in $C$ fall inside the inner-ball with radius $d_{in}$), in which case $C$ is regarded as ``in the semantic domain'' specified by the pre-condition $r.P$.

Similarly, we say a column $C$ is ``\underline{\textit{triggered by}}'' $r$, if $r.S(C) = \text{true}$, meaning that the post-condition $r.S$ is producing non-empty results as detected errors in $C$ (i.e., there exists some $v \in C$ that fall outside of the outer-ball with radius $d_{out}$).

% revised, remove for space
% \iftoggle{full}
% {
    \begin{example}
        \label{ex:cover-trigger}
        We revisit Example~\ref{ex:inference}. Recall that $r_4$ in Table~\ref{tab:rule_example}  can be used to detect the error \codeq{farimont} in $C_4$ of Figure~\ref{fig:example-table-columns}.    
        We say that $C_4$ is ``covered by'' $r_4$, as $C_4$'s semantic domain (\codeq{city}) matches the domain specified in the pre-condition of $r_4$ (an inner-ball centered at \codeq{seattle}). No other columns in Figure~\ref{fig:example-table-columns} can be ``covered by''  $r_4$ (as they are not  columns relating to \codeq{city}, or $r_4.P(C) =$ false). 
    
        Because $r_4$ in Table~\ref{tab:rule_example} can detect an error \codeq{farimont} in $C_4$, we say $C_4$ is also ``triggered by'' $r_4$ (since $r_4.S(C_4) = $ true). On the other hand, if we remove \codeq{farimont} from $C_4$,  the resulting column will no longer be ``triggered by'' $r_4$, as no more error can be detected. % ($r_4.S(C'_4) = $ false).
    \end{example}
%}

Given a large corpus $\mathcal{C}$ and $r$, we can analyze $r$'s behaviour on  $\mathcal{C}$, using a \textit{contingency table}~\cite{kateri2014contingency, everitt1992contingency} shown in Table~\ref{tab:contingency}, where: 
\begin{itemize}[leftmargin=*]
    \item $\left | \mathcal{C}^r_{C, {T}} \right | = \{ C | C \in \mathcal{C}, r.P(C) = \text{true}, r.S(C) = \text{true}\}$ denotes the number of columns in $\mathcal{C}$ that are both covered by and triggered by $r$ ($C$ is ``in the semantic domain'' of $r$, with errors detected). 
    \item $\left | \mathcal{C}^r_{C, \overline{T}} \right |= \{ C | C \in \mathcal{C}, r.P(C) = \text{true}, r.S(C) = \text{false}\}$ denotes the the number of columns in $\mathcal{C}$ that are covered by, but not triggered by $r$ ($C$ is ``in domain'' for $r$, with no errors detected). 
    \item Similarly, $\left | \mathcal{C}^r_{\overline{C}, T} \right |$ and $\left | \mathcal{C}^r_{\overline{C}, \overline{T}} \right |$ correspond to the number of columns in $\mathcal{C}$ that are not covered by $r$ ($C$ is ``not in domain'' for $r$), but with and without detection in the post-condition $r.S(C)$, respectively.
    % \item $\left | \mathcal{C}^r_{\overline{C}, T} \right |= \{ C | C \in \mathcal{C}, r.P(C) = \text{false}, r.S(C) = \text{true}\}$ to denote the the number of columns in $\mathcal{C}$ that are not covered by, but triggered by $r$ ($C$ is ``not in domain'' for $r$, with errors detected). 
    % \item $\left | \mathcal{C}^r_{\overline{C}, \overline{T}} \right |= \{ C | C \in \mathcal{C}, r.P(C) = \text{false}, r.S(C) = \text{false}\}$ to denote the the number of columns in $\mathcal{C}$ that are not covered by or triggered by $r$ ($C$ is ``not in domain'' for $r$, with no errors detected). 
\end{itemize}

Note that  the subscript $C$ and $T$ in these notations would correspond to ``cover'' and ``trigger'', respectively. 
We can see that the top-left %and bottom-left entries 
entry of Table~\ref{tab:contingency}  (covered and triggered), corresponds to  set of columns that would be predicted as having errors by $r$.

Using the contingency table, we perform statistical analysis to: (1) find suitable inner/outer-balls in $r$ that can naturally separate ``in-domain'' vs. ``out-of-domain'' columns, and (2) set each $r$'s confidence by the percentage of false-positives it reports among the covered columns. We will explain each in turn below.

\underline{(1) Find suitable inner/outer-balls using effect-size (Cohen's h).} 
Recall that in Section~\ref{subsec:rule_cand_generation}, we exhaustively enumerate \sdca candidates with different parameters (inner/outer-ball, centroid, etc.), and the hope is that using an unsupervised analysis of the corpus $\mathcal{C}$, we can identify good \sdca that are suitably parameterized.

\begin{figure}[t]
    \vspace{-2mm}
    \centering
    \includegraphics[width=0.27 \linewidth]{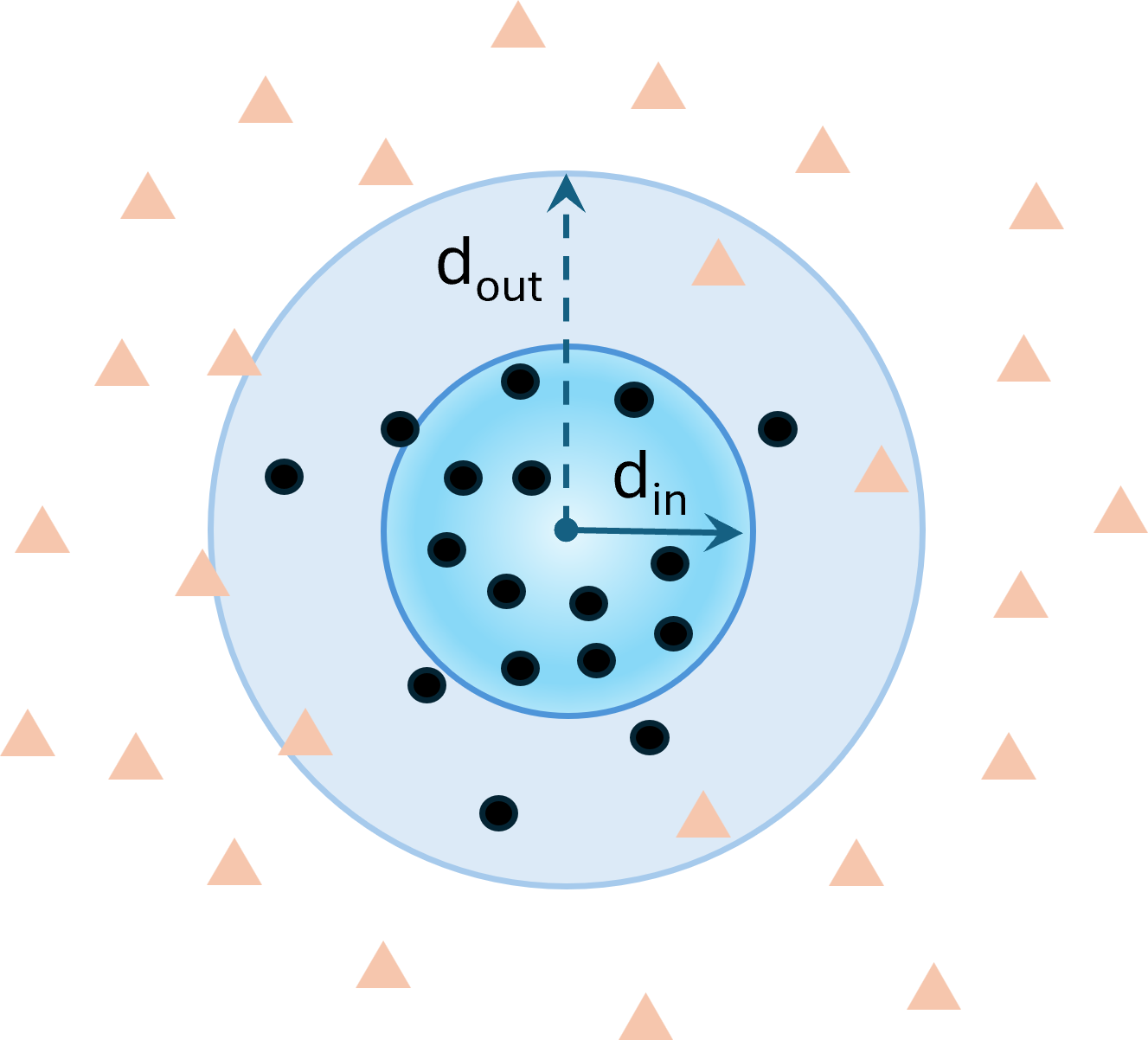}
     \vspace{-2mm}
    \caption{A visualization of inner/outer-balls that find ``natural separation'' of semantic-domains, in the universe of all values. Black-dots represent ``in-domain'' values, triangles represent ``out-of-domain'' values.}
     \vspace{-2mm}
    \label{fig:ideal-balls}
\end{figure}

\begin{table}[t]
%\vspace{-2mm}
    \caption{Contingency table for an example \sdca $r$, where we perform statistical tests to determine $r$'s efficacy.}
\vspace{-2mm}
    \label{tab:contingency}
    \scalebox{0.8}
    {
        \centering
        \begin{tabular}{|>{\centering\arraybackslash}p{4cm}|>{\centering\arraybackslash}p{4cm}|>{\centering\arraybackslash}p{4cm}|} \hline
              & \makecell{cols \underline{covered by} $r$ \\ (``in domain'' columns)} &             
              \makecell{cols \underline{not covered by} $r$ \\ (``out of domain'' columns)}  \\ \hline
            
            \makecell{cols \underline{triggered by} $r$ \\ (error detected) } & $\left | \mathcal{C}^r_{C, T} \right |=10$ & $\left | \mathcal{C}^r_{\overline{C}, T} \right |=160,000$ \\ \hline
    
            \makecell{cols \underline{not triggered by} $r$ \\ (no error detected)} & $\left | \mathcal{C}^r_{C, \overline{T}} \right |=990$ & $\left | \mathcal{C}^r_{\overline{C}, \overline{T}} \right |=40,000$ \\ \hline
            
        \end{tabular}
    }
\vspace{-4mm}
\end{table}

As we analyze these candidates, we know that a good \sdca $r$ for a semantic domain $t$ should ideally have an inner-ball with radius $d_{in}$ tightly enclose most ``in-domain'' values, and an outer-ball with radius $d_{out}$ that can filter out most ``out-of-domain'' values, like visualized in Figure~\ref{fig:ideal-balls}, where dots and triangles represent in-domain and out-of-domain values, respectively.

More specifically, given a table corpus $\mathcal{C}$, when we compute the contingency table for a candidate $r$ using $\mathcal{C}$, like shown in Table~\ref{tab:contingency},  an ideal $r$ with a suitable inner-ball/outer-ball should ``cover'' a good number of columns in $\mathcal{C}$ (e.g., a domain like \codeq{city} will cover many columns in $\mathcal{C}$), reflected by  a large $\left | \mathcal{C}^r_{C, {T}} \right | + \left | \mathcal{C}^r_{C, \overline{T}} \right |$ on the left of the contingency table, and at the same time should rarely ``trigger'' on columns in $\mathcal{C}$, reflected by a small $\left | \mathcal{C}^r_{C, {T}} \right |$ at top-left of the table ($r$ should rarely trigger on columns in $\mathcal{C}$, because the columns we harvested from relational sources are generally clean and error-free -- our manual analysis suggests that over 98\% columns in $\mathcal{C}$  are without errors). In effect, we are looking for $r$ whose ratio $\rho(r) = \left | \mathcal{C}^r_{C, {T}} \right |$ / $\left( \left | \mathcal{C}^r_{C, {T}} \right | + \left| \mathcal{C}^r_{C, \overline{T}} \right| \right)$ is small.

In contrast, if the inner/outer-ball are too small or too large (compared to the ideal balls in Figure~\ref{fig:ideal-balls}), the separations between ``in-domain'' vs. ``out-of-domain'' are no longer clean, and the ratio $\rho(r)$ will be indistinguishable from the same ratio  for the vast majority of  columns that are ``out-of-domain'' (the two entries on the right of Table~\ref{tab:contingency}), written as: $\overline{\rho}(r) = \left | \mathcal{C}^r_{\overline{C}, {T}} \right |$ / $\left( \left | \mathcal{C}^r_{\overline{C}, {T}} \right | + \left | \mathcal{C}^r_{\overline{C}, \overline{T}} \right | \right)$. 

%we will ``over-trigger'' on columns in $\mathcal{C}$ that are error-free, leading to a large $\left | \mathcal{C}^r_{C, {T}} \right |$ and a large $\rho$. If the inner-ball/outer-ball are too  large, they can no longer cleanly separate ``in-domain'' vs. ``out-of-domain'' values, making 
%``in-domain'' indistinguishable from ``out-of-domain'' columns. 

Motivated by this observation, we perform statistical tests on $\rho(r)$ and $\overline{\rho}(r)$ in the contingency table (Table~\ref{tab:contingency}), to test their  \textit{effect size}~\cite{KP12}, which is a principled measure of the magnitude of difference between $\rho(r)$ and $\overline{\rho}(r)$ -- namely, the larger the difference between the two ratios, the better $\rho(r)$ can ``stand out'' from the background noise ratio $\overline{\rho}(r)$, indicating a clean in-domain vs. out-of-domain  separation using $r$. We use \textit{Cohen's h}~\cite{cohen2013statistical} to evaluate the effect size of $r$, defined as:
    \begin{equation}\label{eqt:cohenh}
    h(r) = 2 \left( \arcsin{\sqrt{\rho(r)}}
    -\arcsin{\sqrt{\overline{\rho}(r)}} \right)
    \end{equation}    
    % \begin{equation}\label{eqt:cohenh}
    % \small
    % h = 2(\arcsin{\sqrt{\frac{\left | \mathcal{C}^r_{\overline{C}, T} \right |}{\left | \mathcal{C}^r_{\overline{C}, T} \right | + \left | \mathcal{C}^r_{\overline{C}, \overline{T}} \right |}}}
    % -\arcsin{\sqrt{\frac{\left | \mathcal{C}^r_{C, T} \right |}{\left | \mathcal{C}^r_{C, T} \right | + \left | \mathcal{C}^r_{C, \overline{T}} \right |}}})
    % \end{equation}    
Cohen's h has precise statistical interpretations, where $h \geq 0.8$ indicates large effect size~\cite{cohen2013statistical}, which we use to identify \sdca candidates with suitably parameterized inner/outer-balls that would correspond to natural domains (e.g., Figure~\ref{fig:ideal-balls}).

% revised, remove for space
% \iftoggle{full}
% {
    \begin{example}
        \label{ex:cohens-h}
        The constraint $r_4$ in our running example has a contingency table shown in Table~\ref{tab:contingency}. It ``covers'' 1000 columns in $\mathcal{C}$ (columns relating to \codeq{cities}), of which 10 are ``triggered'' (e.g., Column $C_4$ in Figure~\ref{fig:example-table-columns}). We can calculate $\rho(r_4) = \frac{10}{(990+10)} = 0.01$, which is substantially smaller than the background ratio $\overline{\rho}(r_4) = \frac{160,000}{(160,000 + 40,000)} = 0.8$. Cohen's h would confirm that $r_4$ indeed has a large ``effect-size'' of $h(r_4) = 2.01$. %, that would indeed corresponds to a good constraint. 
        %which is considered relatively large. Indeed, we can observe that it is triggered on only 1\% of the columns it covers, compared to an 80\% triggering frequency on the columns it does not cover. 
    \end{example}
%}

Furthermore, we perform  ``\emph{statistical significance test}'' (complementary to effect size in what is known as power-analysis~\cite{statistical-tests}), using the standard Chi-squared tests~\cite{chi-squared} on the contingency table (Table~\ref{tab:contingency}). We discard $r$ whose $p$ value is not significant at 0.05 level.

% A \sdca is pruned from $R_{can}$ if its effect size is below a threshold.
% In our experiment, this threshold is set to 0.8 by default, which corresponds to a large effect size in statistical terms \cite{cohen2013statistical}. 

\underline{(2) Estimate $r$'s confidence (Wilson's score intervals).} Recall that each \sdca $r=(P, S, c)$ has a confidence score $c$ (shown with examples in the last column of Table~\ref{tab:rule_example}), which is the probability of $r$ not producing false-positives among the ``in-domain'' columns it covers, that we calibrate using $\mathcal{C}$ with precise statistical interpretations as follows.

Specifically, note that the ratio $\hat{c} = \left | \mathcal{C}^r_{C, \overline{T}} \right |$ / $\left( \left | \mathcal{C}^r_{C, {T}} \right | + \left| \mathcal{C}^r_{C, \overline{T}} \right| \right)$ calculated from our contingency table is exactly an unbiased estimator of $c$. 
However, because both $\left | \mathcal{C}^r_{C, {T}} \right |$ and $\left | \mathcal{C}^r_{C, \overline{T}} \right |$ (the left two entries of Table~\ref{tab:contingency}) can be ``rare events'' with small counts, whose $\hat{c}$ ratio   is therefore susceptible to  over- and under-estimation on small samples. In order to guard against this, we produce a ``safe'' lower-bound of $c$ (as it is better to under-estimate $c$ than over-estimate it, to avoid false-positives),  using binomial confidence interval of $\hat{c}$, and specifically we use \textit{Wilson score interval}~\cite{wilson1927probable} to produce a lower-bound estimate\footnote{Note that because the corpus $\mathcal{C}$ is not perfectly clean, the true number of false-triggers is bound to be smaller than the current estimate of {\tiny $\left | \mathcal{C}^r_{C, {T}} \right |$} using $\mathcal{C}$. However, because this is in the denominator of $\rho(r)$, it does not affect our lower-bound analysis, as an over-estimate of {\tiny $\left | \mathcal{C}^r_{C, {T}} \right |$} still leads to a conservative lower-bound of $c$, which is what we want.} of the confidence $c$ of a candidate $r$ as:
    \begin{equation}\label{eqn:wilson-confidence}
    \small
     c = 1 - \frac{\left | \mathcal{C}^r_{C, T} \right | + \frac{1}{2}z^2}{\left | \mathcal{C}^{r}_{C} \right | + z^2} - \frac{z}{\left | \mathcal{C}^{r}_{C} \right | + z^2}\sqrt{\frac{\left | \mathcal{C}^r_{C, T} \right | \cdot \left | \mathcal{C}^{r}_{C,\overline{T}} \right |}{\left | \mathcal{C}^{r}_{C} \right |} + \frac{z^2}{4}}
    \end{equation}
where $\left | \mathcal{C}^{r}_{C} \right | = \left | \mathcal{C}^{r}_{C, T} \right | + \left | \mathcal{C}^{r}_{C, \overline{T}} \right |$, and $z=1.65$ is the normal interval width at 95\% confidence level. 

%$r$'s confidence is calculated using a lower-bound analysis of the fraction of true-positive detection over , in the   false-positive columns it reports, so that we can estimate the accuracy of each $r$ for downstream optimization. 

%Recall that our ratio $\rho(r) = \left | \mathcal{C}^r_{C, {T}} \right |$ / $\left( \left | \mathcal{C}^r_{C, {T}} \right | + \left| \mathcal{C}^r_{C, \overline{T}} \right| \right)$ is exactly an unbiased estimator of the percentage of false-positives reported $r$ among its covered columns on the training corpus $\mathcal{C}$ (as it is the ratio between triggered and covered), assuming $\mathcal{C}$ is error-free. In practice, even though $\mathcal{C}$ are extracted from relational sources and are generally clean, a small fraction of its columns may contain real errors (our analysis on multiple corpora suggest this to be \textasciitilde2\%). Regardless of the exact error rate of $\mathcal{C}$, $\rho(r)$ produces a safe ``upper-bound'' for the real percentage, as the true number of false-triggers in the numerator is bound to be smaller than  $\left | \mathcal{C}^r_{C, {T}} \right |$ estimated from the contingency table. 

\iftoggle{full}
{
    \underline{Discussion.} We note that while heuristic estimates (e.g., directly using $\hat{c}$ to estimate $c$) can still be used in our end-to-end optimization framework, we confirm in our ablation studies that adopting a more principled statistical analysis (Wilson's interval to lower-bound confidence, and Cohen's h for effect-size)  does provide quality benefits over heuristic estimates.
}

\subsection{\sdca Optimizations by LP-Relaxation}
\label{subsec:selectionWithFPRGuarantees}
Let $R_{all}$ be the set of all candidate \sdcas that meet the statistical tests performed in Section~\ref{subsec:rule_quality_eval} (still a large set in tens of thousands, with overlapping coverage and varying degrees of quality). We now describe the  key final step in offline-training, where we perform holistic optimization like sketched in Definition~\ref{def:prob_def}, to select an optimal set $R \subseteq R_{all}$ with FPR and recall guarantees.

%Recall that in Definition~\ref{def:prob_def}, we want to find a core set of \sdcas $R$, such that (1) we can bound the time it takes to perform inference using $R$ (by simply limiting the size of $|R|$), and (2) we can guarantee the quality of predictions made by $R$, in terms of both its recall, written as recall$(R)$, and its false-positive rate, written as FPR$(R)$. 

We will first describe how to estimate FPR$(r)$ and recall$(r)$ below. %, before explaining our LP-relaxation algorithms for with probabilistic guarantees.  

\underline{Estimating FPR.} Recall that the FPR of a constraint $r$ is defined as FPR$(r) = \frac{\text{r-false-positive-columns}}{\text{total-negative-columns}}$, or the number of false-positives $r$ produces, over the total number of negative (error-free) columns. While we don't have labeled data to count these events for $r$ precisely (which would be hugely expensive if we were to label each $r$), we can approximate these events using a large corpus $\mathcal{C}$ in an unsupervised data-driven manner. %\footnote{Observe that another related metric, precision$(r)$, would be defined as $\frac{\text{r-true-positive-columns}}{\text{total-predicted-columns}}$, which is not easy to estimate unsupervised from $\mathcal{C}$.} 

Specifically, since $\mathcal{C}$  is extracted from relational sources, its columns are generally clean and free of errors (for instance, in our manual analysis of a sample of 2400 table columns randomly sampled from spreadsheets and relational tables, we found the error rate of  $\mathcal{C}$ to be  around 2\%, like we will explain in Section~\ref{subsec:exp_dataset}). We can therefore use $|\mathcal{C}|$ to approximate the number of total negative columns in $|\mathcal{C}|$. Also recall that we can estimate false-positives of $r$, based on  $\mathcal{C}^{r}_{C, T}$ in the contingency table (estimated using  $\mathcal{C}$), so putting the two together we can then estimate FPR$(r)$ as {\scriptsize $\frac{\left |\mathcal{C}^{r}_{C, T} \right |}{|\mathcal{C}|}$}.

\underline{Estimating recall.} The recall of  $r$, written as Recall$(r)$, is  the total number of true-positive errors that $r$ can detect.\footnote{We use the absolute version of recall over the  relative version for its simplicity, the two versions differ only by a universal denominator (the total-number-of-positive-columns), and are therefore equivalent in our context.} Since we also don't have labeled data to estimate recall for each $r$, we use 
\textit{distant-supervision} \cite{mintz2009distant, surdeanu2010simple, HH18} to approximate it. 

Specifically, we construct a synthetic corpus for that purpose, written as $\mathcal{C}_{syn} =\{ C(v^e) = C \cup \{v^{e} \}  | C\in \mathcal{C}, C' \in \mathcal{C}, v^{e} \in C'\}$, where each column in $\mathcal{C}_{syn}$ is constructed as $C(v^e) = C \cup \{v^{e}\}$, with $C$ being a randomly sampled column in $\mathcal{C}$, $v^{e}$  being a randomly sampled value from a different column $C'$, such that when $v^{e}$ is inserted into $C$ to produce $C(v^e)$, $v^{e}$  is likely an error in the context of $C(v^e)$. 
Like in distant supervision~\cite{mintz2009distant, surdeanu2010simple, HH18}, this then allows us to compute the set of errors that $r$ can detect in $\mathcal{C}_{syn}$, as 
\begin{equation}
\label{eqn:syn-recall}
   D(r) = \{ C(v^e) | C(v^e)  \in \mathcal{C}_{syn}, r(C(v^e) ) = v^e \} 
\end{equation}
where $r(C(v^e) ) = v^e$ indicates that $r$ can detect the same $v^e$ as constructed in column $C(v^e)$.
We then simply use Recall$(r) = |D(r)|$, as the estimated recall of $r$.
\iftoggle{full}
{
    Although there is a small chance that $v^e$ might not be an actual error in the context of $C(v^e)$, in practice, it leads to only a very small difference in recall estimation. For example, in a manual examination of 100 randomly sampled synthetic columns, we identify only 3 cases where $v^e$ could not be identified as an error/outlier, confirming our assumption that inaccuracies so introduced is minimal (e.g., 3\%), which have negligible impact for the final \sdcas selected by the algorithm as we will show empirically in our experiments.
}

\underline{\css (\cssa)}.
%\subsection{\sdca Selection with Confidence Approx.}
%\label{subsec:coarse_select}
We are now ready to instantiate the high-level problem sketched in Definition~\ref{def:prob_def} as follows.

\begin{definition}\label{def:cs}
[\textbf{\css (\cssa)}]. Given all \sdca candidates $R_{all}$, find a set $R \subseteq R_{all}$ such that its recall Recall$(R)$ is maximized, subject to a constraint that the FPR$(R)$ should not exceed $B_{FPR}$, and a cardinality constraint that the size of $R$ should not exceed $B_{size}$, written as:
{ \small
    \begin{align}
        \text{(\cssa)} \qquad{} \max_{R \subseteq R_{all}} & \left| \bigcup_{r \in R} D(r) \right|          \label{eqn:ccs_obj} \\
        %\text{maximize }    & _{R \subseteq R_{all}} \left | D(R, \mathcal{C}_{syn}) \right |          \nonumber \\
        \text{s.t. }        & \left | R \right | \leq B_{size}  \label{eqn:ccs_size} \\
                            & \sum_{r \in R} \text{FPR}(r) \leq B_{FPR} \label{eqn:ccs_fpr}
    \end{align}
}
\end{definition}

Note that in the objective function Equation~\eqref{eqn:ccs_obj}, we use Recall$(R) = \left| \bigcup_{r \in R} D(r) \right|$ to instantiate the objective function  of the original problem in Definition~\ref{def:prob_def} (Equation~\eqref{eqn:general_obj}), since Recall$(R)$ over a set of constraints $R$ can be calculated as the union of  errors detected by each $r \in R$.

Observe that because individual $r \in R_{all}$ can often have overlapping coverage (e.g., different embedding methods, and different CTA-classifiers that can detect columns of a type, say \codeq{city}, are all present in $R_{all}$), this union term in Equation~\eqref{eqn:ccs_obj} can therefore take the overlaps of recall into consideration when we optimize for the best solution set $R$.

Also note that in Equation~\eqref{eqn:ccs_fpr},  we use $\sum_{r \in R}{\text{FPR}(r)}$ 
in place of FPR$(R)$ in Equation~\eqref{eqn:general_fpr} of Definition~\ref{def:prob_def}, because it can be verified that  FPR$(R) \leq \sum_{r \in R}{\text{FPR}(r)}$ (using an argument similar to union-bound), so that imposing the constraint in Equation~\eqref{eqn:ccs_fpr} esures that  the original constraint $\text{FPR}(R) \leq B_{FPR}$  is also satisfied.

We show that the CSS problem in Definition~\ref{def:cs} is hard and hard to approximate, using a reduction from maximum coverage. A proof of this can be found in \iftoggle{full}{Appendix~\ref{apx:proofs}}{\cite{full}}.

\begin{theorem}\label{thm:crs_np_hard}
The \cssa problem is NP-hard and cannot be approximated with a factor of $(1-1/e)$, unless $NP \subseteq DTIME(n^{O(\log \log n)})$.
\end{theorem}

% \begin{proofsketch}
%     We prove that \cssa is NP-hard by reducing from maximum coverage (MC) problem. Since MC cannot be approximated with a factor larger than $(1-1/e)$ unless $NP \subseteq DTIME(n^{O(\log \log n)})$ \cite{feige98}, the same conclusion applies to CSS.  Complete proofs of the theorems in this paper can be found in Appendix~\ref{apx:proofs}.
% \end{proofsketch}

Despite its hardness, we develop \cs
% \iftoggle{full} % revised, remove for space
% {
    in Algorithm~\ref{alg:cs} 
%}
to solve \cssa using LP-relaxation and randomized rounding~\cite{raghavan1987randomized}, which has an approximation ratio of $(1-1/e)$  (matching the inapproximability result). 
Specifically, we first transform \cssa  into a CSS-ILP problem:
{ \small
    \begin{align}
    %\small
    (\text{CSS-ILP})~~\text{maximize }    & \sum_{C_j \in \mathcal{C}_{syn }} y_j                 \\
    \text{s.t. }        & \sum_{r_i \in R_{all}} x_i \leq B_{size}           \\
                        & \sum_{r_i \in R_{all}} \text{FPR}(r_i) \cdot x_i \leq B_{FPR} \\
                        & \sum_{r_i \in K_j} x_i \geq y_j \;\;\; \forall C_j \in \mathcal{C}_{syn}    \\
                        &  x_i, y_j \in \{0, 1\}            \label{eqn:cssilp_integrality}
    \end{align}
}
Here, we use an indicator variable $x_i$ for each $r_i \in R_{all}$, where $x_i = 1$ indicates $r_i$ is selected into $R$, and 0 otherwise. Let $D(R) =  \bigcup_{r \in R} D(r)$ be the union of all errors detected by $R$.
We use another indicator variable $y_j$ for each column $C_j \in \mathcal{C}_{syn}$, where $y_j = 1$ indicates $C_j \in D(R)$, and 0 otherwise.
Finally, for each $C_j \in \mathcal{C}_{syn}$, we define $K_j \subseteq R_{all}$ as the set of constraints that can detect the error constructed in $C_j$. It can be shown that the CSS-ILP problem so constructed, has the same solution as the original CSS problem.

From CSS-ILP, we construct its LP-relaxation~\cite{raghavan1987randomized}, referred to as CSS-LP, by dropping its integrality constraint in Equation~\eqref{eqn:cssilp_integrality}. The resulting CSS-LP is a linear program that can be solved optimally in polynomial-time, yielding \emph{fractional solution} for each $x_i$. Finally, we use randomized-rounding 
% revised, removed for space
% \iftoggle{full}
% {
    like shown at the end of  Algorithm~\ref{alg:cs},
%}
to turn the fractional $x_i$  into integral solutions $R$.

% revised, remove for space
% \iftoggle{full}
% {
    We show a proof in \iftoggle{full}{Appendix~\ref{apx:proofs}}{\cite{full}} that the  solution from Algorithm~\ref{alg:cs} provides the following guarantees.

    \begin{algorithm}[t]
    \small
        \caption{\cs}\label{alg:cs}
        \Input{All candidate \sdca $R_{all}$}
        \Output{The selected set $R \subseteq R_{all}$}
        \begin{algorithmic}[1]
        \STATE Transform a CSS problem instance  into a CSS-ILP instance
        \STATE Transform a CSS-ILP instance into a LP-relaxation version CSS-LP 
        \STATE $\{x_i\} \gets $ optimal solutions to CSS-LP, solved using LP solvers
        \STATE $R \gets \{\}$
        \FOR{ $r_i \in R_{all}$ }
            \STATE $R \gets R \cup \{r_i\}$, with probability $x_i$
        \ENDFOR
        \RETURN $R$
        \end{algorithmic}
    \end{algorithm}
%}

% \yeye{sep}
% Next, we introduce algorithm \cs, which solves \cssa and produces a solution with an approximation ratio of $(1-1/e)$. The pseudo-code is shown in Algorithm~\ref{alg:cs}. 
% Intuitively, \cs first converts the input into an instance of an integer linear programming problem, referred to as CSS-ILP.
% It then computes the optimal \emph{fractional solution} of the corresponding linear programming relaxation, termed SSLP. 
% With the fractional solution, it determines which constraints should be included in $R$ using a randomized rounding scheme. 

% % \begin{example} \label{emp:cs}
% %     Continuing our running example in Example~\ref{emp:detetable}. 
% %     When converting this example to CSS-ILP, we have $K_1 = \{r_1\}$ since only $r_1$ reports the ground-truth error of $C_1$. Similarly,  $K_2 = \{r_2\}$, $K_3 = \{r_2, r_3\}$, and $K_4 = \{r_4\}$. 
% %     Note that $K_4$ does not include $r_2$ since $r_2.S(C_4)$ (which is \codeq{mankanto}) is not equal to $O(C_4)$ (which is \codeq{saskachewan}).
% %     Assume that after solving the SSLP, we obtain that $x_1=0.4$, $x_2=0.7$, $x_3=0.1$, and $x_4=0.4$. Then, \cs initializes $R = \{\}$ and adds $r_1$ into $R$ with a probability of $0.4$, adds $r_2$ with a probability of $0.7$, and so on.
% % \end{example}
    
% The main result of \cs is given in Theorem~\ref{thm:cs}.

% \iftoggle{full}
% {
    \begin{theorem} \label{thm:cs}
    Let $R$ be the solution returned by Algorithm~\ref{alg:cs}, and $E(\cdot)$ denote expectation. Then the following hold:  $E(\left | R \right |) \leq B_{size}$, $E(\sum_{r \in R} \text{FPR}(r)) \leq B_{FPR}$, and  $E(\left | D(R) \right |)  \geq (1-1/e)OPT$ where $OPT$ is the optimal value. 
    \end{theorem}
    
    Note that   in expectation, our approximation ratio  matches the inapproximability in Theorem~\ref{thm:crs_np_hard}.
%}

%A proof of this result can be found in our technical report\iftoggle{full}{Appendix~\ref{apx:proofs}}{\cite{full}}.

% \begin{proofsketch}
%     We first show that the original \cssa instance and its transformed CSS-ILP instance are equivalent, and achieve the maximum objectives simultaneously. 
%     Denote the solution obtained by solving SSLP as $X = \{x'_i\}$ and $Y = \{y'_j\}$.
%     Since each $r_i$ is selected into $R$ with a probability of $x'_i$, we can show that 
%     $E(\left | R \right |) = E(\sum_{r_i \in R_{all}} x'_i) \leq B_{size}$
%     and 
%     $E(\sum_{r \in R} {FPR}(r)) = E(\sum_{r_i \in R_{all}} {FPR}(r_i) \cdot x'_i) \leq B_{FPR}$.
    
%     To prove that $E(\left | D(R) \right |)  \geq (1-1/e)OPT$, we first show that for each $C_j \in \mathcal{C}_{syn}$, the probability of $C_j \in D(R)$ is at least $(1-1/e) y'_j$.
%     Therefore, 
%     $E(\left | D(R) \right |)  \geq (1-1/e)\sum_{C_j \in \mathcal{C}_{syn}} y'_j \geq (1-1/e)OPT$.
% \end{proofsketch}

\underline{\fss (\fssa)}.
%\subsection{\sdca Selection with Confidence Approx.}
\label{subsec:fine_select} 
While \cssa  reduces $R_{all}$ into $R$ from the perspective of set-based optimization, we find in our evaluation, that the confidence produced by the solution $R \subseteq R_{all}$ from CSS, to deviate substantially from the true calibrated confidence (Equation~\eqref{eqn:wilson-confidence} in Section~\ref{subsec:rule_quality_eval}),  if we use the entire $R_{all}$. This is because \cssa only focuses on the set-based optimization, without considering how well the selected $R$ can approximate the calibrated confidence from the original  $R_{all}$, which leads to poor confidence ranking of predicted results, that negatively affects the result quality (e.g., when evaluated using area under precision-recall curves).

% One problem with the formulation of \cssa, however, is that it does not take into consideration the consistency of the confidence scores for the reported values before and after the selection process. 
% Specifically, let $\text{conf}(C, R')$ denote the confidence of $C$'s ground-truth error given a set $R'$ of \sdcas. The confidence using the set obtained by \cssa (i.e., $\text{conf}(C, R)$) may be significantly lower than the confidence using the original set of \sdcas (i.e., $\text{conf}(C, R_{all})$). 
% Since we primarily focus on end-user data cleaning scenarios, such as cleaning Excel spreadsheets, where high precision (e.g., 95\%) and good ranking on the detection results are essential, it is crucial to preserve consistent confidence scores during the selection process.

% \begin{example}\label{emp:fs_motivation}
%     Continuing the running example in Example~\ref{emp:cs}. Assume that $R_{all} = \{r_1, r_2, r_3, r_4\}$. Consider column $C_3$. $\text{conf}(C_3, R_{all})$ equals $r_3.c = 0.97$, since the confidence of $r_3$ is the highest among all \sdcas that report the ground-truth error in $C_3$. However, after \cssa, assume that \cs chooses $R = \{r_1, r_2\}$. Since only $r_2$ reports the ground-truth error in $C_3$, $\text{conf}(C_3, R)$ becomes $r_2.c = 0.90$, decreasing by 0.07 compared to its confidence before \cssa.
% \end{example}

To address this inadequacy, we propose an improved version of \cssa that ensures confidence approximation in the selection process, which we call \emph{\fss (\fssa)}.
Define $\text{diff}(C, R, R_{all}) = \text{conf}(C, R_{all}) - \text{conf}(C, R)$ as the difference in predicted confidence for any $C \in \mathcal{C}_{syn}$, between using $R_{all}$ and $R$.   
 We define \fssa as follows:
 
\begin{definition}\label{def:fs}
[\textbf{\fss (\fssa)}].
Given all \sdca candidates $R_{all}$, find a set $R \subseteq R_{all}$  to maximize the number of columns detected in $D(R)$, whose predicted confidence using $R$ does not deviate from its true confidence by  $\delta$ (or $\text{diff}(C, R, R_{all}) \leq \delta$), subject to a constraint that the FPR$(R)$ should not exceed $B_{FPR}$, and the size of $R$ should not exceed $B_{size}$, written as:
\begin{align*}
\small
    \text{(\fssa)} \qquad{}  \max_{R \subseteq R_{all}}   & \left | \{ C \,|\, C \in D(R), \text{diff}(C, R, R_{all}) \leq \delta \} \right |          \nonumber \\
    \text{s.t. }        & \left | R \right | \leq B_{size}  \\
                        & \sum_{r \in R} \text{FPR}(r) \leq B_{FPR} \nonumber 
\end{align*}
\end{definition}

%Unlike \cssa, which focuses solely on detecting ground-truth errors, \fssa seeks to maximize the number of detectable columns  where the confidence in reported errors remains consistent before and after the selection process.
Observe that when we set $\delta = 1$, \fssa reduces to \cssa because the confidence approximation requirement of $\text{diff}(C, R, R_{all}) \leq \delta$ is trivially satisfied, making it an advanced variant of \cssa.
% Since \cssa is an NP-hard problem, \fssa is NP-hard as well. 

We propose algorithm \fs to solve \fssa, also with quality guarantees. The pseudo-code for \fs is similar to that of \cs
% revised, removed for space
% \iftoggle{full}
% {
    (Algorithm~\ref{alg:cs})
%}, 
with two key modifications: (1) for each column $C_j \in \mathcal{C}_{syn}$, its indicator variable $y_j = 1$ if $C_j \in \{ C \in D(R) \,|\, \text{diff}(C, R, R_{all}) \leq \delta \}$, and 0 otherwise; (2) for each $C_j \in \mathcal{C}_{syn}$, we set $K_j \subseteq R_{all}$ as the set that can detect the error constructed in $C_j$, with the required confidence approximation specified by $\delta$.
%we set $K_j = \{r \in R_{all} \,|\, r.S(C_j) = O(C_j)$ and $r.c \geq \text{conf}(C_j, R_{all}) - \delta\}$, which corresponds to the set of \sdcas that reports the ground-truth error with a consistent level of confidence. 

% \begin{example}
%     Following the running example setting in Example~\ref{emp:detetable}. Assume that $\delta = 0.01$ and $R = \{r_1, r_5\}$. Then, $D(R) = \{c_3, c_4, c_9\}$, while $dcc(R, R_{all}, \delta, \mathcal{C}_{syn}) = \{c_4, c_9\}$. Even though $R$ can still report $c_3$ since $o(c_3, \{r_1\})= v_{3a}$, $c_3$ does not belong to $dcc(R, R_{all}, \delta, \mathcal{C}_{syn})$ because $\text{conf}(r_1) < cs(c_3, R_{all}) - \delta = 0.97 - 0.01$. Due to the same reason, when constructing the RSILP in \fs, $S_3 = \{r_3\}$ does not contain $r_1$, even though $r_1 \in S_3$ when when constructing the RSILP in \cs (see Example~\ref{emp:cs}).
% \end{example}

We show in  
\iftoggle{full}
{Appendix~\ref{apx:proofs}}
{our technical report~\cite{full}} 
that the \fs approach has a $(1-1/e)$ approximation ratio in expectation, with all constraints also satisfied in expectation, like stated below.
% \begin{proofsketch}    
\begin{theorem} \label{thm:fs}
Let $R \subseteq R_{all}$ be the solution produced by \fs, and $E(\cdot)$ denote expectation. Then the following hold: $E(\left | R \right |) \leq B_{size}$, $E(\sum_{r \in R} {FPR}(r)) \leq B_{FPR}$, and $E(\left | \{ C \in D(R) \,|\, \text{diff}(C, R, R_{all}) \leq \delta \}\right | )  \geq (1-1/e)OPT$ where $OPT$ is the optimal value. 
\end{theorem}

We compare \fs and \cs for their effectiveness in our experiments.

\section{Experiment}
\label{sec:exp}
We perform extensive evaluations using real errors from real data.
Our code, data, and labeled benchmarks are available for future research\footnote{\url{https://github.com/qixuchen/AutoTest}}.

%\yeye{terminology change to apply throughout (including in appendix): Excel, PBI, rule}

% \begin{table}[t]
% \scalebox{0.85}
% {
%     \setlength{\tabcolsep}{3pt}
%     \centering
%     \begin{tabular}{|c|c|c|} \hline
%         & \rttrain & \sttrain  \\ \hline
%         total \# of col     & 247976 & 297099 \\ \hline
%         mean \# of vals (per col)     & 7252.90 & 559.59 \\ \hline
%         median \# of vals (per col)     & 484 & 54 \\ \hline
%         mean \# of dist. vals  (per col)     & 95.89 & 56.79 \\ \hline
%         median \# of dist. vals  (per col)    & 18 & 14 \\ \hline
        
%     \end{tabular}
% }
%     \caption{Training table corpora: detailed statistics}
%     \label{tab:corpora_stat}
% \end{table}

\subsection{Experimental Setup}
\label{subsec:exp_setup}
\textbf{Benchmarks.}
\label{subsec:exp_dataset} To test the effectiveness of different algorithms on real tables ``in the wild'', we focus on real relational tables and spreadsheet tables in our evaluation, and create two benchmarks containing real errors from real tables, described below. % which we refer to as \sttest and \rttest, respectively. 

\underline{\sttest (\sttesta)}. We randomly sample 1200 spreadsheet columns, extracted from real spreadsheets (\code{.xlsx} files crawled from the web), as our \sttesta test set.\footnote{We sample non-numerical columns for testing only, since it is usually trivial to identify non-conforming values (e.g., strings) in numerical columns.} Each column is carefully labelled and cross-checked by human-labellers, as either \emph{clean} (no data errors are present); or \emph{dirty}, in which case all erroneous values in the column are marked in ground-truth for evaluation. A total of 47 columns (3.9\%) contain real errors. 

\underline{\rttest (\rttesta)}. We also sample 1200 relational table columns from real tables extracted from BI models (\code{.pbix} files crawled from the web), as our second test set. Each column is similarly labeled as clean, or dirty, with erroneous values identified in the ground-truth. A total of 40 columns (3.3\%) are identified to contain errors.

%these potential errors and provide a a true/false/not sure label for each of them.

\underline{Existing data-cleaning benchmarks.} To test the applicability of learned \sdcas on existing data-cleaning benchmarks, we further compile 9 commonly-used datasets from prior studies~\cite{mahdavi2019raha, mahdavi2020baran, heidari2019holodetect, ni2023automatic}, which are: \textit{adults}, \textit{beers}, \textit{flights}, \textit{food}, \textit{hosptial}, \textit{movies}, \textit{rayyan}, \textit{soccer} and \textit{tax}. 
%for a total of 81 columns. 
We reuse existing ground-truth in our evaluation. %Each raw column is accompanied with a clean version, which can be used to decide the ground truth.

%It is worth mentioning that the above benchmarks are used to measure the performance of all algorithms (including our algorithms) but our algorithms are trained on unlabelled datasets (i.e., the datasets without any ground-truth errors) (which will be described later). 

\textbf{Evaluation metrics.}
We evaluate the quality of all algorithms, using  standard precision/recall, where precision \begin{small}
$P=\frac{\text{num-of-correct-predicted-errors}}{\text{num-of-total-predicted-errors}}$
\end{small}, and recall 
\begin{small}
$R=\frac{\text{num-of-correct-predicted-errors}}{\text{num-of-total-true-errors}}$
\end{small}.

Since each algorithm have different score thresholds to make predictions at different confidence levels, we plot precision-recall curves (PR-Curves) of all algorithms, and summarize the overall quality of PR-Curves using two standard metrics: 

(1) \underline{Precision-Recall Area-Under-Curve (PR-AUC)}~\cite{baeza1999modern}, which measures the area under the PR-curve, where higher is better; 

(2) \underline{F1-score at Precision=0.8 (F1@P=0.8)}~\cite{bishop2006pattern}, which measures the F1 score (the harmonic mean of precision and recall) at high precision ($P=0.8$). Note that a high level of precision is crucial in our setting (e.g., to win user trust in end-user data cleaning), which is why we use this metric to complement PR-AUC. 

\subsection{Methods Compared}
We compare with the following methods on quality and efficiency.

\label{subsec:methods_compared}
\begin{itemize}[leftmargin=*]
    \item {\textbf{Column-type detection methods}}. Our first group of baselines directly invoke existing column-type detection methods.  For each method below, we compute the domain evaluation score $f_t(v)$ of type $t$ (Definition~\ref{def:func}), for each value $v$ in column $C$, and use the standard z-score on the resulting distribution of $f_t(v)$ to identify potential errors~\cite{hellerstein2013quantitative}. We vary the z-score threshold to plot PR-curves for each method.
%    For each domain compiled in \at, we create a corresponding baseline that reports errors based on local value distribution within the column. Specifically, given a column $C$, the baseline regards a value $v \in C$ as an error if its \textit{outlier score} (to be detailed below) is at least three standard deviations away from the mean outlier score of values in $C$.
%    Specifically, the following baselines are created. 
    \begin{itemize}
    \item {\underline{CTA methods: \textit{Sherlock}~\cite{sherlock}, \textit{Doduo}~\cite{doduo}}}.
    We use the CTA-classifier as domain evaluation function $f_t(\cdot)$ to compute a score distribution for each column $C$.  
    %Let $t$ be the semantic type $C$ is classified to by the classifier. The outlier score of $v$ is its prediction score produced by the classifier on $t$. 
    \item {\underline{Embedding domains: \textit{Glove}~\cite{PSM14}, \textit{Sentence-BERT}~\cite{RG19}}}. We use the embedding distance as the domain evaluation function $f_t(\cdot)$ to compute a score distribution. 
    \item {\underline{Function domains: \textit{DataPrep}~\cite{Dataprep-clean}, \textit{Validator}~\cite{validator}}}. We use the boolean result returned by a type-validation function (true=1, false=0)  as the domain evaluation function $f_t(\cdot)$.
    \item {\underline{Pattern domains: \textit{Regex}}}. We use whether a value $v$ matches an inferred regex pattern of a column $C$ (match = 1, non-match = 0) as the domain evaluation function $f_t(\cdot)$. %\footnote{We use the same patterns generated by \at for a fair comparison.}

    \end{itemize}

    \item {\textbf{GPT-4}}. 
    Language models such as GPT have shown strong abilities in diverse tasks~\cite{gpt}. Since our task can also be formulated as a natural-language task,  we invoke GPT-4\footnote{Version gpt-4-0125, accessed via OpenAI API in 2024-06.} with extensive prompt optimization, including (1) select few-shot examples~\cite{gpt}; and (2) use chain of thought (COT)~\cite{wei2022chain} (to require GPT-4 to reason about its detection with possible repairs, so that it can stay truthful with fewer false-positives). We report 4 variants here based on prompts used, which are: \underline{few-shot-with-COT}, \underline{few-shot-no-COT}, \underline{zero-shot-with-COT}, and \underline{zero-shot-no-COT}.

% \iftoggle{full}
% {
%    \item {\textbf{GPT-4 with fine-tuning}}. In addition to prompting GPT-4, we also fine-tune a new error-detection model using GPT-4 as the base-model, utilizing the same training data available to \at (Section~\ref{sec:autotest}, with 80K training columns), in the hope that the fine-tuned model can perform better in the specific task of error-detection. We performed hyper-parameter search, and find batch-size 32, learning rate 0.1 to produce the best-performing model with fine-tuning. %assess the performance of a model fine-tuned using the synthetic dataset for rule selection, i.e., $\mathbf{C}_{syn}$.     For each reported column, we ask the GPT models to identify the most likely outlier and to provide a confidence score.
% }

    \item {\textbf{Katara}} \cite{chu2015katara}.  Katara performs data cleaning by mapping table columns to Knowledge-Bases (KB) like YAGO, to identify columns of type \code{city},  \code{country}, etc., so that errors can be detected. This is similar in spirit to ours, but is limited to symbolic knowledge-bases, and are based on heuristic mapping with static thresholds (not trained/calibrated).
    
    \item {\textbf{Auto-Detect}} \cite{HH18}. This approach detects errors due to incompatible data patterns, based on co-occurrence statistics. While it also leverages a corpus to produce predictions, it is only applicable to patterns, limiting its coverage. %We report all columns where at least one pair of values is found to be incompatible. Following \cite{HH18}, the score for each reported column is determined by the pair with the highest estimated precision. The identified outlier is considered correct if one of the value in this pair agrees with the ground truth.
    
    \item {\textbf{Outlier detection methods}}. There is a large literature on outlier detection, we select \underline{\textit{RKDE}} \cite{kim2012rkde},  \underline{\textit{PPCA}} \cite{tipping1999ppca} and \underline{\textit{IForest}} \cite{liu2008iforest} for comparison, which are shown to be the best-performed methods in an empirical study~\cite{domingues2018comparative}. We also include three classical methods: \underline{\textit{SVDD}} \cite{tax2004support},  \underline{\textit{DBOD}} \cite{knox1998algorithms} and \underline{\textit{LOF}} \cite{breunig2000lof}, that compared with in an earlier study~\cite{HH18} similar to our problem setting.
    
    \item {\textbf{Commercial}}. We also test our benchmarks on two commercial software targeting non-technical end-users, that can automatically detect errors in tables. We refer to these two systems as \underline{Vendor-A} and \underline{Vendor-B} in our experiments. %\footnote{No confidence scores are explicitly produced by these vendors, so all returned errors are assumed to have the same confidence score.}
%    \item \underline{\textbf{Vendor-X}}. This is Excel but it is better to not name the system in the paper.
    
    \item {\textbf{Auto-Test}}. This is our proposed method. We report 3 variants of \at, which are (1) \underline{\ar}, which uses the entire set of candidate constraints $R_{all}$ after quality-based pruning  (Section~\ref{subsec:rule_quality_eval}),
    (2) \underline{\cs} 
    % revised, remove for space
    % \iftoggle{full}
    % {
        (Algorithm~\ref{alg:cs}),
    % }
    % {
    %     (Section~\ref{subsec:selectionWithFPRGuarantees}),
    % }
    and (3) \underline{\fs} (Section~\ref{subsec:selectionWithFPRGuarantees}).
    
    %By default, we set the threshold for confidence and Cohen's h in the rule quality assessment process to 0.9 and 0.8, respectively.
    We invoke the solver in SciPy \cite{scipy} to solve our LP. By default, we set $B_{size}$ to 500,  $B_{FPR}$ to 0.1, and $\delta$ in \fs to $10^{-3}$.

    For training, we use three corpora: (i) 247K relational table columns extracted from relatoinal sources~\cite{auto-bi}, henceforth referred to as \rttrain, and (ii) 297K spreadsheet table columns extracted from real spreadsheets, referred to as \sttrain, and (iii) 298K real table columns extracted from a publically available corpus \tablib \cite{tablib}. We kept \rttrain and \sttrain completely separate from \rttesta and \sttesta.  To test generalizability, we also train on one corpus (e.g., \rttrain) and test on the benchmark from a different source (e.g., \sttesta).
    % For training, we use two corpora: (i) 247K relational tables extracted from  real BI models, henceforth referred to as \rttrain, and (ii) 297K real spreadsheet tables extracted from real spreadsheets, referred to as \sttrain, held completely separate from \rttesta and \sttesta. (Note that we also test generalizability, by train on \rttrain and test on the different \sttesta, and vice versa).
    \iftoggle{full}
    {
        Detailed statistics of the corpora are reported in Table~\ref{tab:corpora_stat}.
    }
    {
    }
\end{itemize}
All experiments are run on a Linux machine with a 64-core, 2.4 GHz CPU and 512 GB memory.

\iftoggle{full}
{
    \begin{table*}[t]
    \vspace{-2mm}
    \caption{Training table corpora: detailed statistics}
    \label{tab:corpora_stat}
    \vspace{-3mm}
        \centering
        \scalebox{0.6}{
            \begin{tabular}{|c|c|C{2.2cm}|C{2.2cm}|C{2.5cm}|C{2.8cm}|} \hline
                Corpus  & total \# of columns & mean \# of vals (per column) & median \# of vals (per column) & mean \# of dist. vals  (per column)& median \# of dist. vals  (per column) \\ \hline
                
                \rttrain     & 247976 & 7252.90 &  484  & 95.89 &  18               \\ \hline
        
                \sttrain   & 297099  &  559.59 & 54  & 56.79 &  14            \\ \hline
                
                \tablib   & 298399  &  630.24 & 36  & 100.90 &  14            \\ \hline
                
            \end{tabular}
        }
    \vspace{-3mm}
    \end{table*}
}

\subsection{Quality Comparisons}
\label{subsec:quality_comparison}

\begin{small}
\begin{table*}[t]
    \captionof{table}{Quality comparisons, reported as (F1@P=0.8, and PR-AUC), on \sttesta and \rttesta.}
    \label{tab:performance_summary}
    \vspace{-3mm}
    \centering
    \setlength{\tabcolsep}{2.5pt} % Adjust the value to increase or decrease the space between columns
    \scalebox{0.67}
    {
        \begin{tabu}{|c|c||c|c|c|c||c|c|c|c|} \hline
        & &  \multicolumn{4}{c||}{ \sttest (\sttesta) } & \multicolumn{4}{c|}{\rttest (\rttesta)} \\ 
        \hline
              & Method  & real &  +5\% syn err. &  + 10\% syn err. &  +20\% syn err. & real & +5\% syn err. & +10\% syn err. & +20\% syn err. \\ \hline \hline
              
        \multirow{3}{*}{Ours} & \ar       & 0.23, 0.38      & 0.36, 0.39       & 0.47, 0.57       & 0.50, 0.66       & \textbf{0.21}, \textbf{0.34}       & \textbf{0.30}, 0.36       & \textbf{0.36}, 0.48       & 0.36, 0.54  \\ \cline{2-10}

        & \fs & \textbf{0.34}, \textbf{0.45}      & \textbf{0.38}, \textbf{0.52}       & \textbf{0.48}, \textbf{0.62}       & \textbf{0.53}, \textbf{0.68}       & \textbf{0.21}, \textbf{0.34}       & \textbf{0.30}, \textbf{0.46}       & \textbf{0.36}, \textbf{0.56}       & \textbf{0.40}, \textbf{0.62}  \\ \cline{2-10}
        
        & \cs  & 0.25, 0.43      & 0.35, 0.52       & 0.41, 0.60       & 0.52, 0.67       & 0.05, 0.31       & 0.25, 0.43       & 0.28, 0.53       & 0.39, 0.61  \\ \hline

        % \rowfont{\color{blue}}
        % \multirow{3}{*}{Ours (\tablib)} & \ar       & 0.33, 0.52      & 0.38, 0.56       & 0.42, 0.59      & 0.50, 0.66     & 0, 0.33      & 0.29, 0.44       & 0.31, 0.47       & 0.34, 0.56  \\ \cline{2-10}
        
        % \rowfont{\color{blue}}
        % & \fs & 0.30, 0.52      &  0.39, 0.56       & 0.45, 0.61      & \textbf{0.53}, \textbf{0.68}     & 0.20, \textbf{0.46}      & \textbf{0.43}, \textbf{0.56}       & \textbf{0.40}, \textbf{0.56}       & \textbf{0.46}, 0.60  \\ \cline{2-10}

        % \rowfont{\color{blue}}
        % & \cs  & 0.23, 0.53      & 0.41, 0.56       & 0.45, 0.60      & 0.52, 0.66     & 0, 0.32      & 0.06, 0.47       & 0.35, 0.52       & 0.38, 0.61  \\ \hline
    
       % \multirow{3}{*}{Ours (\sttrain)} & \ar   & 0, 0.32      & 0.26, 0.40       & 0.28, 0.48       & 0.41, 0.61       & 0, 0.29       & 0.27, 0.40       & 0.21, 0.44       & 0.27, 0.53  \\ \cline{2-10}

       % & \fs & 0.10, 0.36      & 0.21, 0.45       & 0.28, 0.52       & 0.45, 0.64       & 0.03, 0.31       & 0.29, 0.42       & 0.25, 0.47       & 0.27, 0.55   \\ \cline{2-10}
    
       % & \cs & 0, 0.30      & 0.09, 0.39       & 0.09, 0.48       & 0.40, 0.61       & 0, 0.18      & 0.08, 0.33       & 0.10, 0.41       & 0.13, 0.52   \\ \hline
    
        \multirow{7}{*}{\makecell{Column-type detection \\ methods}} & Sherlock        & 0, 0.04      & 0, 0.05       & 0, 0.10       & 0.01, 0.21       & 0, 0.03       & 0, 0.06       & 0, 0.14       & 0, 0.22  \\ \cline{2-10}
    
        & Doduo           & 0.04, 0.06      & 0.06, 0.09       & 0.09, 0.17       & 0.08, 0.31       & 0, 0.03       & 0, 0.05       & 0, 0.10       & 0, 0.20  \\ \cline{2-10}
    
        & Glove           & 0, 0.10      & 0.03, 0.18       & 0.07, 0.26       & 0.06, 0.35       & 0.05, 0.10       & 0.06, 0.13       & 0.03, 0.18       & 0.03, 0.28  \\ \cline{2-10}
    
        & SentenceBERT    & 0.08, 0.14      & 0.12, 0.18       & 0.11, 0.23       & 0.18, 0.36       & 0.09, 0.09       & 0.14, 0.18       & 0.11, 0.19       & 0.09, 0.28  \\ \cline{2-10}
    
        & Regex           & 0.04, 0.25      & 0.06, 0.30       & 0.09, 0.41       & 0.27, 0.51       & 0, 0.14       & 0.03, 0.28       & 0.01, 0.38       & 0.11, 0.48  \\ \cline{2-10}
    
        % Python-domains  & 0.16, 0.29      & 0.09, 0.27       & 0.10, 0.39       & 0.13, 0.49       & 0.08, 0.13       & 0.06, 0.24       & 0.03, 0.40       & 0.03, 0.50  \\ \hline
    
        & DataPrep  & 0.08, 0.22      & 0.09, 0.25       & 0.10, 0.38      & 0.12, 0.49       & 0.05, 0.14       & 0.06, 0.24       & 0.03, 0.40       & 0.03, 0.50 \\ \cline{2-10}
    
        & Validators  & 0.04, 0.29     & 0.03, 0.29       & 0.01, 0.31       & 0.01, 0.44       & 0, 0.03       & 0, 0.05       & 0, 0.30      & 0.03, 0.44  \\ \hline
        
       \multirow{2}{*}{\makecell{Data-cleaning}} & AutoDetect  & 0, 0.18      & 0, 0.17       & 0, 0.18       & 0, 0.25       & 0, 0.09       & 0, 0.12       & 0, 0.15       & 0.01, 0.25  \\ \cline{2-10}
    
        & Katara  & 0, 0.04       & 0, 0.05        & 0, 0.10        & 0, 0.20       & 0, 0.03        & 0, 0.05       & 0, 0.10       & 0, 0.19  \\ \hline

        \multirow{6}{*}{\makecell{Outlier detection methods}} & SVDD  & 0.04, 0.04      & 0.06, 0.06       & 0.09, 0.10       & 0.09, 0.15       & 0.05, 0.04       & 0.06, 0.06       & 0.03, 0.07       & 0.03, 0.12  \\ \cline{2-10}

        & DBOD  & 0, 0.15      & 0, 0.23       & 0, 0.35       & 0, 0.46       & 0, 0.12       & 0, 0.29       & 0, 0.40       & 0, 0.51  \\ \cline{2-10}
        
        & LOF  & 0, 0.08      & 0, 0.12       & 0, 0.18       & 0, 0.24       & 0, 0.04       & 0, 0.12       & 0, 0.16       & 0, 0.22  \\ \cline{2-10}

        & RKDE  & 0.04, 0.20      & 0.06, 0.24      & 0.09, 0.31       & 0.24, 0.40       & 0.05, 0.11       & 0.06, 0.21       & 0.03, 0.27       & 0.12, 0.35  \\ \cline{2-10}

        & PPCA  & 0, 0.14      & 0, 0.15      & 0, 0.19       & 0.17, 0.26       & 0, 0.06       & 0, 0.12       & 0, 0.15       & 0, 0.20  \\ \cline{2-10}

        & IForest  & 0, 0.13      & 0, 0.15       & 0, 0.19       & 0.11, 0.25       & 0, 0.05       & 0, 0.13       & 0.11, 0.17       & 0.12, 0.22  \\ \hline

        \multirow{\iftoggle{full}{6}{4}}{*}{GPT} & few-shot-with-COT      & 0, 0.20      & 0, 0.30       & 0, 0.38      & 0, 0.56       & 0, 0.16       & 0, 0.33       & 0, 0.48       & 0, 0.53   \\ \cline{2-10}
    
        & few-shot-no-COT   & 0, 0.20     & 0, 0.32       & 0, 0.38       & 0, 0.56       & 0, 0.10       & 0, 0.22       & 0, 0.44       & 0, 0.56 \\ \cline{2-10}
    
        & zero-shot-with-COT   &0, 0.15     & 0, 0.28       & 0, 0.34       & 0, 0.53       & 0, 0.16       & 0, 0.26       & 0, 0.43       & 0, 0.52 \\ \cline{2-10}
    
        & zero-shot-no-COT   & 0, 0.11     & 0, 0.23       & 0, 0.25       & 0, 0.43       & 0, 0.08       & 0, 0.21       & 0, 0.40       & 0, 0.46 \\ \cline{2-10}
    
    %    & GPT-finetune-lr1   & 0,0.02     & 0.33, 0.43        & 0.57, 0.67        & 0.58, 0.70        & 0, 0.03        & 0.38, 0.45       & 0.48, 0.61        & 0.46, 0.63  \\ \cline{2-10}
    
    %    & GPT-finetune-lr0.1   & 0, 0.05     & 0, 0.42        & 0.65, 0.65        & 0.75, 0.79        & 0, 0.04        & 0, 0.54       & 0.73, 0.79        & 0.77, 0.82  \\ \hline
    
    \iftoggle{full}
    {
       & GPT-finetuned   & 0, 0.05     & -        &  -       &  -       & 0, 0.04        &  -      &  -        &  -  \\ \hline
    }
    {
    \hline
    }

          \multirow{2}{*}{Commercial}  & Vendor-A   &  0, 0.18    &   0, 0.20      &  0, 0.22       &  0, 0.27       &  0, 0.02       &  0, 0.05      &  0, 0.11       &  0, 0.21 \\ \cline{2-10}
            & Vendor-B   &   0, 0.02   &    0, 0.05     &  0, 0.10       &  0, 0.21       &   0, 0.02      &  0, 0.05      &  0, 0.11       &  0, 0.21  \\ \hline
        
        \end{tabu}
    }
    % \vspace{-4mm}
    \vspace{-3mm}
\end{table*}
\end{small}

\iftoggle{full}
{
    \begin{figure*}%[t]
    \centering
    \begin{tabular}{c c}
    \begin{minipage}{6.5cm}
    \centering
    \includegraphics[width = \linewidth]{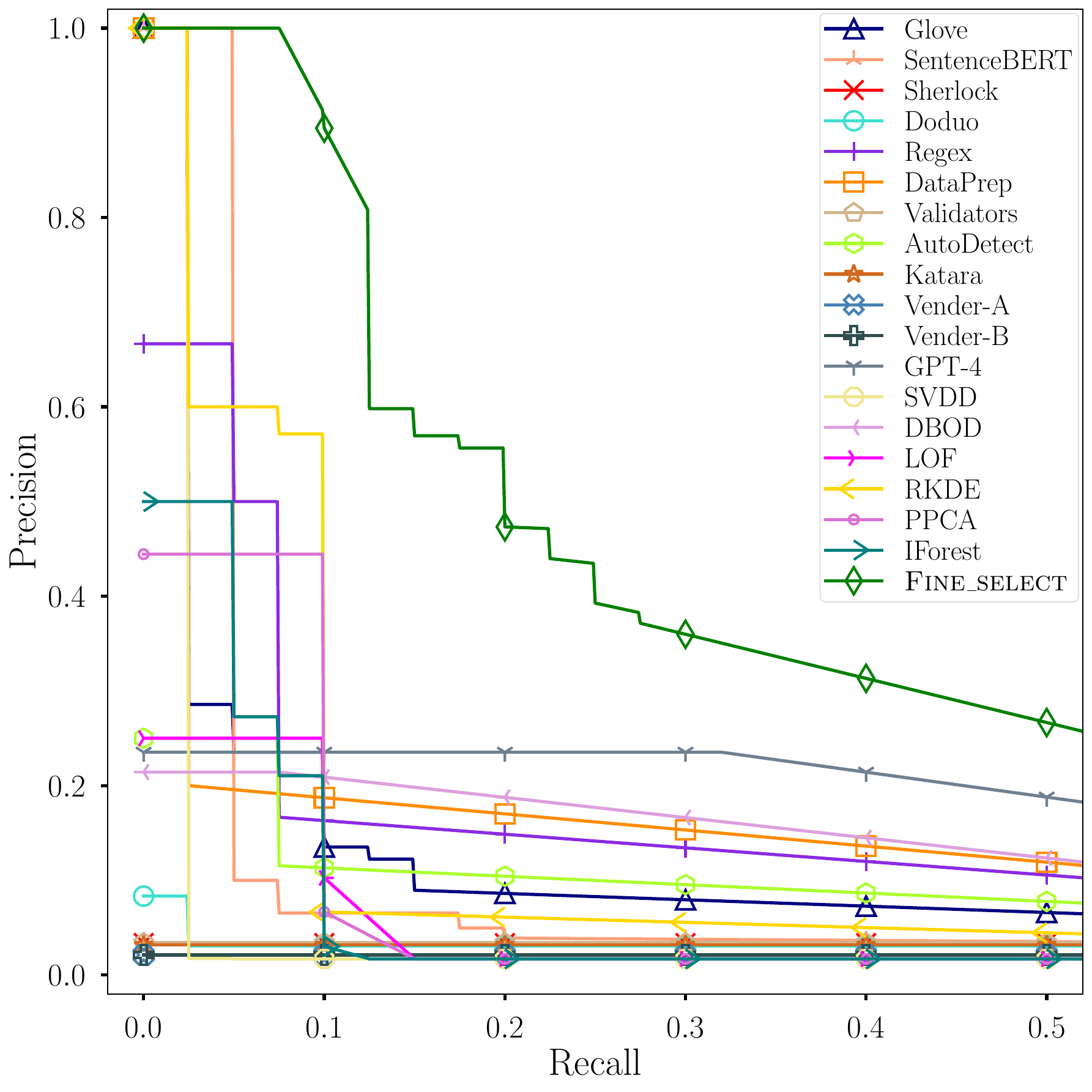}
    \end{minipage}
    &
    \begin{minipage}{6.5cm}
    \centering
    \includegraphics[width = \linewidth]{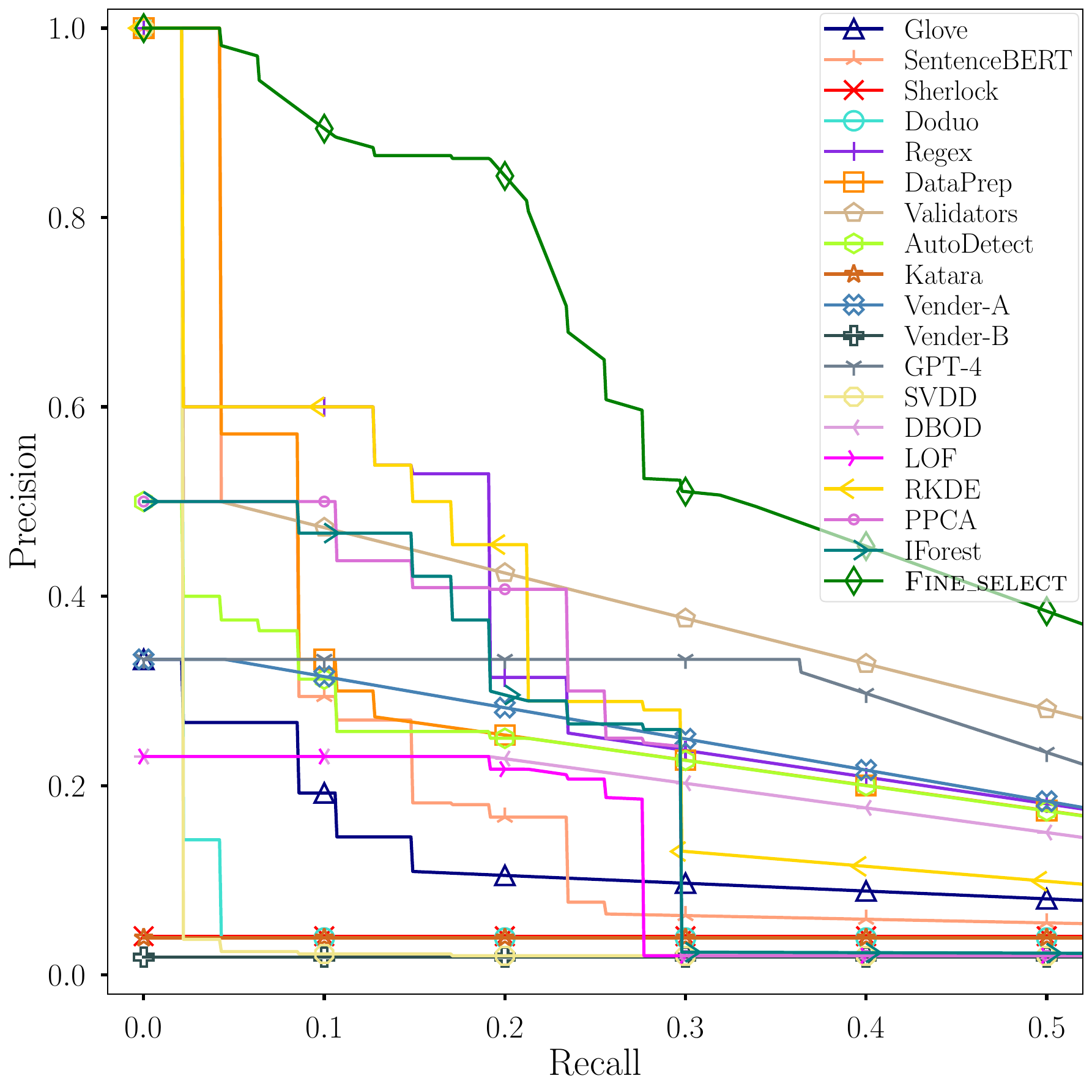}
    \end{minipage}
    % \vspace{-4mm}
    \\
    \begin{minipage}{6.5cm}
    \centering
    \captionof{figure}{PR curves of all methods on the 1200-column  \rttest (\rttesta) (Trained on \rttrain)
    }
    \label{fig:pr_benchmark_pbi_benchmark_rule_PBICSV}
    \end{minipage}
    &
    \begin{minipage}{6.5cm}
    \centering
    \captionof{figure}{PR curves of all methods on the 1200-column \sttest (\sttesta) (Trained on \rttrain)
    } 
    \label{fig:pr_benchmark_excel_benchmark_rule_PBICSV}
    \end{minipage}
    \\
    \end{tabular}
    \end{figure*}
}
{
    \begin{figure*}[t]

    \centering
    \begin{tabular}{c c c}
    % \hspace{-10mm}
    \begin{minipage}{4.2cm}
    \centering
    \includegraphics[width=\linewidth]{figures/AutoTest_experiment/pr_benchmark/pr_benchmark_pbi_benchmark_rule_PBICSV.pdf}
    \end{minipage}
    &
    % \hspace{-5mm}
    \begin{minipage}{4.2cm}
    \centering
    \includegraphics[width=\linewidth]{figures/AutoTest_experiment/pr_benchmark/pr_benchmark_excel_benchmark_rule_PBICSV.pdf}
    \end{minipage}
    &
    \begin{minipage}{4.8cm}
    \centering

    \includegraphics[width=\linewidth]{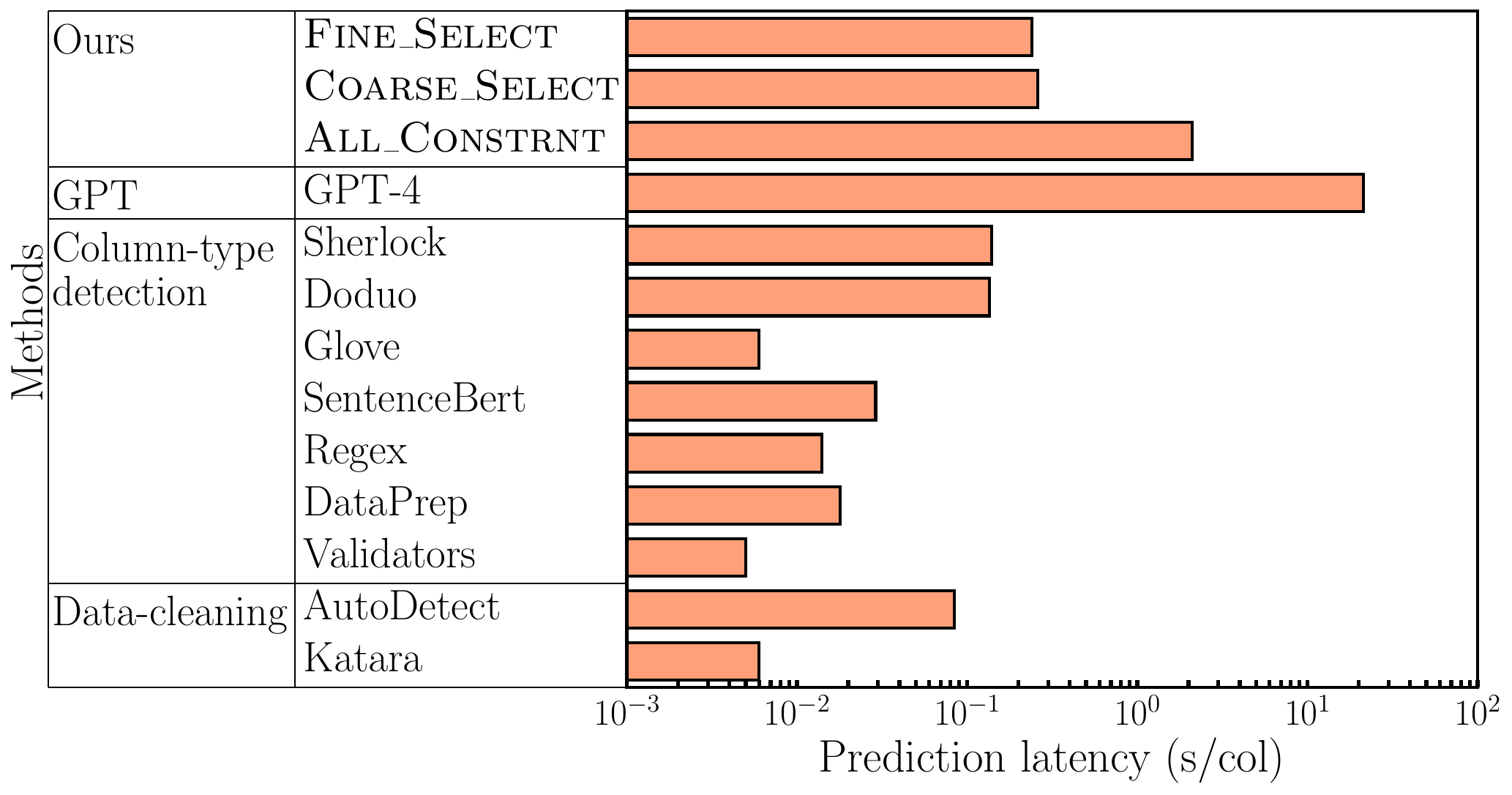}
    
    \end{minipage}
    % \vspace{-4mm}
    \\
    \begin{minipage}{4.2cm}
    \centering
    \vspace{-4mm}
    \captionof{figure}{PR curves on the 1200-column  \rttesta, trained on \rttrain.}
    \label{fig:pr_benchmark_pbi_benchmark_rule_PBICSV}
    \end{minipage}
    &
    \begin{minipage}{4.2cm}
    \centering
    \vspace{-4mm}
    \captionof{figure}{PR curves on the 1200-column \sttesta, trained on \rttrain.}
    \label{fig:pr_benchmark_excel_benchmark_rule_PBICSV}
    \end{minipage}
    &
    \begin{minipage}{4.8cm}
    \centering
    \vspace{-4mm}
    \captionof{figure}{Online prediction latency: average time to process one column (all methods)}
    \label{fig:efficiency_comaprison}
    \end{minipage}
    \\
    \end{tabular}
    \vspace{-3mm}
    \end{figure*}
}

\iftoggle{full}{
    \begin{figure*}%[t]
    \centering
    \begin{tabular}{c c}
    \begin{minipage}{6.5cm}
    \centering
    \includegraphics[width=\linewidth]{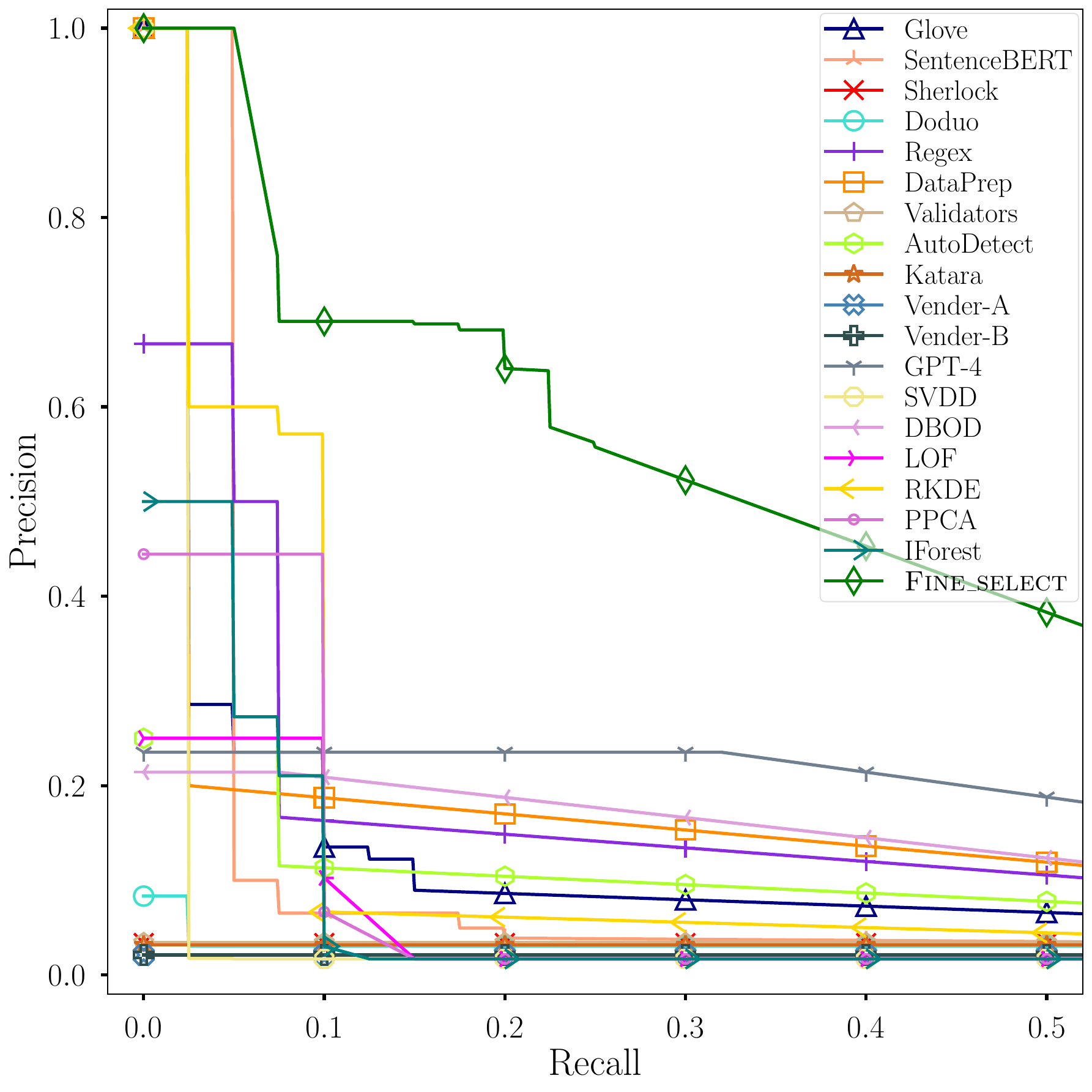}
    \end{minipage}
    &
    \begin{minipage}{6.5cm}
    \centering
    \includegraphics[width=\linewidth]  {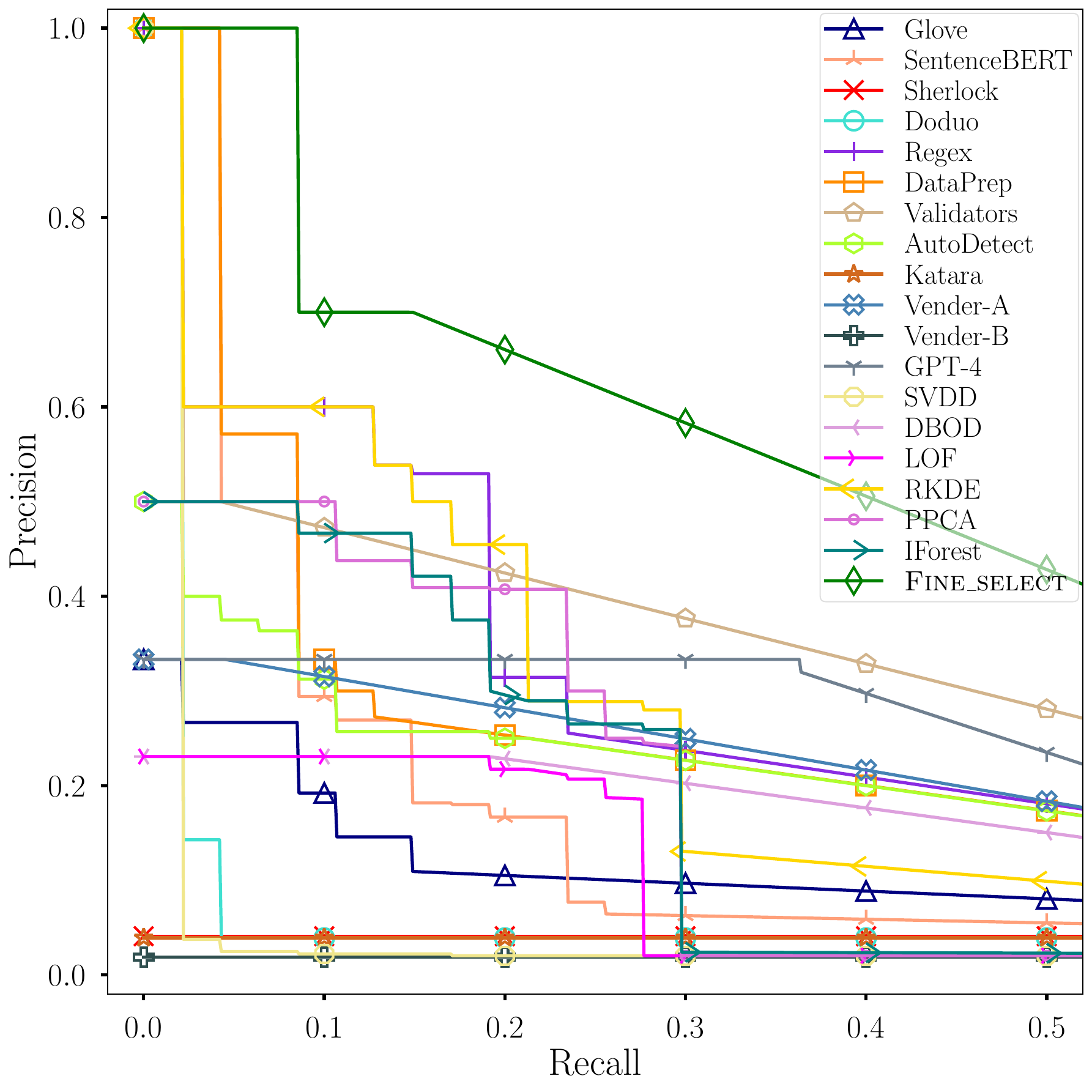}
    \end{minipage}
    \\
    \begin{minipage}{6.5cm}
    \centering
    \captionof{figure}{PR curves on the 1200-column \rttesta, trained on \tablib.
    }
    \label{fig:pr_benchmark_pbi_benchmark_rule_Tablib}
    \end{minipage}
    &
    \begin{minipage}{6.5cm}
    \centering
    \captionof{figure}{PR curves on the 1200-column \sttesta, trained on \tablib.}
    \label{fig:pr_benchmark_excel_benchmark_rule_Tablib}
    \end{minipage}
    \\
    \end{tabular}
    \end{figure*}

}
{
}

\underline{Quality comparison with real errors}. In Figure~\ref{fig:pr_benchmark_pbi_benchmark_rule_PBICSV} and \ref{fig:pr_benchmark_excel_benchmark_rule_PBICSV}, we compare the PR-curves of all methods, on two benchmarks \rttest (\rttesta) and \sttest (\sttesta), respectively.
To avoid clutter in these figures, we show the best method \fs from the \at family, trained using \rttrain 
\iftoggle{full}
{
    .
}
{
    (additional results can be found in our full technical report~\cite{full}). 
}
Similarly, for methods in GPT-4 family, we also only show few-shot-with-COT since it performs the best, as can be seen in Table~\ref{tab:performance_summary}.
%For better readability, we only show \fs for our approaches, and only plot the curve with fewshot and COT for GPT-4-based approaches, as suggested by Table~\ref{tab:performance_summary}, which indicates that this setting yields the best performance.

The proposed \fs substantially outperforms all other methods. It is worth noting that \fs trained using \rttrain not only performs well on \rttesta, but also on \sttesta, demonstrating strong generalizability  (spreadsheet vs. relational tables).

Among all baselines, SentenceBERT,  DataPrep, and Regex perform better than other domain-based baselines, while RKDE performs better than other outlier detection baselines, but these methods still lag significantly behind the proposed \fs.  Note that while GPT-4
\iftoggle{full}
{
    and fine-tuned GPT-4 
}
{}
can detect many data errors (around 80\%), it also produce a large number of false-positives (especially on columns involving code-names, abbreviations, and proprietary vocabularies that are not standard English), which affects its quality.

%and fails to maintain high precision even among the top-ranked columns. This indicates that users would need considerable effort to distinguish TPs in the reported columns. The fine-tuned GPT-4 model tends to overfit to synthetic errors during training, resulting in poorer performance on both benchmarks compared to directly prompting without fine-tuning.

In Table~\ref{tab:performance_summary}, we further summarize the PR-curves using two metric numbers:  (1) F1@P=0.8, and (2) PR-AUC, both of which show a picture similar to what we observe on the PR-curves, where \fs outperforms alternatives methods. 

\iftoggle{full}
{
To understand the generalizability of our proposed method, we additionally use a different table corpora called TabLib~\cite{tablib} as the training data, and report the resulting PR curves in Figure~\ref{fig:pr_benchmark_pbi_benchmark_rule_Tablib} and Figure~\ref{fig:pr_benchmark_excel_benchmark_rule_Tablib}. We observe similar trends as in Figure~\ref{fig:pr_benchmark_pbi_benchmark_rule_PBICSV} and Figure~\ref{fig:pr_benchmark_excel_benchmark_rule_PBICSV}, where \at consistently outperforms all 21 alternative baselines, demonstrating the generalizability of our approach to different underlying corpora.
}

\underline{Quality comparison with real and synthetic errors}. In addition to testing on the real \rttesta and \sttesta, we further report 3  settings for each of the benchmark in Table~\ref{tab:performance_summary}, where we inject synthetic errors (using values randomly sampled from other columns), at 5\%/10\%/20\% levels, on top of real errors. We observe that  \fs continues to dominate all other methods, confirming its effectiveness across a spectrum of error rates.

\begin{figure*}
    \centering
    \vspace{-2mm}
    \includegraphics[width=1 \linewidth]{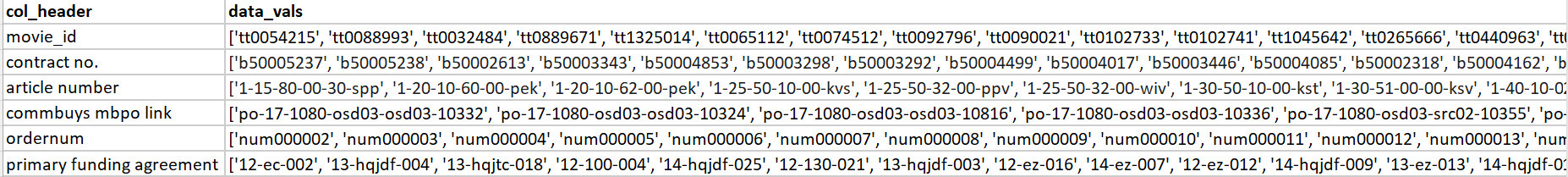}
    \vspace{-6mm}
    \caption{Examples columns with specialized meanings, but are still ``covered'' by SDCs: each row here corresponds to a real data column, with its column-header and data-values listed. Many of these columns convey specialized meanings (e.g., specialized contract no., article number, etc.), which are nevertheless covered by our pattern-based SDCs, as our method learns a generalized notation of what a reliable pattern-domain may look like, which transcends specific meanings in each column.
    }
    \vspace{-2mm}
    \label{fig:specialized-ex-pattern}
\end{figure*}

\underline{Coverage of specialized content.}  In addition to precision/recall, a question we want to explore is whether the generated \sdcas only cover well-known concepts commonly represented on the web. Figure~\ref{fig:specialized-ex-pattern} shows a sampled analysis, with examples of real columns that have specialized meanings and structures (corresponding to contract-number, article-number, etc.), some of which are likely unique to a specialize domain or few datasets. Our pattern-based SDCs can nevertheless reliably install pattern-based SDCs for such columns, as the method learns a generalized notation of what a  reliable domain-pattern should look like, which transcends specific meanings conveyed in the data, therefore providing ``coverage'' even when the underlying domains may be highly specialized.

\iftoggle{full}
{
We show additional quality results, such as training using different corpora (\sttrain), in Appendix~\ref{apx:train-on-excel}.   
}
{
We show additional results, e.g., training using different corpora, in~\cite{full} in the interest of space.
}

%\yeye{mention about generalizability (train pbi, test excel). Mention about train on excel results in appendix.} 
%For each testing dataset, we mark the method that achieves the best performance in bold. We observe that our methods greatly outperform the competitors in all settings, and in most of the setting the best performance is achieved by \fs. 

%We compared our methods against competitors on testing datasets with different levels of error rates. The testing datasets include the PBI and Excel benchmarks, each with approximately 2\% columns containing errors. Besides, for each benchmark, we also inject synthetic ``dirty'' columns to create testing datasets with 5\%, 10\% and 20\% error rates, respectively. 

\subsection{Efficiency Analysis}

\iftoggle{full}
{
    \begin{figure*}[ht]
    \centering
    \begin{tabular}{c c c}
    \begin{minipage}{5cm}
    \centering
    \includegraphics[width=\linewidth]{figures/AutoTest_experiment/latency/efficiency_comparison.pdf}
    \end{minipage}
    &
    \begin{minipage}{4cm}
    \centering
    % \vspace{-3mm}
    \includegraphics[width=\linewidth]{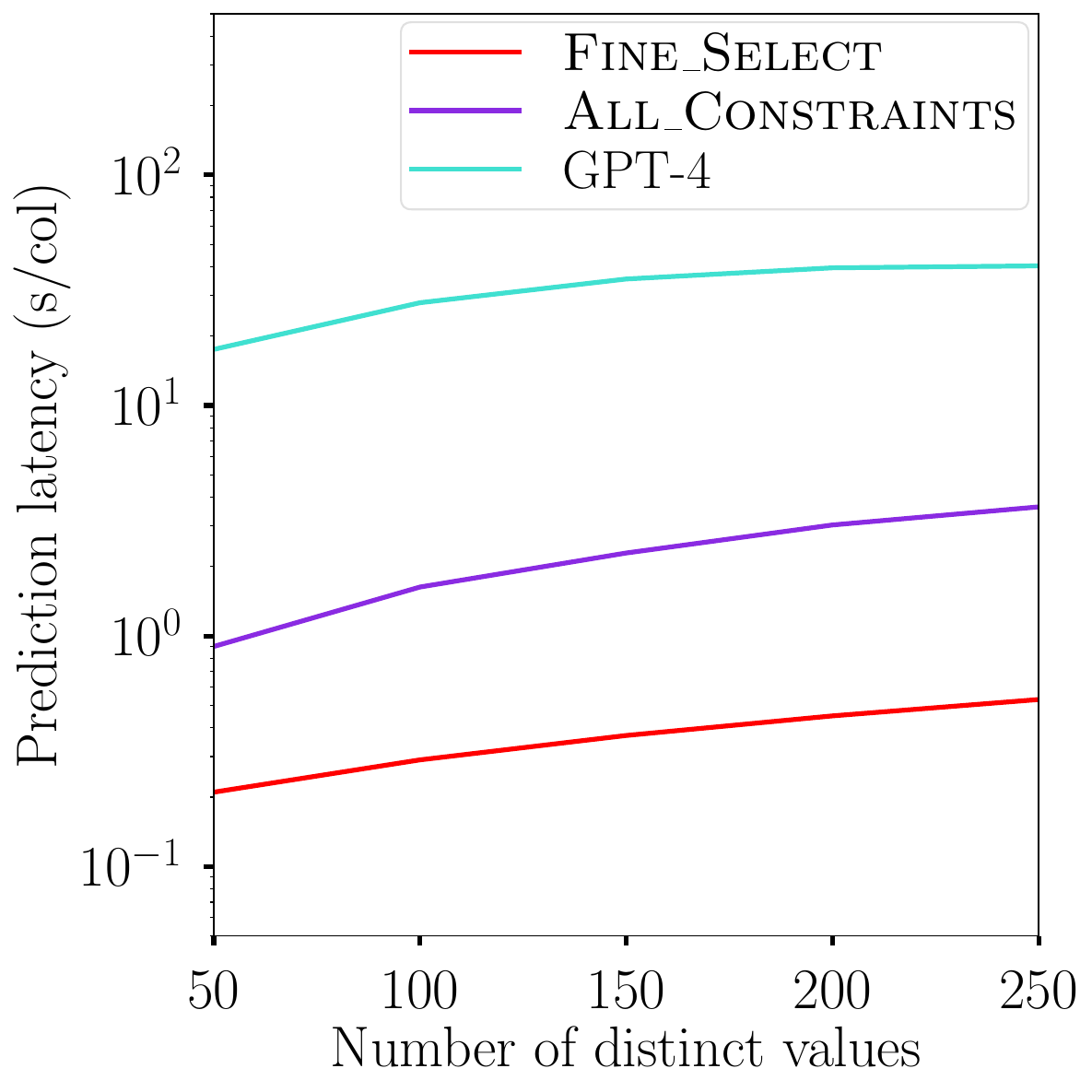}
    \end{minipage}
    &
    \begin{minipage}{4cm}
    \centering
    % \vspace{-2mm}
    \includegraphics[width=\linewidth]{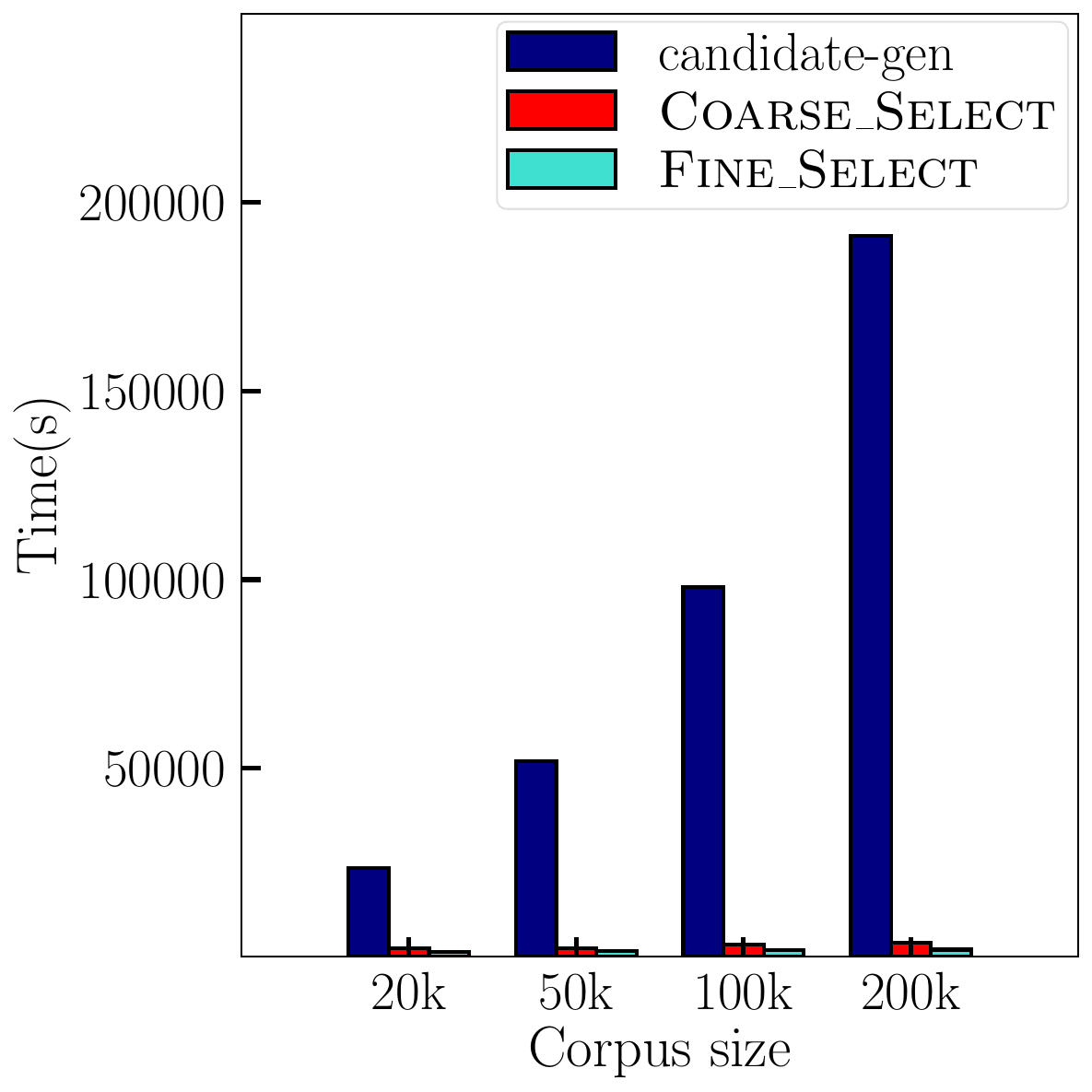} 
    \end{minipage}
    % \vspace{-6mm}
    \\
    \begin{minipage}{5cm}
    \captionof{figure}{Online prediction latency: average time to process one column (all methods)}
    \label{fig:efficiency_comaprison}
    \end{minipage}
    &
    \begin{minipage}{4cm}
    \captionof{figure}{Online prediction latency: vary column size}
    \label{fig:latency_vary_col_len}
    \end{minipage}
    &
    \begin{minipage}{4cm}
    \captionof{figure}{Offline training time: vary training corpus size}
    \label{fig:offline_training_time}
    \end{minipage}
    \end{tabular}
    \end{figure*}
}
{
    % \begin{figure}[th]
    %     \centering
    %     \iftoggle{coarse-select}{
    %         \includegraphics[scale = 0.21]{figures/AutoTest_experiment/latency/efficiency_comparison.pdf}
    %     }
    %     {
    %         \includegraphics[scale = 0.21]{figures/AutoTest_experiment/latency/efficiency_comparison_nocs.pdf}
    %     }
    %     \vspace{-4mm}
    %     \captionof{figure}{Online prediction latency: average time to process one column (all methods)}
    %     \label{fig:efficiency_comaprison}
    % \end{figure}
}

% \begin{figure}
%     \centering
%     \includegraphics[scale = 0.24]{figures/AutoTest_experiment/latency/offline_training_time.eps}
%     \captionof{figure}{Offline training time: vary training corpus size}
%     \label{fig:offline_training_time}
% \end{figure}

\underline{Online prediction latency}.
Figure~\ref{fig:efficiency_comaprison} shows the average latency of making predictions for one column. The proposed \fs takes around 0.2 seconds on average, which is interactive and suitable for user-in-the-loop scenarios. \ar in comparison, is an order of magnitude slower, showing the benefit \fs in compressing and selecting most beneficial \sdcas.  GPT-4 is the slowest as it takes over 20 seconds for one column on average.

\iftoggle{full}
{
    Figure~\ref{fig:latency_vary_col_len} shows the time it takes to process one column on the y-axis, with varying the number of distinct values in a column on the x-axis. %The running time for all methods grow when the number of distinct values in the input column grow, column contains more values. 
    \fs is 5 to 7 times faster than \ar, which is important as end-user scenarios (e.g. spreadsheets) require interactive speeds. GPT-4 is 80 times slower in comparison.

    \underline{Offline training efficiency}. Figure~\ref{fig:offline_training_time} shows the offline training time of our methods, when we vary the size of the training data. We break down the execution time into (1) "candidate-gen": which is the time it takes to enumerate possible candidates and assess their individual quality (Section~\ref{subsec:rule_cand_generation}), 
    (2) \cs, which is the coarse-grained selection (Section~\ref{subsec:fine_select}), (3) \fs, which is the fine-grained selection (Section~\ref{subsec:fine_select}). 
    We observe that "candidate-gen" dominates the overall latency in the offline stage, which grows linearly with the corpus size, and takes about 50 hours on a \emph{single} machine for a corpus with 200,000 columns (can be easily \emph{parallelized} on Map-Reduce-like systems). The latency of \cs and \fs also increases with larger corpora, but negligible (less than 4,000 seconds on a corpus with 200,000 columns) compared to the candidate-gen step.
}
{
    Additional results on latency, including offline latency analysis, can be found in~\cite{full}.
}

\subsection{Sensitivity Analysis}
\label{subsec:sensitivity}

We analyze the sensitivity of \at to different parameters.

\begin{table*}[t]
%\vspace{-15mm}
    \caption{Quality and latency comparison for \fs, as we vary the constraint count budget ($B_{size}$)}
    \vspace{-3mm}
    \label{tab:rule_count_latency_quality}
    \centering
    \scalebox{0.67}
    {
        \begin{tabular}{|c||c|c|c|c|c||c|c|c|c|c|} \hline
                        &  \multicolumn{5}{c||}{\sttesta} & \multicolumn{5}{c|}{\rttesta} \\ \hline 
        Constraint count budget ($B_{size}$)     & 100       & 200   & 500     & 1000      &  \ar (26673) & 100       & 200   & 500     & 1000      & \ar (26673) \\ \hline \hline

        % maximum recall  & 0.55      & 0.61      & 0.63      & 0.64      & 0.73
        %                 & 0.11      & 0.29      & 0.37      & 0.39      & 0.60 
        %                  \\ \hline
                        
        Quality: F1@P=0.8 & 0.29      & 0.31      & \underline{0.34}      & \textbf{0.35}      & 0.23
                        & 0.09      & 0.11      & \textbf{0.21}      & \textbf{0.21}      & \textbf{0.21}                    \\ \hline
    
        Quality: PR-AUC          & 0.42      & 0.41      & \underline{0.44}      & \textbf{0.46}      & 0.38
                        & 0.22      & 0.27      & \textbf{0.34}      & 0.31      & \textbf{0.34}                  \\ \hline \hline
        Latency: second  per column & 0.13      & 0.16      & 0.21      & 0.23      & 1.44
                        & 0.12      & 0.18      & 0.24      & 0.26      & 2.10
                         \\ \hline
                                    
        \end{tabular}
    }
\end{table*}

\begin{small}
\begin{table*}[t]
    \vspace{-3mm}
    \captionof{table}{Sensitivity to different training corpora}
    \label{tab:corpora_sensitivity}
    \vspace{-3mm}
    \centering
    \setlength{\tabcolsep}{2.5pt} % Adjust the value to increase or decrease the space between columns
    \scalebox{0.75}
    {
        \begin{tabular}{|c||c|c|c|c||c|c|c|c|} \hline
        &  \multicolumn{4}{c||}{ \sttest (\sttesta) } & \multicolumn{4}{c|}{\rttest (\rttesta)} \\ 
        \hline
          & real &  +5\% syn err. &  + 10\% syn err. &  +20\% syn err. & real & +5\% syn err. & +10\% syn err. & +20\% syn err. \\ \hline 
              
        \rttrain & 0.34, 0.45      & 0.38, 0.52       & 0.48, 0.62       & 0.53, 0.68       & 0.21, 0.34       & 0.30, 0.46       & 0.36, 0.56       & 0.40, 0.62 \\ \hline
        
       \sttrain & 0.05, 0.30      & 0.18, 0.43       & 0.28, 0.52       & 0.45, 0.64       & 0.02, 0.29       & 0.25, 0.43       & 0.25, 0.47       & 0.27, 0.55   \\ \hline

        \tablib & 0.15, 0.45      &  0.34, 0.54       & 0.45, 0.61      & 0.53, 0.68     & 0.13, 0.41     & 0.37, 0.54       & 0.40, 0.56       & 0.46, 0.60  \\ \hline

        \end{tabular}
    }
\end{table*}
\end{small}

\underline{Sensitivity to the number of constraints.}
Table~\ref{tab:rule_count_latency_quality} shows the effect of varying the number of constraints $B_{size}$ in \fs, %\footnote{Results for \cs are similar}, 
using  \ar (with 26673 constraints) and GPT-4 as reference points. \fs shows strong efficiency benefit (7-10x faster) over \ar, while having the same or even better quality with just 500 constraints (e.g., \fs shows even higher PR-AUC and F1@P=0.8 than \ar  on \rttesta, likely because it is forced to select high-quality \sdcas).

\underline{Sensitivity to training corpora.}
We summarize the performance of using \rttrain, \sttrain and \tablib as the training corpus in Table \ref{tab:corpora_sensitivity}. Our results show that the performance with \rttrain and \tablib follows a similar trend, both are better than \sttrain. This can be attributed to the fact that \sttrain are crawled from human-generated spreadsheet tables, which tend to be noisier than the machine-generated tables in \rttrain and \tablib. This observation suggests that the quality of the training corpus plays a critical role in the effectiveness of mined \sdcas.

%We observe that the time required to check a column grows along with the number of rules applied. 
%On average, \ar requires around two 2 seconds to evaluate a column. However, by selecting a smaller set of 500 rules, the evaluation time becomes about 10 times faster on both benchmarks, as checking a single column requires only 0.2 seconds.

\iftoggle{full}
{
    \underline{Sensitivity to FPR budgets.}
    %\yeye{to discuss: this does not seem to show clear trends, move to full?}
    In Figure~\ref{fig:fine_select_vary_FPR}  and \ref{fig:coarse_select_vary_FPR}, we vary the FPR budget $B_{FPR}$ from 0.02 to 0.1 and plot the corresponding PR curves for \fs and \cs. For comparison, we also include the PR curve for \ar. The coordinate of the rightmost turning point on each PR curve indicates the overall precision and recall levels of all the prediction results for each setting.
    Our results show that adjusting the FPR budget provides a trade-off between precision and recall. A lower FPR budget allows only the highest-quality rules to be selected, which leads to higher precision in the reported columns. However, this comes at the cost of detecting fewer errors due to the reduced number of selected rules.

    \begin{figure}
    \centering
    \begin{tabular}{c c}
    \begin{minipage}{6cm}
    \centering
    \includegraphics[width=0.6\linewidth]{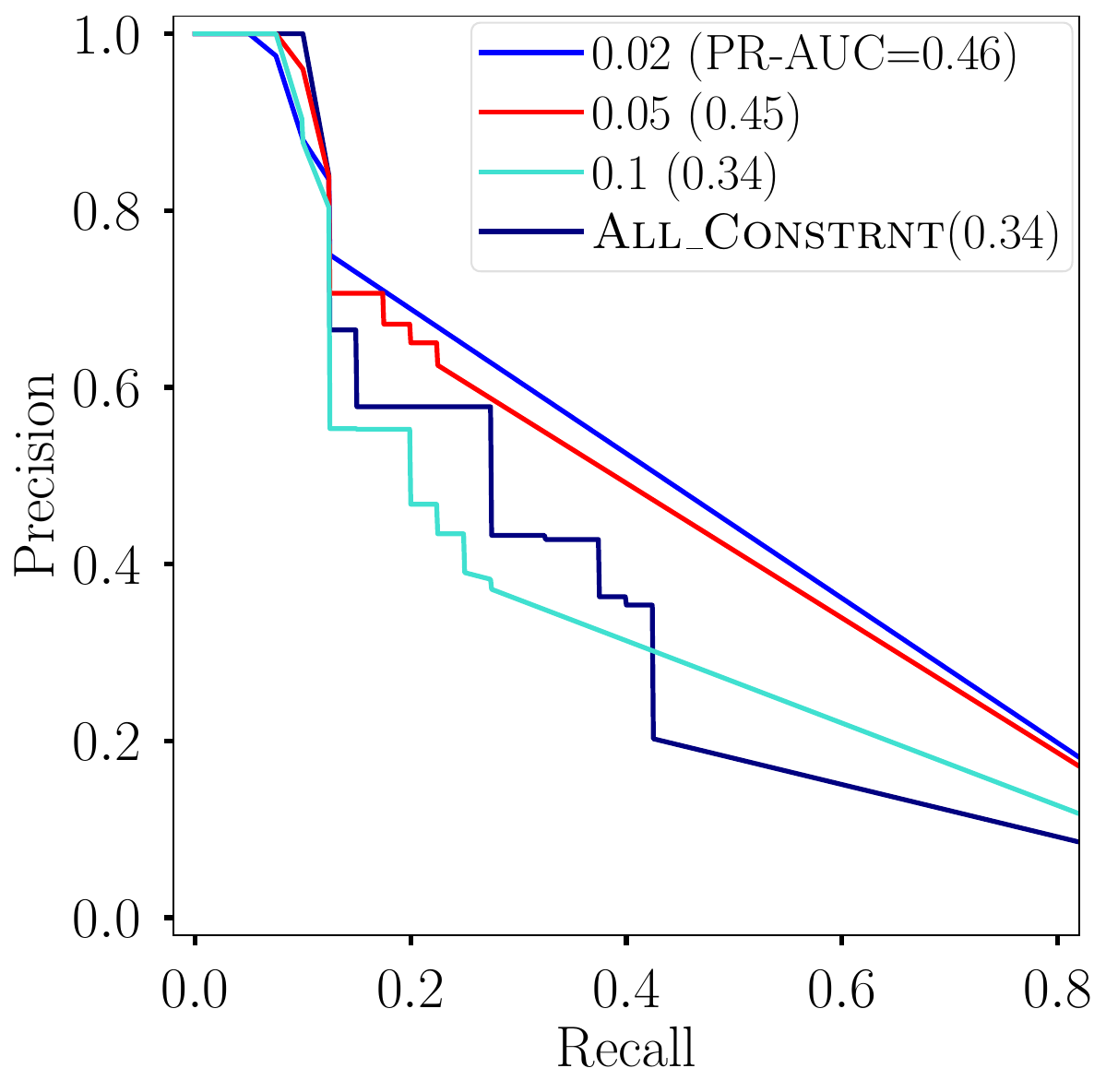}
    \end{minipage}
    &
    \begin{minipage}{6cm}
    \centering
    \includegraphics[width=0.6\linewidth]{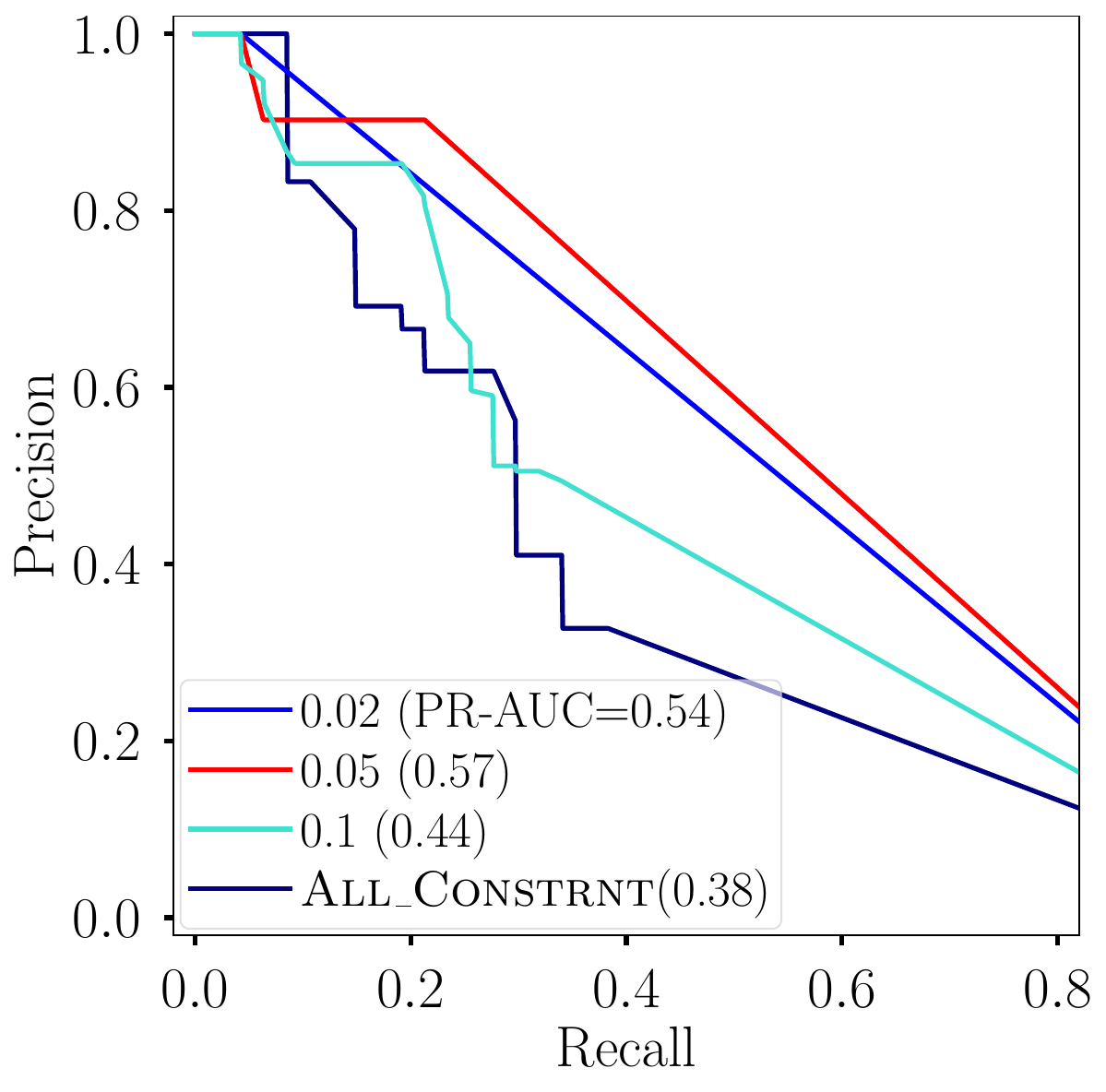}
    \end{minipage}
    \\
    (a) \rttesta
    &
    (b) \sttesta
    % \vspace{-3mm}
    \\
    \multicolumn{2}{c}{
    \begin{minipage}{8cm}
    \captionof{figure}{\fs: vary FPR budget}
    \label{fig:fine_select_vary_FPR}
    \end{minipage}
    }
    \end{tabular}
    \end{figure}

    \begin{figure}[]
    \centering
    \begin{tabular}{c c}
    \begin{minipage}{6cm}
    \centering
    \includegraphics[width=0.6\linewidth]{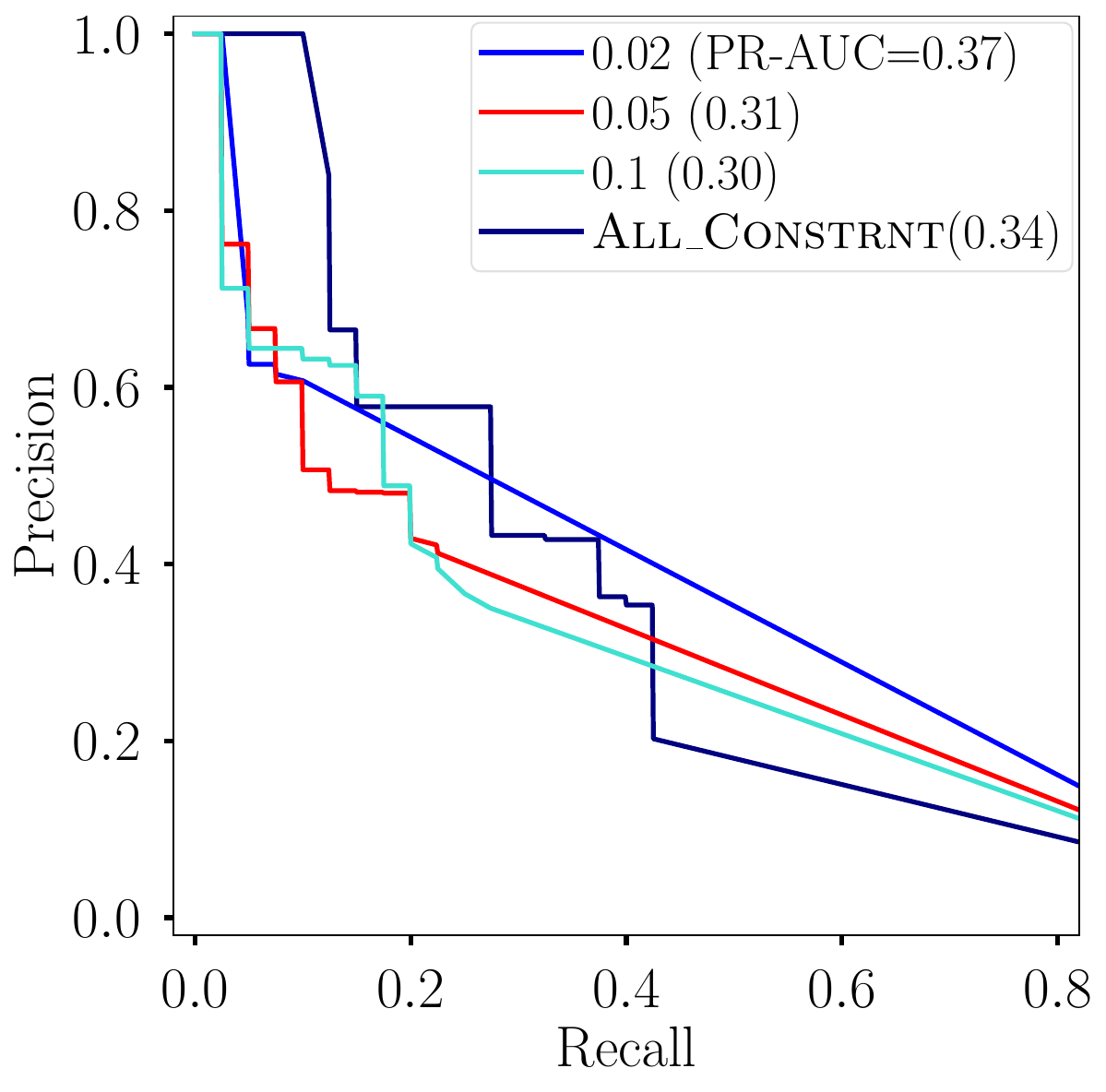}
    \end{minipage}
    &
    \begin{minipage}{6cm}
    \centering
    \includegraphics[width=0.6\linewidth]{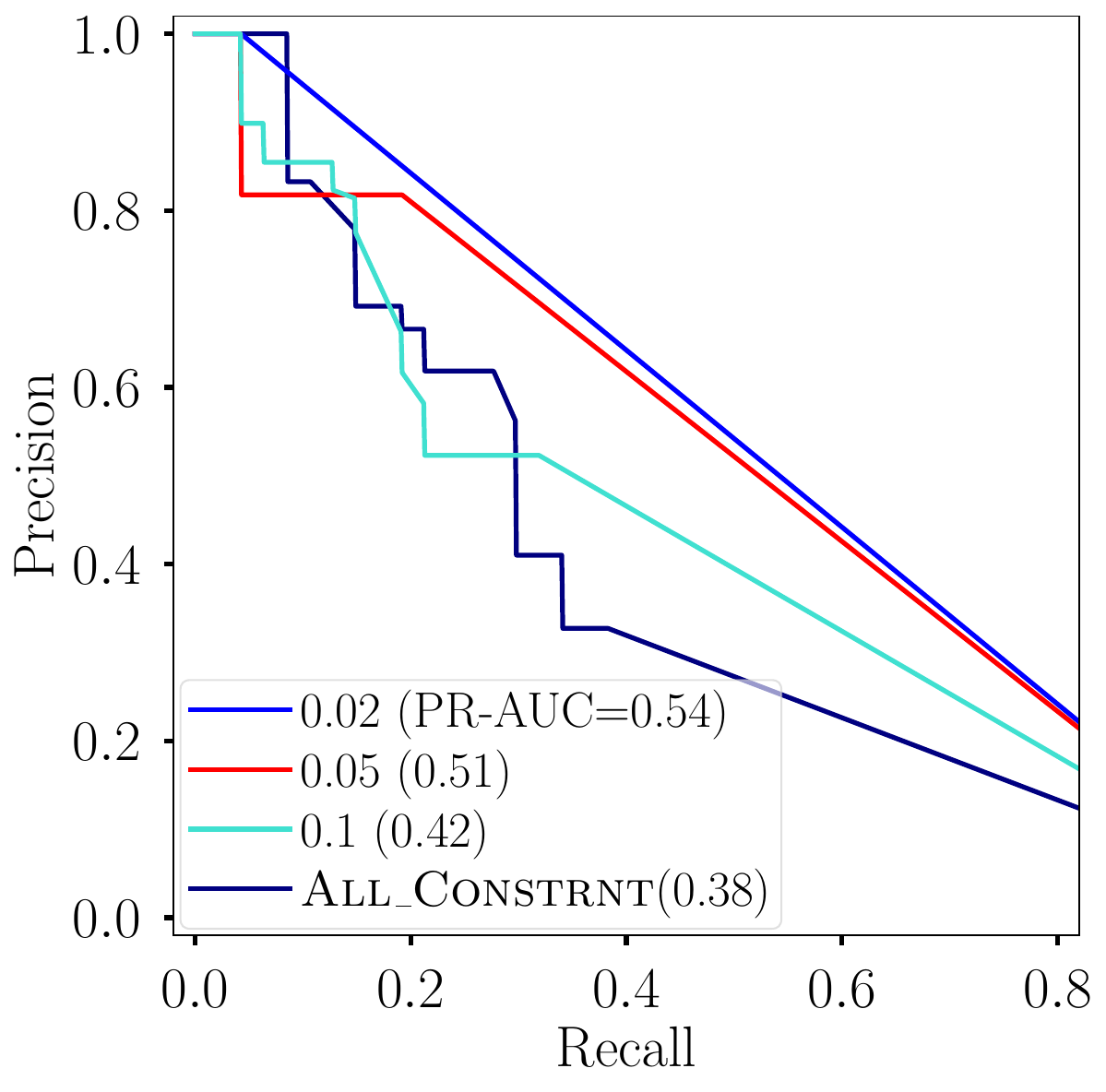}
    \end{minipage}
    \\
    (a) \rttesta
    &
    (b) \sttesta
    % \vspace{-3mm}
    \\
    \multicolumn{2}{c}{
    \begin{minipage}{8cm}
    \captionof{figure}{\cs: vary FPR budget}
    \label{fig:coarse_select_vary_FPR}
    \end{minipage}
    }
    \end{tabular}
    \end{figure}
}

\iftoggle{full}
{
    \underline{Sensitivity to constraint size budgets.} 
    We plot the PR curves of \fs and \cs with $B_{size}$ varying from 100 to 1000 in Figure~\ref{fig:fine_select_vary_rule_count} and \ref{fig:coarse_select_vary_rule_count}, respectively. 
    We observe that increasing the size budget enhances the performance, as reflected on the PR curves.
    We also notice from Figure~\ref{fig:fine_select_vary_rule_count} (a) and (b) that \fs matches or even surpass the performance of \ar when using 500 to 1000 rules. This  demonstrates the effectiveness of our rule selection process in achieving high performance even with a substantially reduced number of rules.
    %Table~\ref{tab:rule_count_latency_quality} shows the trade-off between error detection quality and latency when varying $B_{size}$ in \fs. When the number of rules increases, the quality-related metrics (i.e., F1-score at a 0.8 precision level and PR-AUC) both increases, with the cost of higher latency. We find that 500 rules is a sweet spot where the growth of error detection quality begins to plateau, while the latency remains small. 

    \begin{figure}[]
    \centering
    \begin{tabular}{c c}
    \begin{minipage}{6cm}
    \centering
    \includegraphics[width=0.6\linewidth]{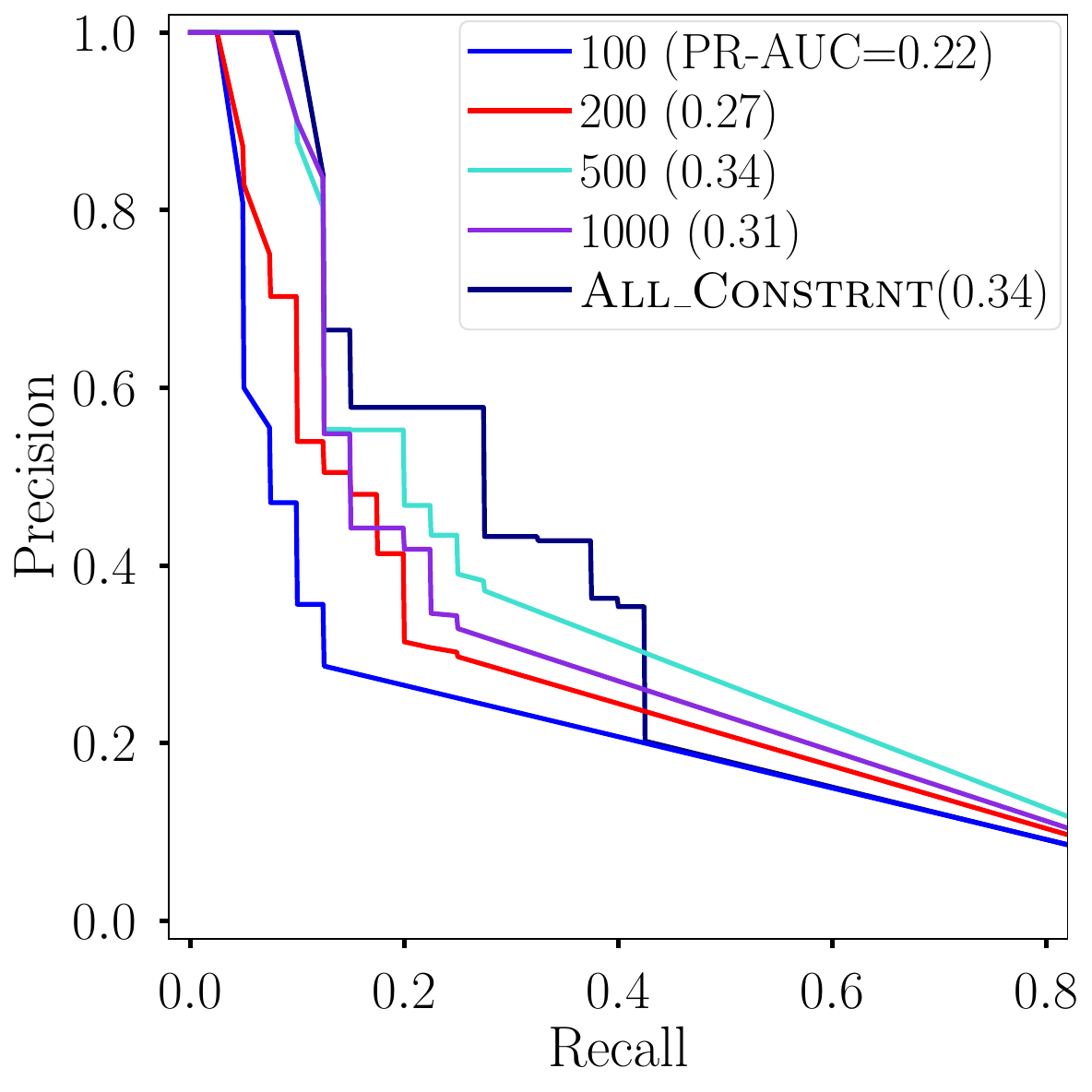}
    \end{minipage}
    &
    \begin{minipage}{6cm}
    \centering
    \includegraphics[width=0.6\linewidth]{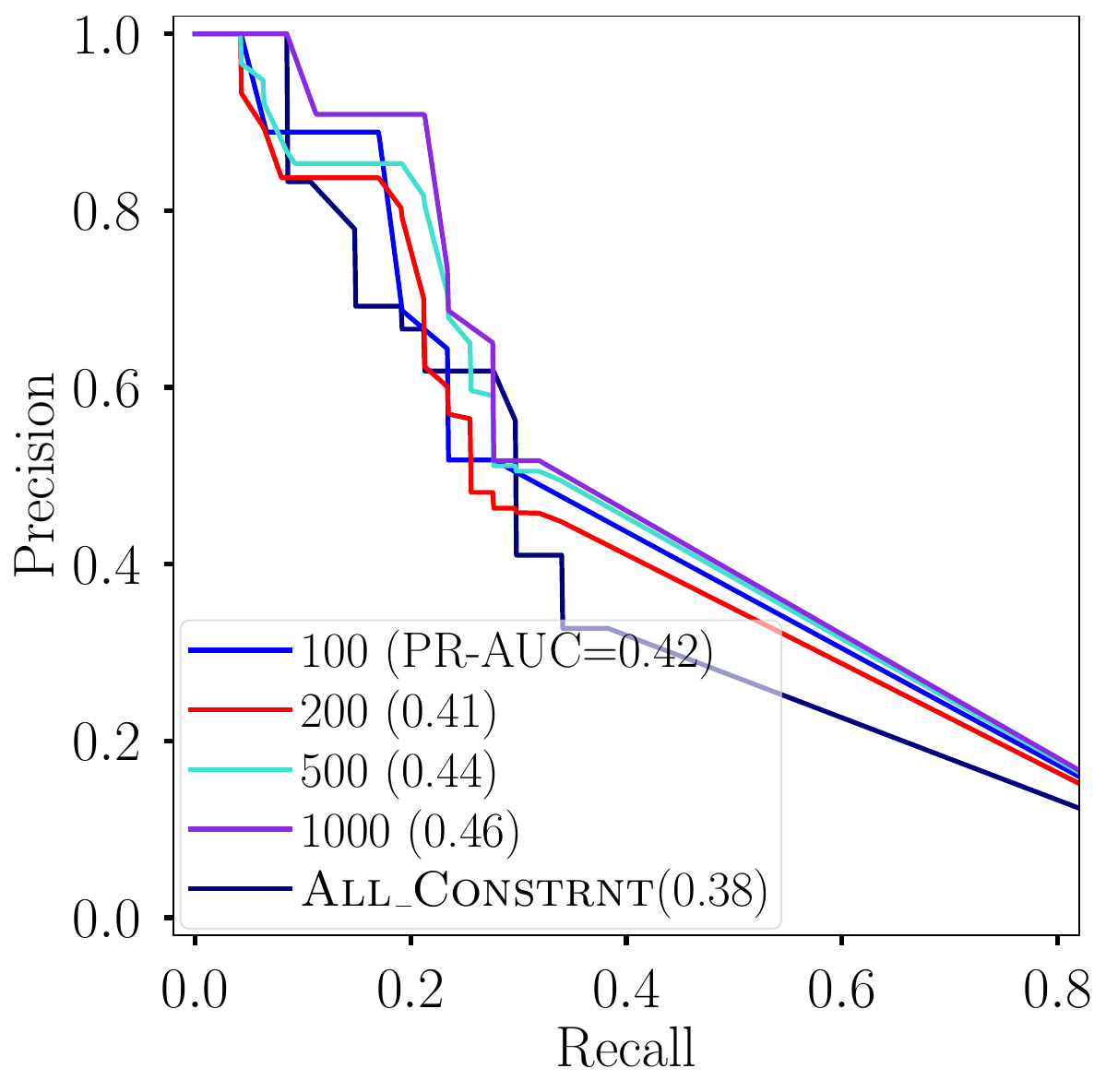}
    \end{minipage}
    \\
    (a) \rttesta
    &
    (b) \sttesta
    % \vspace{-3mm}
    \\
    \multicolumn{2}{c}{
    \begin{minipage}{8cm}
    \captionof{figure}{\fs: vary rule count budget}
    \label{fig:fine_select_vary_rule_count}
    \end{minipage}
    }
    \end{tabular}
    \end{figure}

    \begin{figure}[]
    \centering
    \begin{tabular}{c c}
    \begin{minipage}{6cm}
    \centering
    \includegraphics[width=0.6\linewidth]{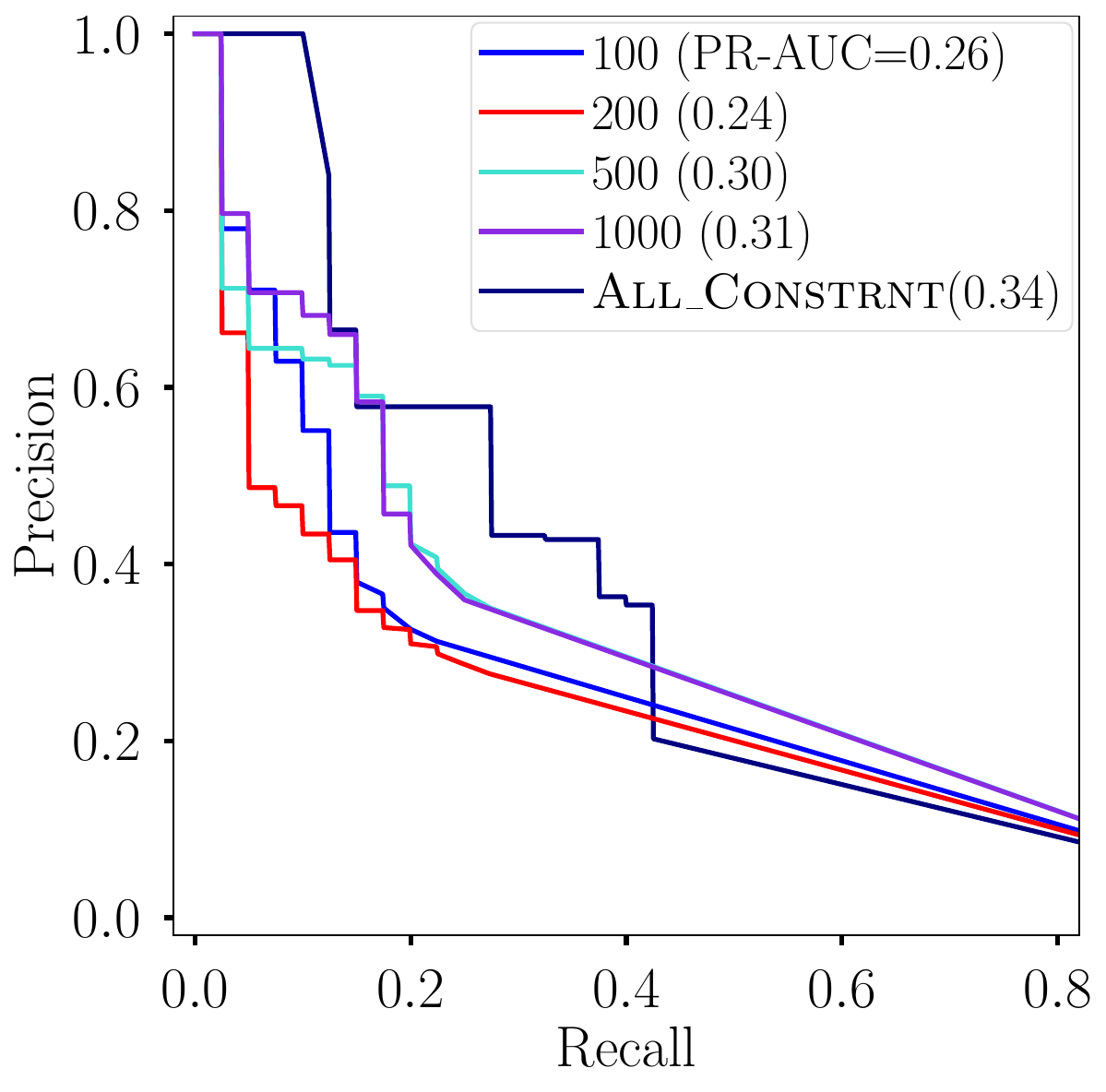}
    \end{minipage}
    &
    \begin{minipage}{6cm}
    \centering
    \includegraphics[width=0.6\linewidth]{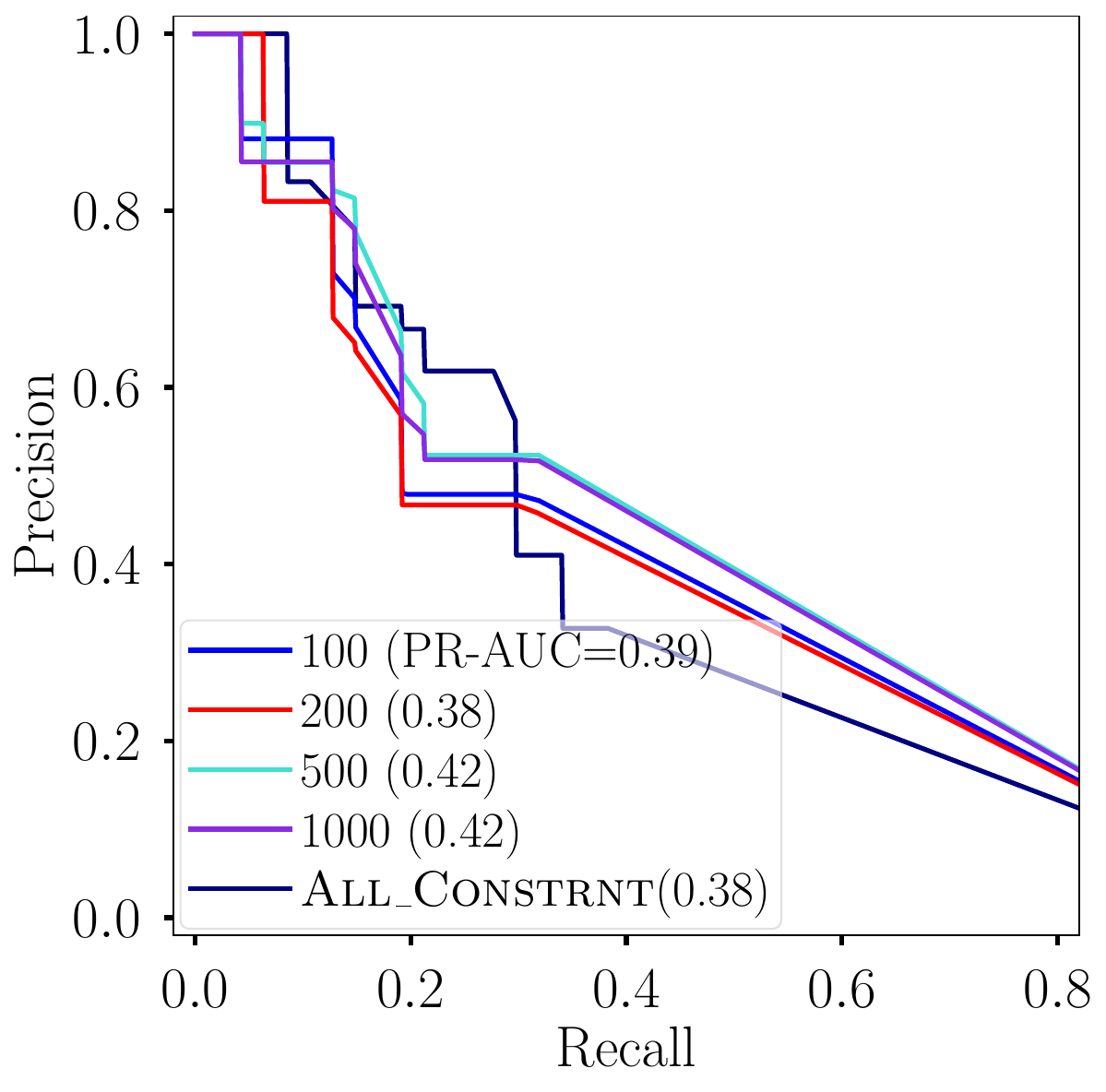}
    \end{minipage}
    \\
    (a) \rttesta
    &
    (b) \sttesta
    % \vspace{-3mm}
    \\
    \multicolumn{2}{c}{
    \begin{minipage}{8cm}
    \captionof{figure}{\cs: vary rule count budget}
    \label{fig:coarse_select_vary_rule_count}
    \end{minipage}
    }
    \end{tabular}
    \end{figure}
}

\iftoggle{full}
{
    \begin{figure}[ht]
    \centering
    \begin{tabular}{c c}
    \begin{minipage}{6cm}
    \centering
    \includegraphics[width=0.6\linewidth]{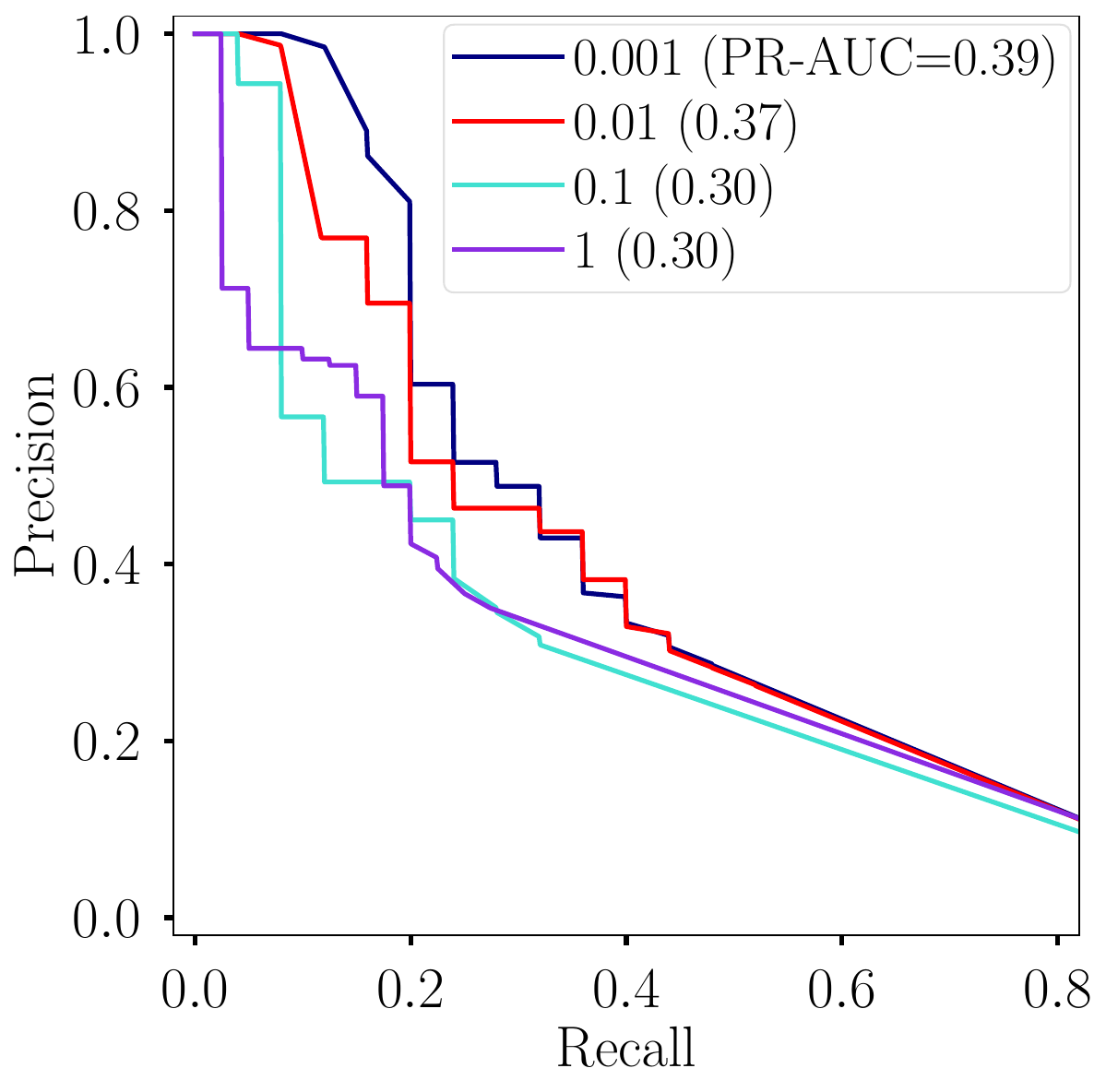}
    \end{minipage}
    &
    \begin{minipage}{6cm}
    \centering
    \includegraphics[width=0.6\linewidth]{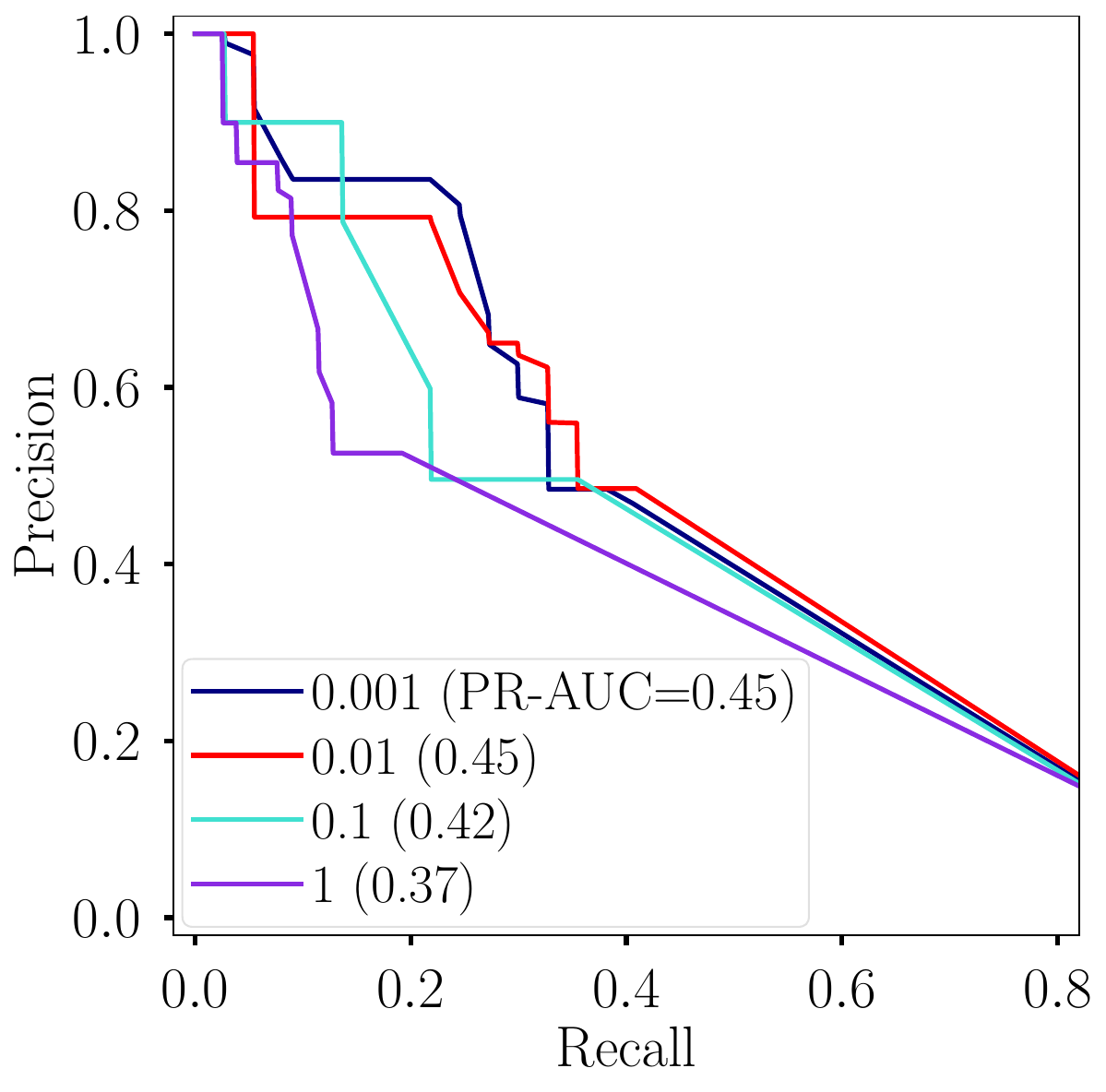}
    \end{minipage}
    \\
    (a) \rttesta
    &
    (b) \sttesta
    % \vspace{-3mm}
    \\
    \multicolumn{2}{c}{
    \begin{minipage}{8cm}
    \captionof{figure}{\fs: vary parameter $\delta$}
    \label{fig:fine_select_vary_delta}
    \end{minipage}
    }
    \end{tabular}
    \end{figure}
}
{
}

\iftoggle{full}
{
    \begin{figure}[ht]
    \centering
    \begin{tabular}{c c}
    \begin{minipage}{6cm}
    \centering
    \includegraphics[width=0.6\linewidth]{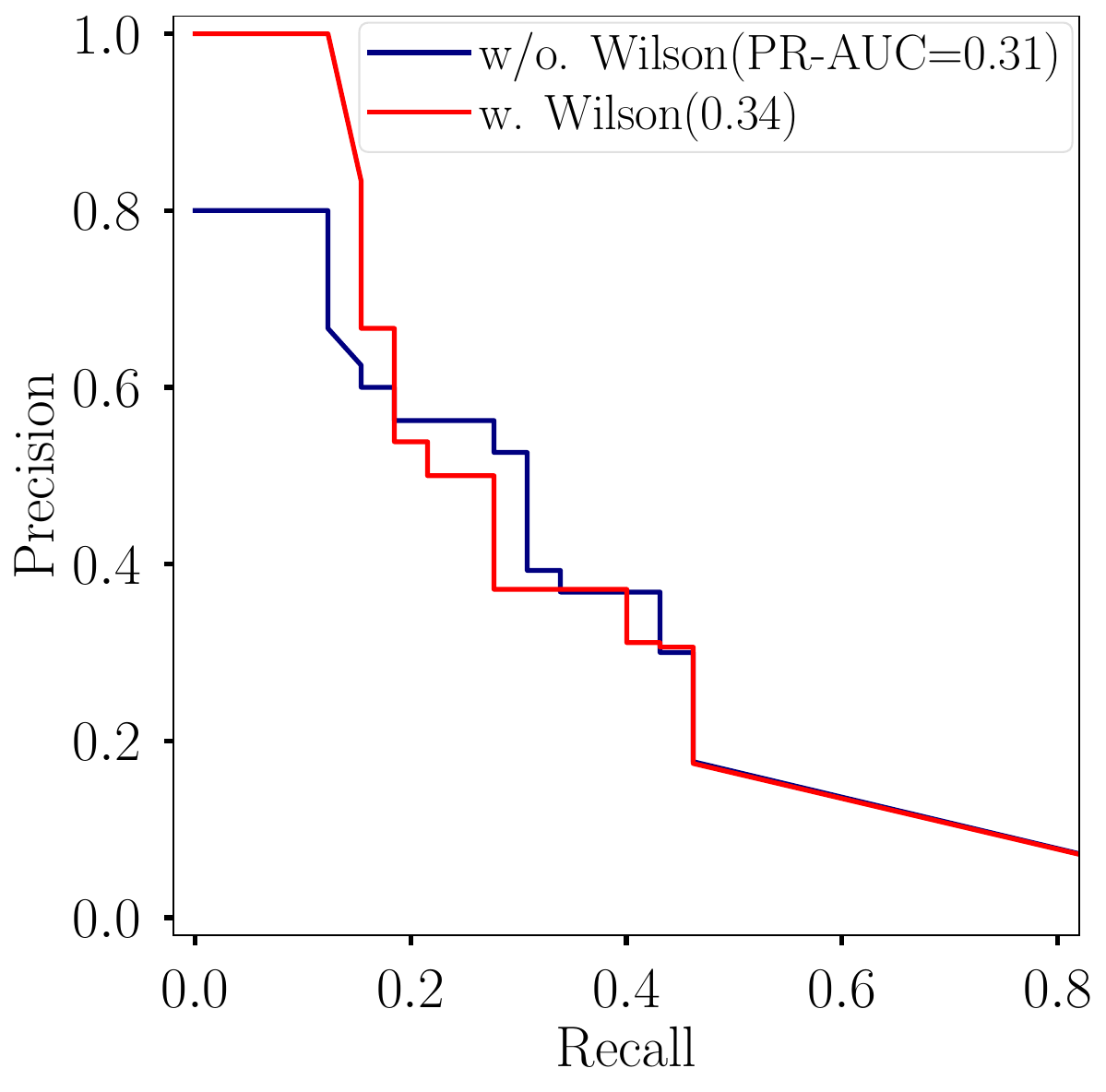}
    \end{minipage}
    &
    \begin{minipage}{6cm}
    \centering
    \includegraphics[width=0.6\linewidth]{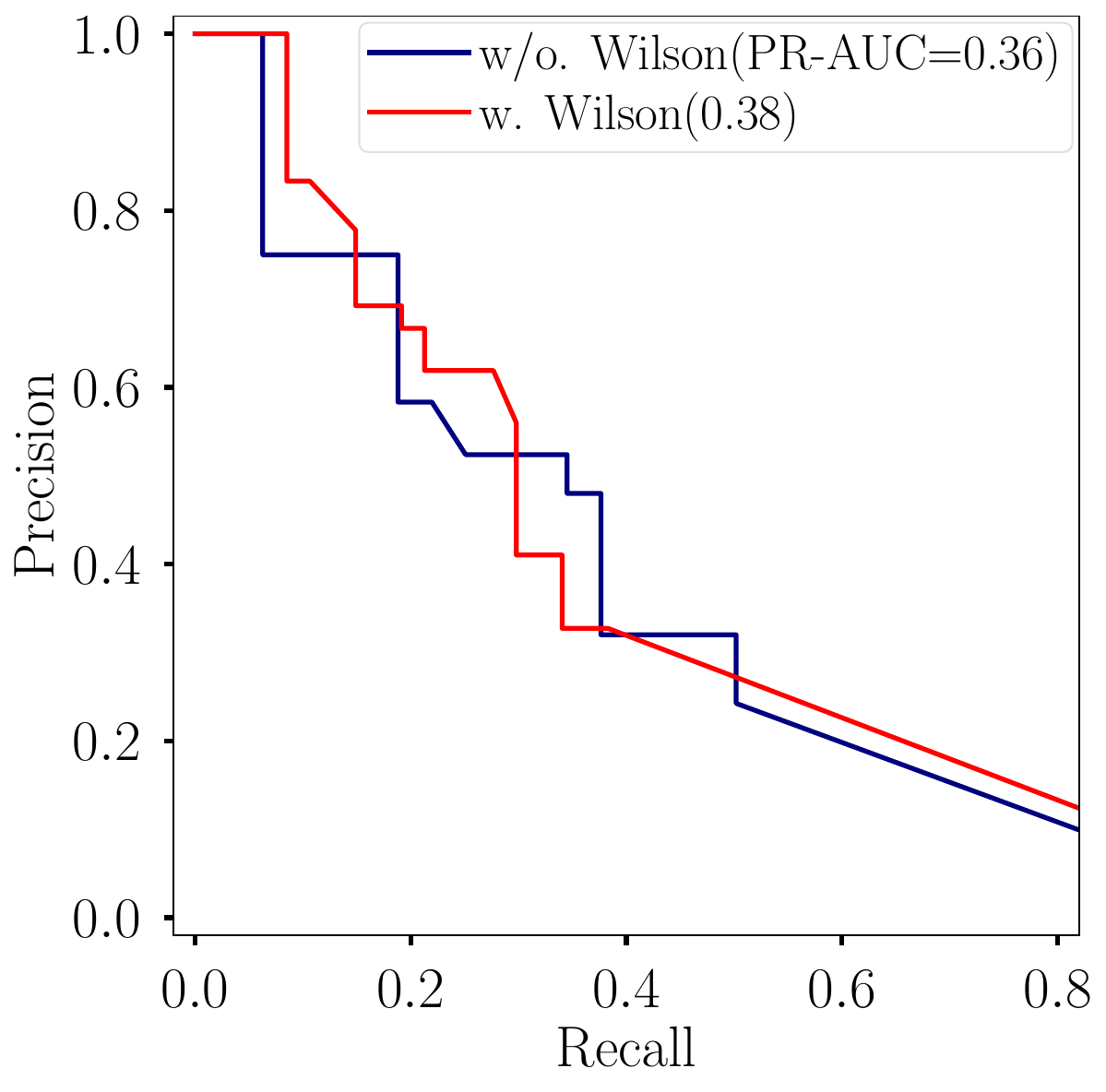}
    \end{minipage}
    \\
    (a) \rttesta
    &
    (b) \sttesta
    % \vspace{-3mm}
    \\
    \multicolumn{2}{c}{
    \begin{minipage}{8cm}
    \captionof{figure}{Sensitivity to Wilson score interval}
    \label{fig:wilson_ablation}
    \end{minipage}
    }
    \end{tabular}
    \end{figure}
}
{
}

\iftoggle{full}
{
    \begin{figure}[ht]
    \centering
    \begin{tabular}{c c}
    \begin{minipage}{6cm}
    \centering
    \includegraphics[width=0.6\linewidth]{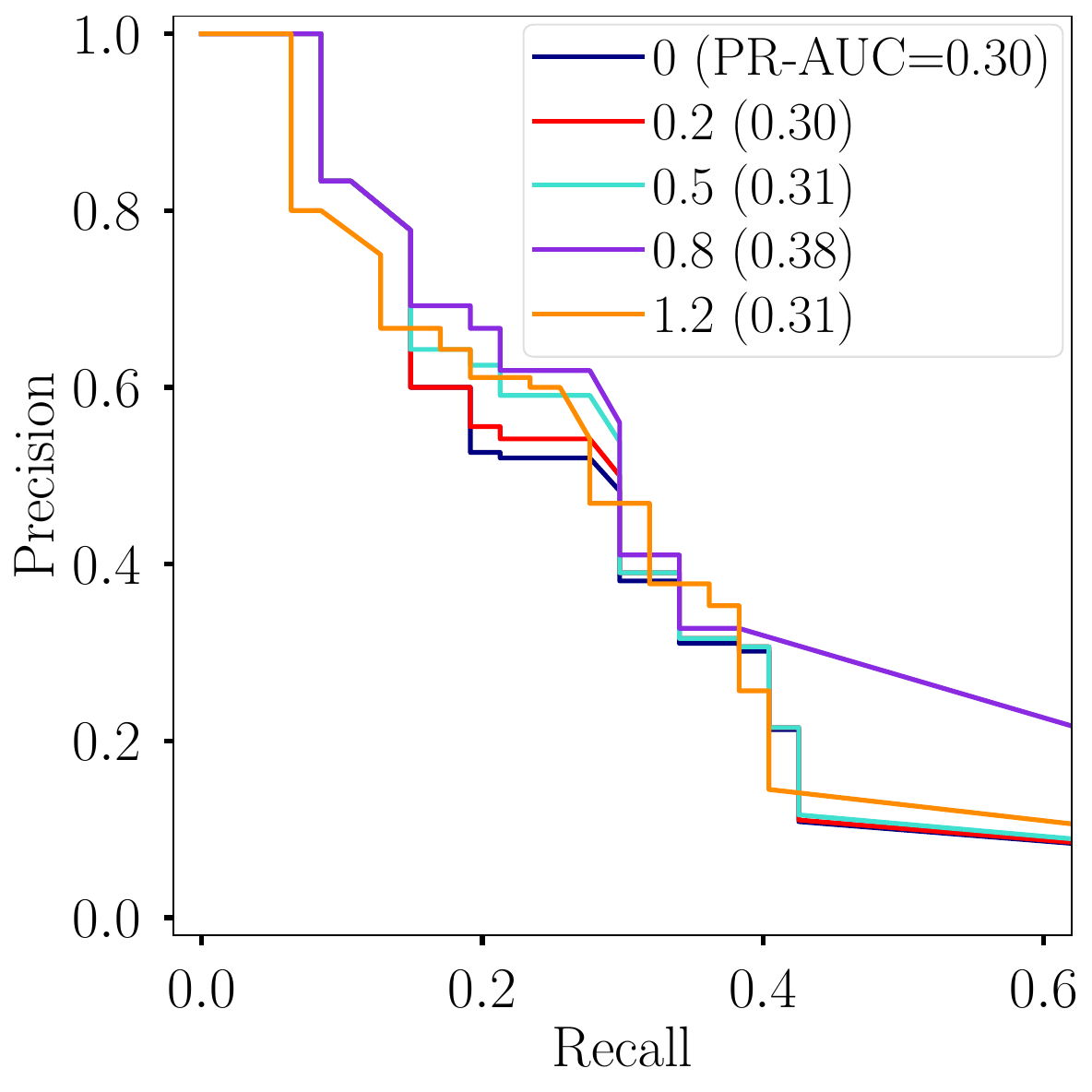}
    \end{minipage}
    &
    \begin{minipage}{6cm}
    \centering
    \includegraphics[width=0.6\linewidth]{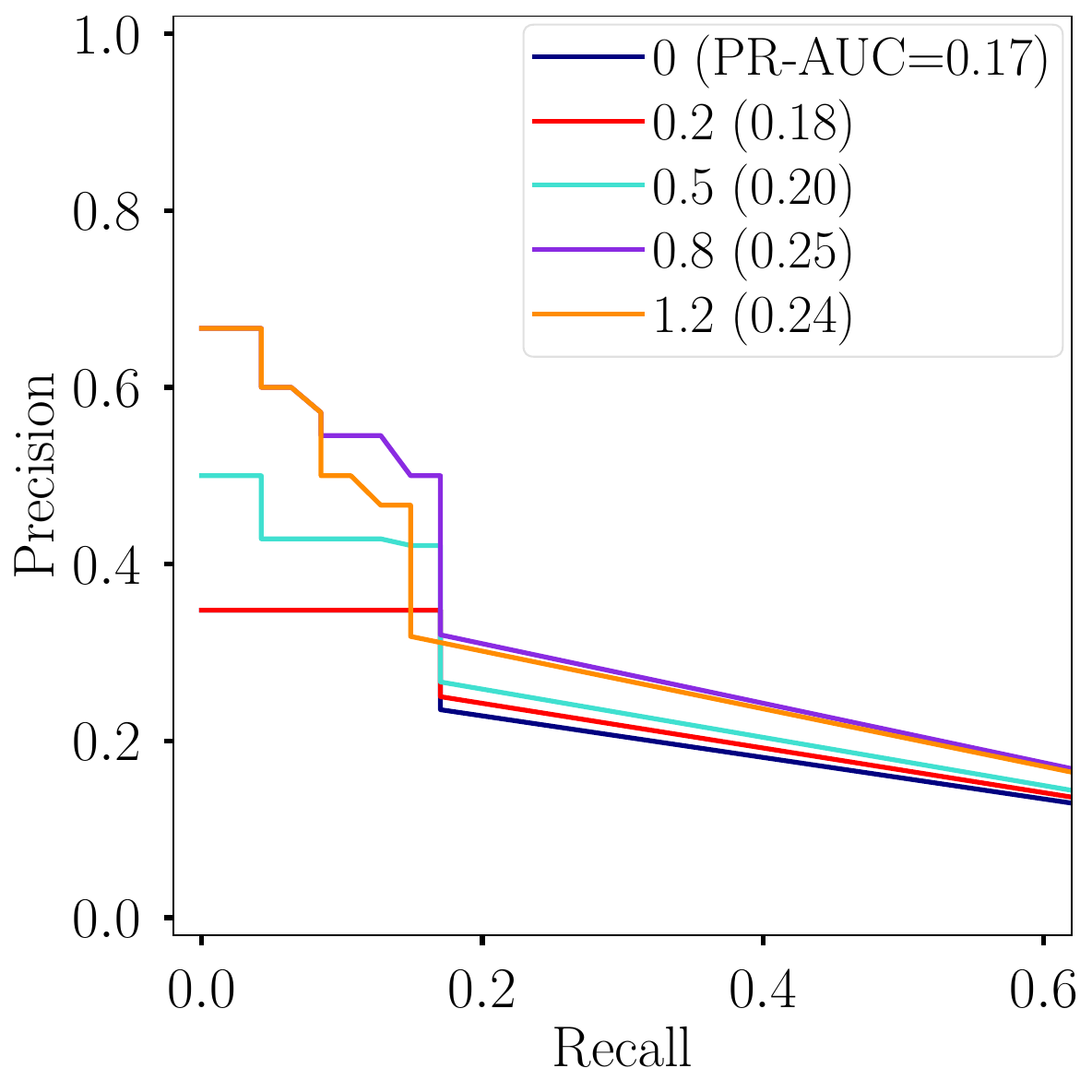}
    \end{minipage}
    \\
    \begin{minipage}{6cm}
    (a) Trained on \rttrain, tested on \sttesta
    \end{minipage}
    &
    \begin{minipage}{6cm}
    (b) Trained on \sttrain, tested on \sttesta
    \end{minipage}
    % \vspace{-3mm}
    \\
    \multicolumn{2}{c}{
    \begin{minipage}{8cm}
    \captionof{figure}{Sensitivity to Cohen's h threshold}
    \label{fig:cohenh_ablation}
    \end{minipage}
    }
    \end{tabular}
    \end{figure}
}
{
}

\begin{figure}[t]
\vspace{-1mm}
\centering
\begin{tabular}{c c}
\begin{minipage}{6cm}
\centering
\includegraphics[width=0.6\linewidth]{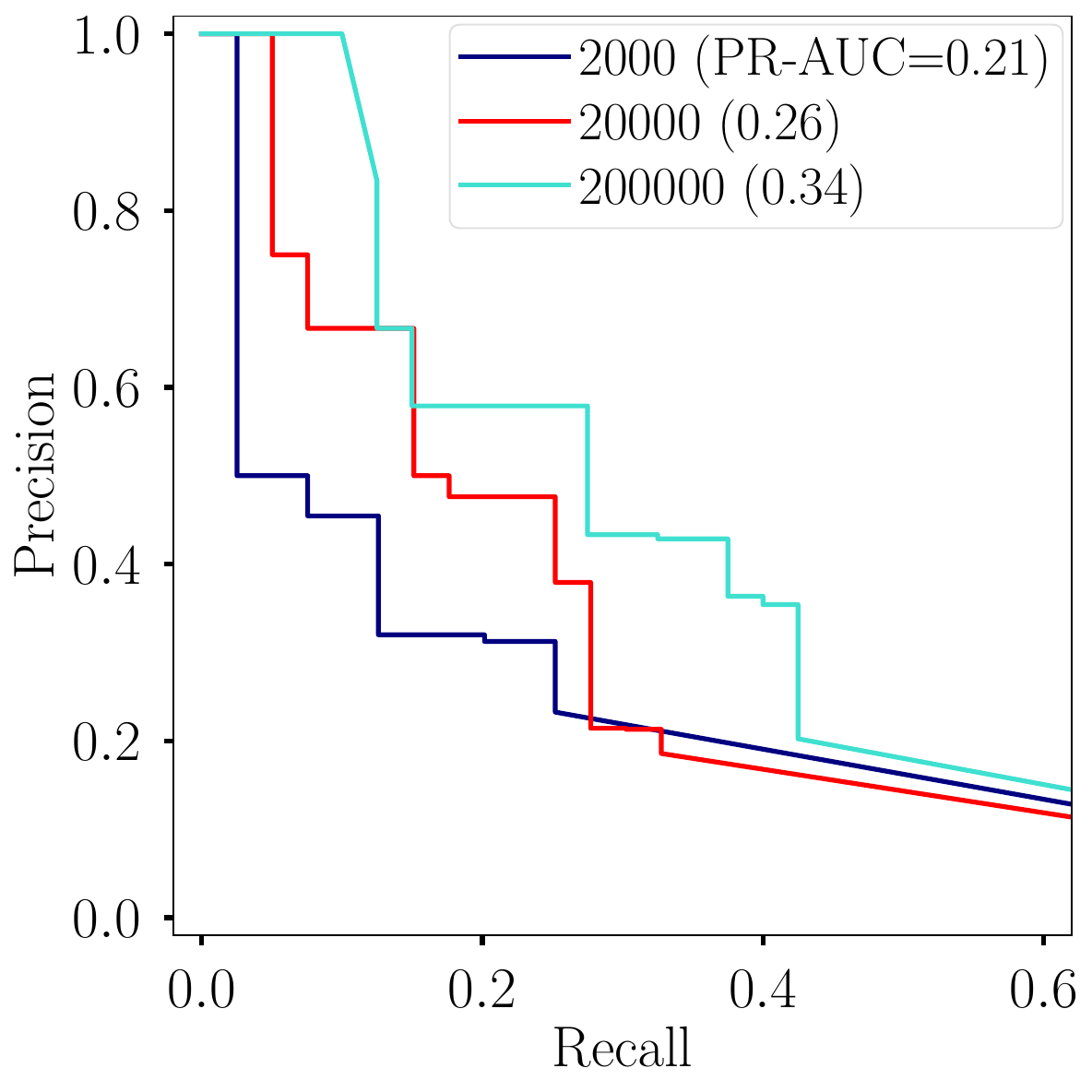}
\end{minipage}
&
\begin{minipage}{6cm}
\centering
\includegraphics[width=0.6\linewidth]{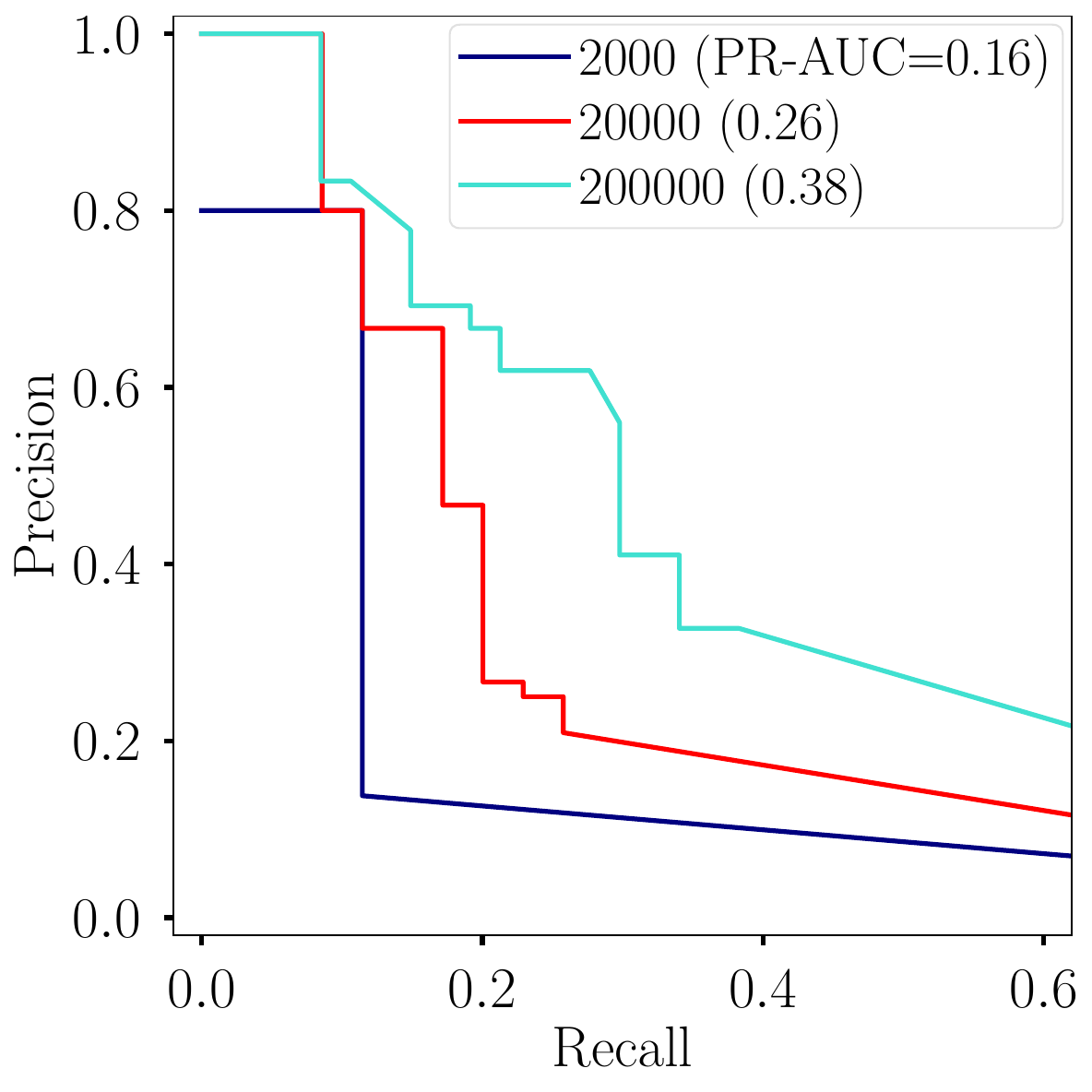}
\end{minipage}
\\
(a) \rttesta
&
(b) \sttesta
% \vspace{-3mm}
\\
\multicolumn{2}{c}{
\begin{minipage}{8cm}
\vspace{-3mm}
\captionof{figure}{Sensitivity to training corpus size}
\label{fig:sensitivity_corpus_size}
% \vspace{-5mm}
\end{minipage}
}
\end{tabular}
\vspace{-4mm}
\end{figure}

\iftoggle{full}
{
    \underline{Sensitivity to parameter $\delta$.} 
    Figure~\ref{fig:fine_select_vary_delta} shows the quality effect when varying parameter $\delta$ in \fs from $0.001$ to $1$. A lower $\delta$ typically leads to better PR curves, as it provides better consistency of confidence scores, leading to better ranking of reported errors. 
    This shows the importance of \fs (since $\delta \geq 1$ would correspond to the simpler \cs formulation).
}

\iftoggle{full}
{
    %\subsection{Ablation Studies}
    %\label{subsec:ablation}
    %In this section, we present ablation studies on the metrics used to assess rule quality.
    \underline{Sensitivity to Wilson score interval.} Figure~\ref{fig:wilson_ablation} plots the PR-curves of \fs when using the Wilson score interval (i.e., Equation~\eqref{eqn:wilson-confidence}) versus straightforward ratio when estimating the confidence of constraints. Wilson score leads to improved PR-curves, showing the importance of not over-estimating the confidence when data is sparse (Section~\ref{subsec:rule_quality_eval}).
    %Although using Wilson score interval tends to underestimate the confidence, it provides better performance, as evidenced by the better PR curve on both benchmarks. This suggests that avoiding overestimation of confidence can be beneficial for our framework.
}
{
}

\iftoggle{full}
{
    \underline{Sensitivity to Cohen's h thresholds.}
    Figure~\ref{fig:cohenh_ablation} reports the PR curves of \ar when varying the threshold $h$ in Cohen's h. 
    We observe improvements when we increase $h$, on both training corpora, with the best performance obtained when $h=0.8$, which in Cohen's h interpretation, corresponds to a large ``effect size'' in statistical terms \cite{cohen2013statistical}, showing the importance of using ``effect size'' in automated statistical hypothesis tests (Section~\ref{subsec:rule_quality_eval}). We also observe that the performance does not improve, and even slight degenerates, when further increasing the threshold of $h$ to 1.2, probably because a portion of good \sdcas are pruned under this setting.
    %Furthermore, since Excel is a inherently noisier corpus compared to PBI, the impact of employing Cohen's h when training on this dataset is particularly significant, as can be observed from the wide performance gap in Figure~\ref{fig:cohenh_ablation} (b).
}

\underline{Sensitivity to training corpus size.}
We study the effect of varying training corpus size from 2,000 to 200,000 columns in Figure~\ref{fig:sensitivity_corpus_size}. 
On both benchmarks, quality improves with more training data, showing the effectiveness of our data-driven approach.  %Also notice that a larger training corpus also enhances the generalizability of the resulting constraints, as the set of constraints from using 200,000 PBI columns performs significantly better on Excel benchmark compared to the set trained with 20,000 PBI columns. 

\underline{Robustness to low-quality \sdca candidates.}
To test whether \at is robust to low-quality \sdcas, we study the effect of injecting 1000 random hashing \sdcas candidates (in Section~\ref{subsec:rule_cand_generation}). Specifically, a random hashing \sdca has a domain-evaluation function $f^d_{hash}(h_i, v) = h_i(v)$ where $h_i$ is a hash function that randomly maps $v$ to a real number between 0 and 1. Since hash functions do not correspond to any meaningful domain, these \sdcas are inherently of low quality. We found that all adversarial \sdca candidates are rejected by our statistical test, and consequently have no effect on our final results (e.g., with no false positive detections produced). 

\iftoggle{full}
{
}
{
    Additional results such as sensitivity to FPR budget, size budget, Cohen's h, and Wilson, can be found in~\cite{full} in the interest of space.
}

\iftoggle{full}
{

    \subsection{Ablation Study}
    We also study the importance of different components via ablation studies that we discuss below.
    
    \underline{Contribution of different types of column-type detection.}
    Since we use diverse column-type detection methods (CTA-classifier, embedding, etc.) in the same framework,  we study the benefit of each method, by removing one method at a time, which leads to ``no-CTA'', ``no-embedding'', ``no-pattern'' and ``no-function'' 
    \iftoggle{full}
    {
        in Table~\ref{tab:ablation_domain_contribution} and Figure~\ref{fig:domain_ablation}.
    }
    {
        in Table~\ref{tab:ablation_domain_contribution}.
    }
    The results show that each method contributes to the overall quality, underlining the importance of having a framework like \at to unify diverse column-type detection methods for error-detection.
    
    \underline{Contribution of statistical tests.} Table~\ref{tab:ablation_wilson_cohenh} shows both the Wilson score interval (to bound confidence $c$), and Cohen's h  are important. %In the setting where Wilson score interval is removed, the naive method in Equation~\ref{eqt:conf_naive} is used for confidence computation. In the setting where Cohen's h is removed, \sdca candidates in $R_{can}$ are pruned using only confidence and significant level, without considering effect size. 
    We observe that the Wilson interval is especially important for high-precision (as reflected in big drops in F1@P=0.8 when we remove Wilson), while Cohen's h  shows benefit to the overall PR-AUC. 
}

\iftoggle{full}
{
    \begin{table}[]
        \caption{Ablation study: contribution of each type of method, reported as (F1@P=0.8, and PR-AUC).}
        \vspace{-3mm}
        \label{tab:ablation_domain_contribution}
        \centering
        \scalebox{0.8}{
            \begin{tabular}{|c||c||c|} \hline
                         &      \sttesta      & \rttesta   \\ \hline
              \fs & \textbf{0.34}, \textbf{0.45} & \textbf{0.21}, \textbf{0.34} \\ \hline
             no-CTA & 0.34, 0.45 & 0.17, 0.32 \\ \hline
             no-embedding & 0.29, 0.43 & 0.13, 0.30 \\ \hline
             no-pattern & 0.22, 0.40 & 0.18, 0.34 \\ \hline
             no-function & 0.15, 0.38 & 0.17, 0.32 \\ \hline
            
            \end{tabular}
        }
    \end{table}

}

\iftoggle{full}
{
    \begin{table}[t]
        \caption{Ablation study of using Wilson score interval and Cohen's h, reported as (F1@P=0.8, and PR-AUC).}
        \vspace{-3mm}
        \label{tab:ablation_wilson_cohenh}
        \centering
        \scalebox{0.75}
        {
            \begin{tabular}{|c||c||c|} \hline
                        
                     &      \sttesta      & \rttesta   \\ \hline
            \ar & \textbf{0.23}, \textbf{0.38} & \textbf{0.21}, \textbf{0.34} \\ \hline
           no Wilson score interval  & 0.12, 0.36  & 0.18, 0.31   \\ \hline
           no Cohen's h  &  0.23, 0.35   &  0.21, 0.32    \\ \hline

            \end{tabular}
        }
    \end{table}
}

\iftoggle{full}
{
    \begin{figure}[ht]
    \centering
    \begin{tabular}{c c}
    \begin{minipage}{6cm}
    \centering
    \includegraphics[width=0.6\linewidth]{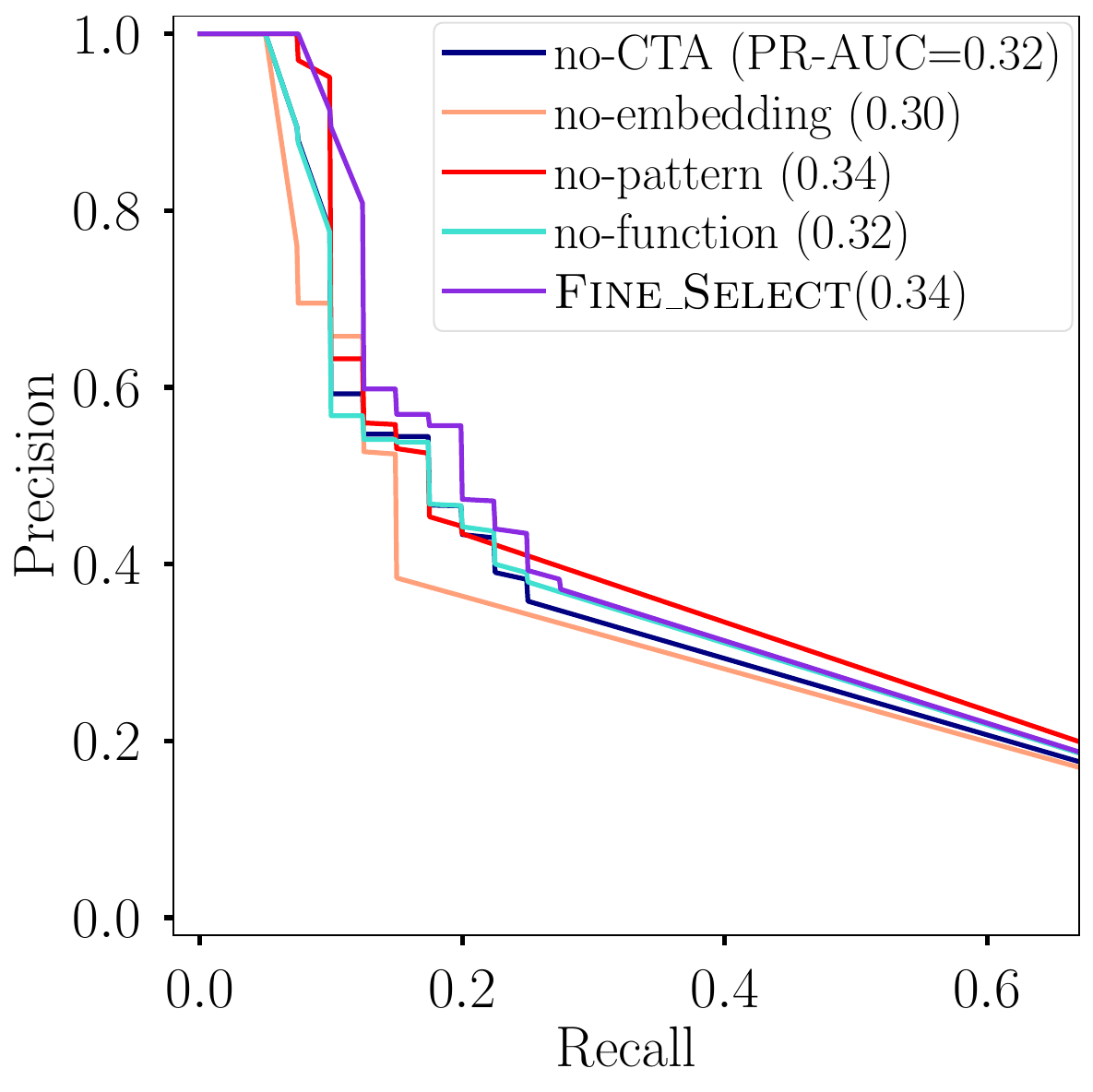}
    \end{minipage}
    &
    \begin{minipage}{6cm}
    \centering
    \includegraphics[width=0.6\linewidth]{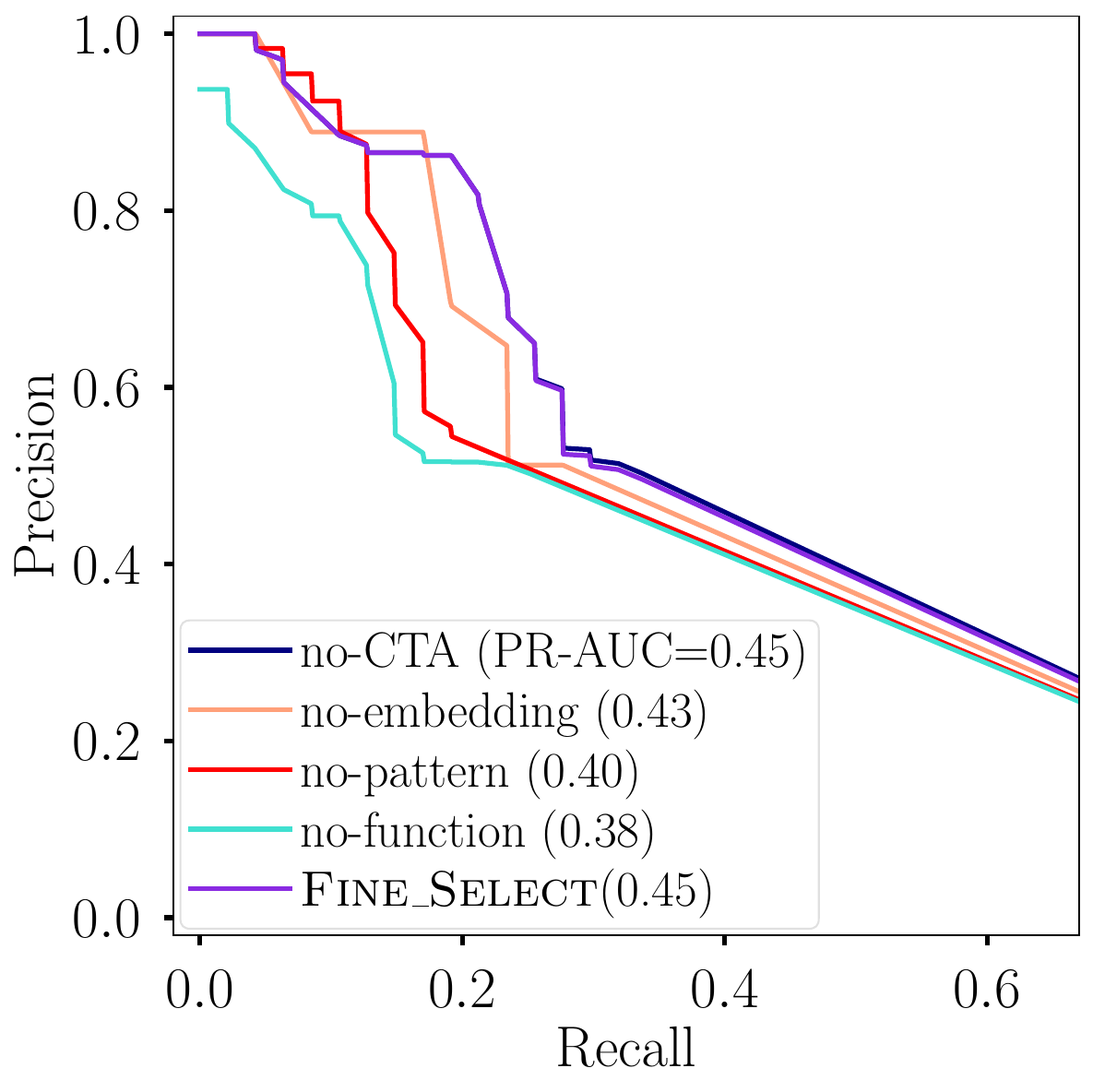}
    \end{minipage}
    \\
    (a) \rttesta
    &
    (b) \sttesta
    % \vspace{-3mm}
    \\
    \multicolumn{2}{c}{
    \begin{minipage}{8cm}
    \captionof{figure}{Ablation study: removing a type of ``column-type detection'' method at a time.}
    \label{fig:domain_ablation}
    \end{minipage}
    }
    \end{tabular}
    \end{figure}
}
{
}

\subsection{Quality on data-cleaning benchmarks} 
\label{subsec:data-cleaning-exp}

\begin{table*}[]
%\vspace{-15mm}
    \caption{Quality results of applying \sdcas learned in \at on existing data-cleaning benchmarks.  Note that the cell-level precision (reported in the last line) is evaluated by strictly comparing our detection, vs. the ground-truth ``clean'' version of the benchmarks, which however underestimates the true precision, because the current ground-truth labels in some of the benchmarks are incomplete that ``miss'' real errors. We therefore also report adjusted precision in parenthesis ``()'' -- for example, our 9-dataset aggregate precision is ``95\% (97\%)'', meaning that precision is 95\% (174/183) if strictly using existing ground-truth, which however increases to 97\% (179/183)  if we use augmented ground-truth that is manually labeled and shared at~\cite{full}. \iftoggle{full}{We report detailed examples of such cases of ``missed errors in ground-truth'' in Table~\ref{tab:case_study_data_cleaning_new_errors}.}{}}
    \vspace{-2mm}
    \label{tab:data_cleaning_summary}
    \centering
    \scalebox{0.56}
    {
    \hspace{-3mm}
        \begin{tabu}{|c|p{4cm}||c||c|c|c|c|c|c|c|c|c|} \hline 
                    
        &  & 9-dataset overall & \textit{adults} & \textit{beers} & \textit{flights} & \textit{food} & \textit{hosptial} & \textit{movies} & \textit{rayyan} & \textit{soccer} & \textit{tax}  \\ \hline \hline
    
         \multirow{2}{*}{ Dataset statistics }  &  \# of total categorical cols & 85 & 9  & 6 & 6 & 10 & 16 & 14 & 8 & 8 & 8  \\ \cline{2-12}

       %  \rowfont{\color{red}}
       % &  \# of cols with errors marked in ground-truth & 44 & 1  & 3 & 4 & 3 & 12 & 8 & 6 & 3 & 4 \\ \cline{2-12}
       
       &  \# of cols covered by existing ground-truth & 36 & 1  & 3 & 4 & 1 & 12 & 0 & 8 & 1 & 6 \\ \hline \hline %\Xhline{1pt} % Another bold horizontal line
       
       \multirow{2}{*}{ Quality: column-level} &  \underline{Coverage}: \# of cols with new constraints by using \sdca  & \textbf{17} & 2  & 2  & 0 & 3 & 4 & 2 & 1 & 2 & 1  \\ \cline{2-12}
       
%       \multirow{3}{*}{ Quality: column-level} &  \underline{Recall}: \# of cols covered by \sdca  & 26 & 2  & 3  & 0 & 4 & 7 & 2 & 2 & 2 & 4  \\ \cline{2-12}
       
       &  \underline{Precision}: \% of new \sdcas that are correct  & \textbf{94\%} & 100\%  & 67\%  & - & 100\% & 100\% & 100\% & 100\% & 100\% & 100\%  \\ \hline

%       &  \underline{Recall}: \# of new cols covered by \sdca not in existing ground-truth  & 12 & 2 & 1 & 0 & 2 & 3 & - & 2 & 2 & 0 \\ \hline  \hline

       \multirow{2}{*}{ Quality: cell-level} &  \underline{True-positives}: \# of detected data errors using \sdcas  & \textbf{183} & 0  & 5 & 0 & 3 & 13 & 161 & 1 & 0 & 0 \\ \cline{2-12}
       &  \underline{Precision}: \% of detected data errors that are correct  & \textbf{95\% (97\%)} & -  & 40\% (60\%) & - & 0\% (33\%) & 92\% (100\%) & 99\% (100\%) & 0\% (100\%) & - & - \\ \hline 
       % &  \underline{Precision}: \% of detection data errors using \sdcas  that are correct  & 97\% & -  & 40\% & - & 33\% & 100\% & 100\% & 100\% & - & - \\ \hline 

%       Latency &  Elapsed time (seconds) to process input table   & 237.14  &  29.00 & 17.92 & 16.99 & 11.17 & 15.99 & 57.78 & 8.40 & 28.04 & 51.85 \\ \hline
        
        \end{tabu}
    }
    \vspace{-2mm}
\end{table*}

In addition to testing using large-scale real benchmarks collected in the wild (\sttesta and \rttesta), we also test \sdcas learned using \at against 9 existing data-cleaning datasets used in prior work~\cite{dc-beskales2013relative, dc-chu2013holistic, dc-ge2020hybrid, dc-khayyat2015bigdansing, mahdavi2020baran, dc-rekatsinas2017holoclean}, listed in Table~\ref{tab:data_cleaning_summary}.  Our goal of this experiment is to test whether our learned \sdcas can identify new constraints to complement existing constraints in these datasets, thereby augmenting existing data cleaning algorithms. % (since \sdcas can be easily expressed using denial constraints and other logic forms). 

Table~\ref{tab:data_cleaning_summary} reports our results in terms of (1) \underline{column-level coverage}, or new constraints that we discover using \sdcas not in existing ground-truth, (2) \underline{column-level precision}, or the fraction of new \sdcas constraints judged as correct, (3) \underline{cell-level true-positives}, or the number of erroneous cells that the new \sdcas can detect, and (4) \underline{cell-level precision}, or the fraction of erroneous cells detected by \sdcas that are correct.

\begin{table*}[t]
    \vspace{-2mm}
    \caption{Details of real \sdca that are automatically applied on existing data-cleaning benchmarks, using \at. Many of these \sdca constraints here offer new mechanisms to identify data errors, that are not possible using existing constraints from these benchmark data (e.g., note that existing constraints from these benchmark data do not cover many columns marked as ``-'', or existing constraints identify errors using complementary mechanisms, such as ``2 letters'' for \code{state} columns, while our \sdca use ML-based CTA ``state-classifiers'' from Sherlock and Doduo, which are more fine-grained in detecting subtle errors).}
    \vspace{-2mm}
    \label{tab:new_recall_details}
    % \vspace{-3mm}
    \centering
    \scalebox{0.57}
    {
        \begin{tabular}{|p{1.1cm}|p{1.3cm}|p{3cm}||p{3cm}||p{4.5cm}|p{4cm}|p{5cm}|} \hline 
                    
        Dataset & Column & Example values & Existing  constraints in benchmark data & \at: New \sdca constraints (pre-condition)  & \at: New \sdca constraints (post-condition) & New errors \sdca can identify if present (not by existing constraints) \\  \hline  \hline 

        \textit{adult} & race & white, black, others, ... & - & 80\% column values have their \textit{Glove} distances to ``red'' < 5.5 & values whose \textit{Glove} distances to ``red'' > 7.5 & \makecell[l]{(Typo): wite, blaack, ... \\ (Incompatible): seattle, male, ...} \\ \hline

        \textit{adult} & sex & female, male & - & 80\% column values have their \textit{Glove} distances to ``male'' < 7.0 & values whose \textit{Glove} distances to ``male'' > 9.5 &  \makecell[l]{(Typo): femele, malle, ... \\ (Incompatible): masculina, finnish, ...}   \\ \hline

        \textit{beers} & city & san francisco, columbus, louisville, ... & brewery id $\rightarrow$ city & 80\% column values have their \textit{Glove} distances to ``hawaii'' < 6.0 & values whose \textit{Glove} distances to ``hawaii'' > 11.0 & \makecell[l]{(Typo): louisvilla, seettle, ... \\ (Incompatible): maine, 9th ave., ...}  \\ \hline

        \textit{beers} & state & or, in, ca, fl,  ... &  brewery id $\rightarrow$ state, state (2 letters) & 80\% column values have their \textit{Sherlock state-classifier} scores > 0.5 & values whose \textit{Sherlock state-classifier} scores $\leq$ 0 & \makecell[l]{(Typo): ax, xk, ... \\ (Incompatible): us, xl, ...}  \\ \hline

    % this is a FP  %  \textit{beers} & ounces & 12.0 ounces, 12.0 oz, 16.0 oz, ... & - & {\color{red}95\% column values return true on function \textit{validate\_url()} }& {\color{red}values that return false on function \textit{validate\_url()} } & - \\ \hline

        \textit{food} & facility type & restaurant, school, grocery store, ... & - & 80\% column values have their \textit{Doduo type-classifier} scores > 4 & values whose \textit{Doduo type-classifier} scores < -1 & \makecell[l]{(Typo): childern's service, koisk, ... \\ (Incompatible): asia, dummy\_type, ...}  \\ \hline

        \textit{food} & city & chicago, schaumburg, lake zurich, ... & - & 80\% column values have their \textit{Glove} distances to ``berlin'' < 5.5 & values whose \textit{Glove} distances to ``berlin'' > 8.0 & 
        \makecell[l]{(Typo): chiago, buffolo, ... \\ (Incompatible): upenn, mcdonald, ...} \\ \hline

        \textit{food} & state & il, ilxa &  city $\rightarrow$ state & 80\% column values have their \textit{Doduo state-classifier} scores > 4 & values whose \textit{Doduo state-classifier} scores < -2 &
        \makecell[l]{(Typo): xx, nt, ... \\ (Incompatible): usa, tottenham, ...}  \\ \hline

        \textit{hospital} & sample & 0 patients, 107 patients, 5 patients, ... &  -  & 93\% column values match pattern ``$\backslash$d+ $\backslash$[a-zA-Z]+'' & values not matching pattern ``$\backslash$d+ $\backslash$[a-zA-Z]+'' & \makecell[l]{(Typo): x patients, 3x patients,... \\ (Incompatible): empty, sample\_size, ...} \\ \hline
        
        \textit{hospital} & state & al, ak &  zip $\rightarrow$ state, county $\rightarrow$ state, state (2 letters)  & 80\% column values have their \textit{Sherlock state-classifier} scores > 0.5 & values whose \textit{Sherlock state-classifier} scores $\leq$ 0 & \makecell[l]{(Typo): ax, xk, ... \\ (Incompatible): us, xl, ...}  \\ \hline

        \textit{hospital} & hospital type & acute care hospitals &  condition, measure name $\rightarrow$ hospital type  & 80\% column values have their \textit{Doduo category-classifier} scores > 4.5 & values whose \textit{Doduo category-classifier} scores < -1.5 & \makecell[l]{(Typo): acute caer, clinix, ... \\ (Incompatible): london, co. kildare, ...} \\ \hline
        
        \textit{hospital} & emergency service & yes, no &  zip $\rightarrow$ emergency service  & 80\% column values have their \textit{Glove} distances to ``no'' < 5.5 & values whose \textit{Glove} distances to ``no'' > 7.0 & \makecell[l]{(Typo):  yxs, nao, ...  \\ (Incompatible): emergency, 95503, ...} \\ \hline

        \textit{movie} & id & tt0054215, tt0088993, tt0032484, ... &  -  & 85\% column values match pattern ``$\backslash$[a-zA-Z]+$\backslash$d+'' & values not matching pattern ``$\backslash$[a-zA-Z]+$\backslash$d+'' & (Incompatible):  iron\_man\_3, dark\_tide, ...  \\ \hline

        \textit{movie} & duration & 109 min, 96 min, 120 min, ... &  -  & 93\% column values match pattern ``$\backslash$d+ $\backslash$[a-zA-Z]+'' & values not matching pattern ``$\backslash$d+ $\backslash$[a-zA-Z]+'' & (Incompatible):  2 hr 30 min, nan, ...   \\ \hline

        \textit{rayyan} & article created\_at & [1/1/71, 4/2/15, 12/1/06, ...] & - & 90\% column values return true on function \textit{validate\_date()} & values that return false on function \textit{validate\_date()} & (Incompatible): nan, june, ... \\ \hline

        \textit{soccer} & position & defender, midfield, goalkeeper, ... &  -  & 80\% column values have their \textit{Sherlock position-classifier} scores > 0.1 & values whose \textit{Sherlock position-classifier} scores $\leq$ 0 & \makecell[l]{(Typo): strikor, forwrad, ... \\ (Incompatible): difensore, goleiro, ...} \\ \hline

        \textit{soccer} & city & cardiff, dortmund, munich, ... &  -  & 80\% column values have their \textit{Sentence-BERT} distances to ``panama'' < 1.2 & values whose \textit{Sentence-BERT} distances to ``panama'' > 1.375  & \makecell[l]{(Typo): cardif, munihei, ... \\ (Incompatible): fl, 744-9007, ...}  \\ \hline

        \textit{tax} & state & ma, nv, ar, ... &  zip $\rightarrow$ state, area code $\rightarrow$ state, state (2 letters)  & 80\% column values have their \textit{Sherlock state-classifier} scores > 0.5 & values whose \textit{Sherlock state-classifier} scores $\leq$ 0 & \makecell[l]{(Typo): ax, xk, ... \\ (Incompatible): us, xl, ...}  \\ \hline
        
        \end{tabular}
    }
    \vspace{-2mm}
\end{table*}

\textbf{Column-level results.}
We can see that our approach can indeed discover new \sdca constraints on 16 columns (not known in existing benchmark ground-truth), from a total of 81 columns, in which 94\% new constraints are correct. 

Table~\ref{tab:new_recall_details} lists all new \sdcas found automatically on existing Data-cleaning benchmarks. Observe that some of these columns do not originally have applicable constraints in benchmark ground-truth (marked by ``-''), in which case our \sdcas auto-applied on such columns would clearly provide value, by enabling new mechanisms for error detection. 
While for the rest of the columns existing constraints do exist, our \sdca can nevertheless still augment them. For example, on column \codeq{city} in dataset \codeq{beers}, although there is an FD constraint between \codeq{brewery id} and \codeq{city}, there are still many errors that cannot be reliably detected by the FD alone -- e.g., the FD constraint cannot find errors for rows with a unique brewery id in the table, while \sdca can help to detect errors such as typos (e.g., \codeq{seatle}) and incompatible values (e.g., \codeq{9th ave}) in such cases.

\begin{table*}[t]
    % \vspace{-4mm}
    \caption{Example new errors detected by \sdcas, marked in underline, that are not known or labeled as errors in existing benchmark ground-truth. For example, in the \codeq{hospital} dataset, a cell with value {\codeq{empty}} in the \codeq{sample} column (with typical values like \codeq{0 patients}, \codeq{107 patients}) are not marked in ground-truth; in the \codeq{food} dataset, a cell with misspelled \codeq{childern} is not marked in the ground-truth, etc. This shows the promise that \sdcas can complement existing constraint-based cleaning to identify additional errors. }
    \label{tab:case_study_data_cleaning_new_errors}
    % \vspace{-3mm}
    \centering
    \scalebox{0.58}{
    \begin{tabular}{|C{1cm}|C{2.5cm}|C{3cm}||C{3.8cm}||C{3.9cm}|C{3cm}||C{4cm}|} \hline
         Dataset & Column & Example column values &  Existing constraints in benchmark ground-truth & New \sdca constraint (pre-condition) &  New \sdca constraint (post-condition) & New errors detected by \sdca (not known in ground-truth) \\ \hline

       \textit{hospital} & sample & [0 patients, 107 patients, 5 patients, ...] &  - & 93\% column values match pattern ``$\backslash$d+ $\backslash$[a-zA-Z]+'' & values not matching pattern ``$\backslash$d+ $\backslash$[a-zA-Z]+''  & { ``\underline{empty}''} \\ \hline

        % \textit{movies} & duration & [109 min, 96 min, 120 min, ... ]& - & 93\% column values match pattern ``$\backslash$d+ $\backslash$[a-zA-Z]+'' & values not matching pattern ``$\backslash$d+ $\backslash$[a-zA-Z]+''  &{\color{red} nan }  \\ \hline
         \textit{food} & facility type & [restaurant, grocery store, catering, ...] &  - &  80\% column values have their \textit{Doduo type-classifier} scores > 4 & values whose \textit{Doduo type-classifier} scores < -1 & { ``\underline{childern's} service facility''}  \\ \hline
        
          \textit{rayyan} & article created\_at & [1/1/71, 4/2/15, 12/1/06, ...] & - & 90\% column values return true on function \textit{validate\_date()} & values that return false on function \textit{validate\_date()} & { ``\underline{nan}'' } \\ \hline

%          \textit{beers} & ounces & 12.0 oz, 16.0 oz, 12.0 ounce, ... &  {\color{red} nan } \\ \hline
    \end{tabular}
    }
   \vspace{-3mm}
\end{table*}

\textbf{Cell-level results.}
At the cell-level, we can see from Table~\ref{tab:data_cleaning_summary} that these automatically-installed new \sdcas alone (without using any other constraints in benchmark ground-truth, which typically require human experts to program), can already identify 183 data values as errors, with an overall precision of 95\%, when evaluated against the ground-truth clean data. 

Interestingly, we would like to highlight that using these new \sdcas enable us to uncover \underline{\textit{new errors not known or labelled in existing ground-truth}}, some of the example errors, and their corresponding \sdcas, are shown in Table~\ref{tab:case_study_data_cleaning_new_errors}. Note that no constraints are programmed on these example columns in the existing benchmarks (indicated by ``-'' in the table), but the \sdcas we automatically apply, can find typos (misspelled \codeq{childern}), and incompatibility (strings like \codeq{empty} and \codeq{nan} mixed in data columns) that are not known in existing ground-truth. We believe this demonstrates that \sdca has the potential to augment existing data-cleaning methods, to identify new and complementary data errors not covered by existing constraints.

\iftoggle{full}
{
    Since these are plausible real errors that are ``missed'' by existing ground-truth data, using a strict evaluation that only compares against existing ground-truth clearly ``under-estimates'' our true precision (because our detections shown in Table~\ref{tab:case_study_data_cleaning_new_errors} will be considered as false-positives, when they are in fact real errors). We therefore manually re-annotate the ground-truth and report new cell-level precision  in the last line of Table~\ref{tab:data_cleaning_summary}, in parenthesis ``()''. For example,  our 9-dataset aggregate precision is ``95\% (97\%)'', meaning that precision is 95\% (174 out of 183 detections are true-positives) if we evaluate strictly using existing ground-truth (with missed errors), which increases to 97\%  (179 out of 183 detections are true-positives) if we use augmented ground-truth that is manually labeled to account for missed errors like shown in  Table~\ref{tab:case_study_data_cleaning_new_errors}.

}
{

We want to stress that in expert-driven data-cleaning scenarios and with experts in the loop, existing data-cleaning formalism such as denial constraints are still way more expressive, such that \emph{\sdcas are not meant to outperform or replace existing methods in such settings} -- this particular experiment is only meant to  show that \sdcas can serves as a new class of constraints (auto-applied to relevant table columns), that may complement existing data cleaning methods. 
    
}

\iftoggle{full}
{
    \textbf{Limitations}. We want to stress that our \sdcas are designed to operate automatically, but on single-columns only (which is when \sdcas can be reliably applied without human-experts). \sdcas cannot  capture rich and complex  constraints  across columns (e.g., denial constraints and CFDs), for which human experts are still better at understanding the nuances, and determine whether a complex multi-column constraint should hold or not (e.g., when a candidate denial constraint between \code{salary} and \code{tax-rate} appears to
    %\code{ID $\rightarrow$ Name} 
    hold on 90\% rows, it is still better for domain experts to determine if this is a true constraint that should really hold and be enforced or not)~\cite{discoverrule-chu2014ruleminer, discoverrule-fan2010discovering, discoverrule-huhtala1999tane, discoverrule-wyss2001fastfds}.  
    
    In other words, in expert-driven data-cleaning scenarios, we believe that \sdca offers a new perspective to \emph{augment} rich constraints extensively studied in the data-cleaning literature, but is not meant to replace  complex constraints that only humans can program.
}

\section{Conclusions and Future Work}
\label{sec:conclusion}
In this work, we propose a new class of data-quality constraints that we argue are overlooked in the literature. We show that such constraints can unify diverse  column-type detection methods in the same framework, and once learned from large table corpora using \at, can reliably apply to new and unseen tables. 

Future directions include integrating \sdcas with existing integrity constraints, and study how best to leverage them in the expert-driven data cleaning scenarios. Testing the coverage of our proposed method on specialized domains and corpora, is another direction of future work.

\section*{Acknowledgement}

We thank three anonymous reviewers for their constructive feedback that helps to improve this paper. Qixu Chen and Raymond Chi-Wing Wong are supported by fund WEB24EG01-H.

\clearpage
\bibliographystyle{ACM-Reference-Format}  
\bibliography{Auto-Test}

%\clearpage

%\balance

\iftoggle{full}
{
    \clearpage
    \appendix
    
    \section{Quality Results: Training using different corpora} 
\label{apx:train-on-excel}

We compare the performance of our algorithms when trained on different corpora in Table~\ref{tab:performance_summary_train_on_excel}, and plot the PR curves on \rttesta and \sttesta when trained using \sttrain in Figure~\ref{fig:pr_benchmark_pbi_benchmark_rule_Excel} and \ref{fig:pr_benchmark_excel_benchmark_rule_Excel}, respectively. 
The results indicate that training on \sttrain results in inferior performance compared to \rttrain across all testing datasets. 
This performance gap can be attribute to the lower quality of \sttrain compared to \rttrain.
Firstly, \sttrain is inherently noisier since the columns were collected from web spreadsheets without further validation, leading to a higher error rate in columns. This obstructs the \sdca learning process since its effectiveness depends on the implicit assumption that the majority of columns are error-free.
Secondly, as can be observed from Table~\ref{tab:corpora_stat}, compared to that in \rttrain, columns in \sttrain are generally shorter and contain fewer distinct values, which also hinders the learning of effective \sdca. It is worth mentioning that columns with longer length and more distinct values are in general more preferred in training, since they gives a more comprehensive overview on what values appear in the corresponding domain.

\begin{table*}[ht]
    \captionof{table}{Summary of algorithm performance when trained on different corpora}
    \centering
    \setlength{\tabcolsep}{2.5pt} % Adjust the value to increase or decrease the space between columns
    \scalebox{0.65}
    {
    \begin{tabular}{|c|c||c|c|c|c||c|c|c|c|} \hline
        & &  \multicolumn{4}{c||}{ \sttest (\sttesta) } & \multicolumn{4}{c|}{\rttest (\rttesta)} \\ 
        \hline
              & Method  & real &  +5\% syn err. &  + 10\% syn err. &  +20\% syn err. & real & +5\% syn err. & +10\% syn err. & +20\% syn err. \\ \hline \hline
    % \begin{tabular}{|c||c|c|c|c||c|c|c|c|} \hline
    %       Method  & Excel & Excel (5\%) & Excel (10\%) & Excel (20\%) & PBI & PBI (5\%) & PBI (10\%) & PBI (20\%) \\ \hline
    
    \multirow{3}{*}{\makecell{Trained on \\ \rttrain}} & \ar       & 0.30, 0.50      & 0.40, 0.51       & 0.47, 0.57       & 0.50, 0.66       & \textbf{0.32}, \textbf{0.41}       & \textbf{0.35}, 0.42       & \textbf{0.36}, 0.48       & 0.36, 0.54  \\ \cline{2-10}

     & \fs & \textbf{0.54}, \textbf{0.58}      & \textbf{0.47}, \textbf{0.57}       & \textbf{0.48}, \textbf{0.62}       & \textbf{0.53}, \textbf{0.68}       & \textbf{0.32}, 0.40       & \textbf{0.35}, \textbf{0.48}       & \textbf{0.36}, \textbf{0.56}       & \textbf{0.40}, \textbf{0.62}  \\ \cline{2-10}

    & \cs  & 0.35, 0.53      & 0.43, 0.56       & 0.41, 0.60       & 0.52, 0.67       & 0.07, 0.27       & 0.21, 0.41       & 0.28, 0.53       & 0.39, 0.61  \\ \hline

    \multirow{3}{*}{\makecell{Trained on \\ \sttrain}} & \ar       & 0, 0.32      & 0.26, 0.40       & 0.28, 0.48       & 0.41, 0.61       & 0, 0.29       & 0.27, 0.40       & 0.21, 0.44       & 0.27, 0.53  \\ \cline{2-10}

    & \fs & 0.10, 0.36      & 0.21, 0.45       & 0.28, 0.52       & 0.45, 0.64       & 0.03, 0.31       & 0.29, 0.42       & 0.25, 0.47       & 0.27, 0.55   \\ \cline{2-10}

    & \cs & 0, 0.30      & 0.09, 0.39       & 0.09, 0.48       & 0.40, 0.61       & 0, 0.18      & 0.08, 0.33       & 0.10, 0.41       & 0.13, 0.52   \\ \hline

    \end{tabular}
    }
    \label{tab:performance_summary_train_on_excel}
\end{table*}

\begin{figure*}[ht]
\centering
\begin{tabular}{c c}
\begin{minipage}{6.5cm}
\centering
\includegraphics[width=\linewidth]{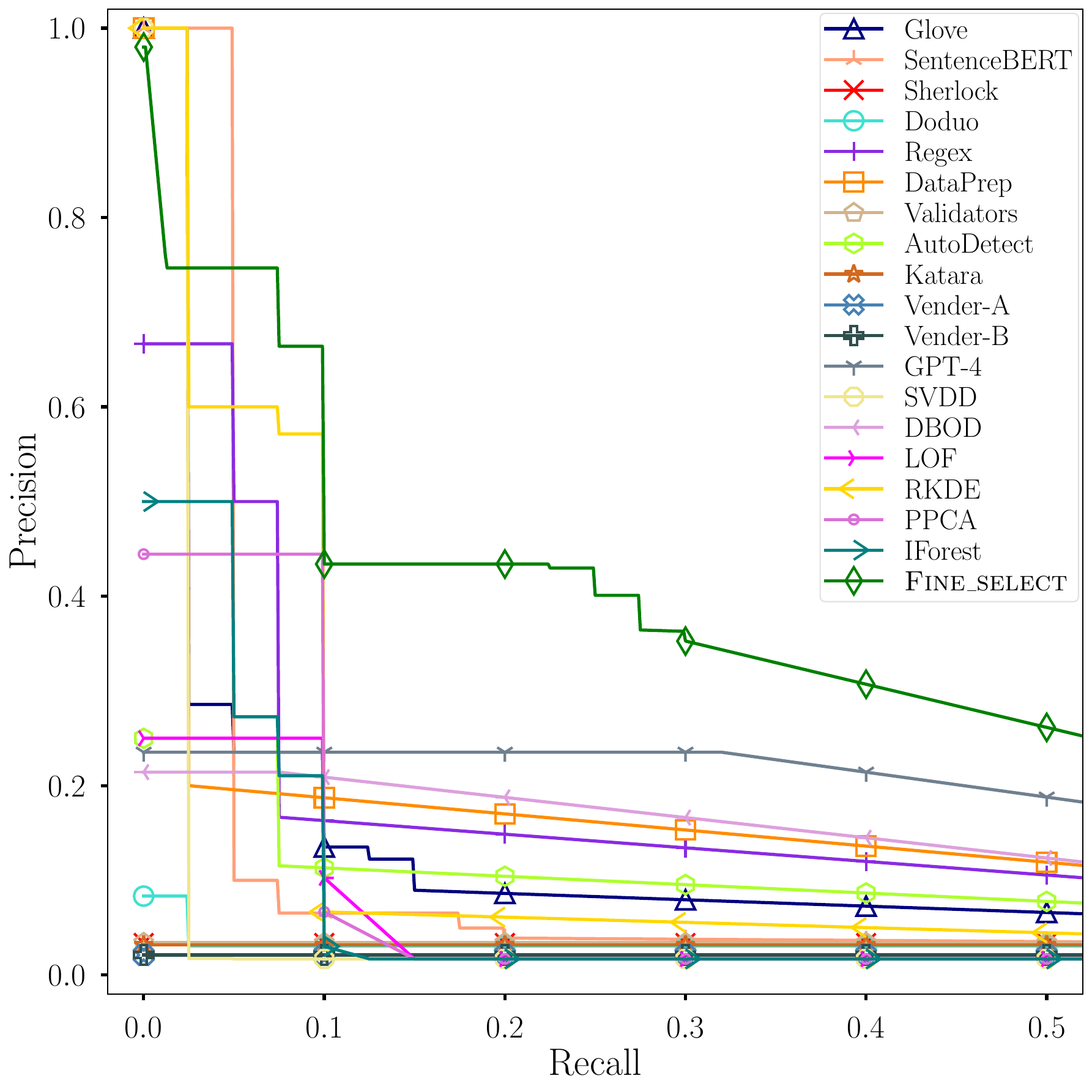}
\end{minipage}
&
\begin{minipage}{6.5cm}
\centering
\includegraphics[width=\linewidth]{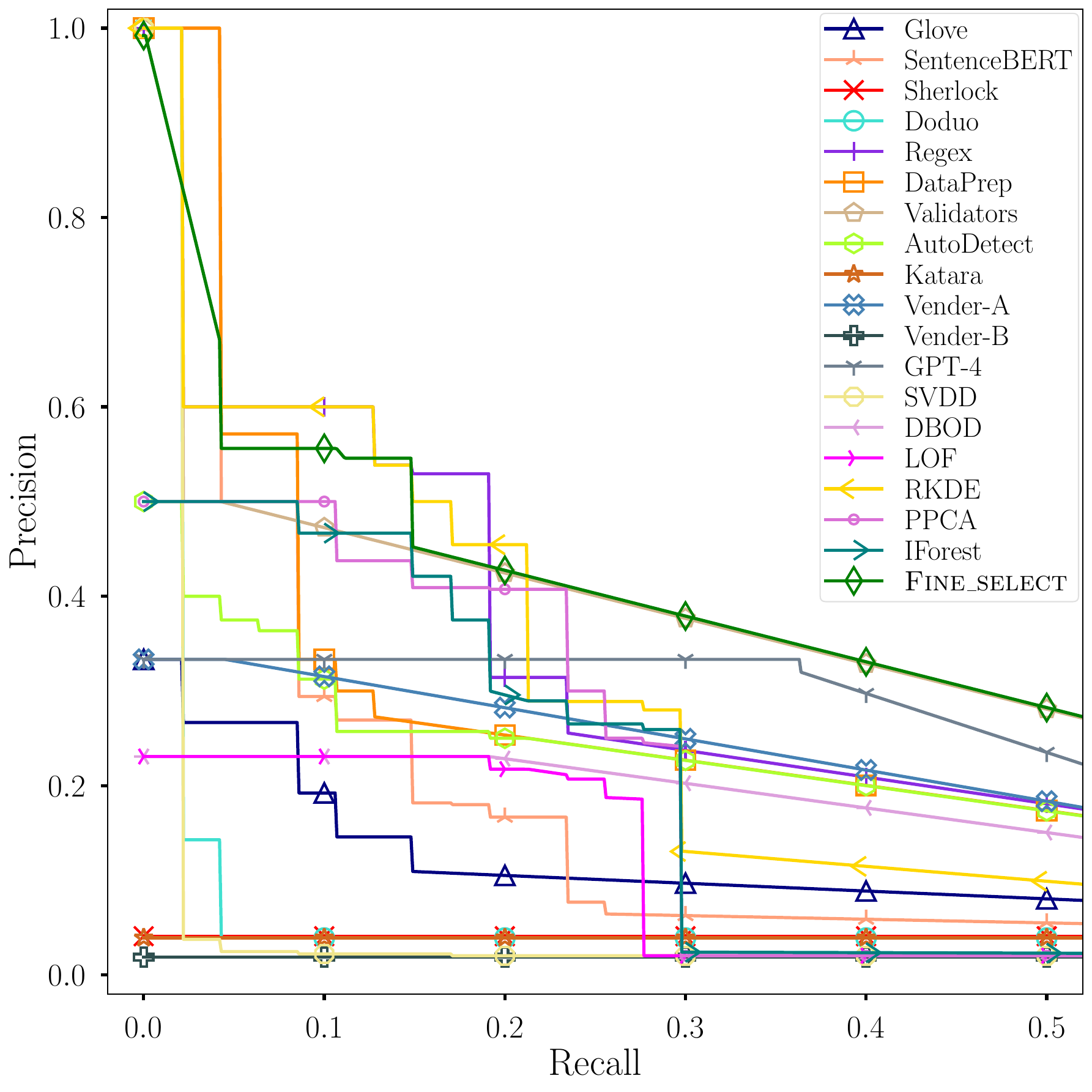}
\end{minipage}
\\
\begin{minipage}{6.5cm}
\centering
\captionof{figure}{PR curves on \rttesta when trained using \sttrain}
\label{fig:pr_benchmark_pbi_benchmark_rule_Excel}
\end{minipage}
&
\begin{minipage}{6.5cm}
\centering
\captionof{figure}{PR curves on \sttesta when trained using \sttrain}
\label{fig:pr_benchmark_excel_benchmark_rule_Excel}
\end{minipage}
\\
\end{tabular}
\end{figure*}

    % \section{Performance Detail on Data Cleaning Benchmarks} 
% \label{apx:detail_data_clean}

\section{Implementation Details}
\label{apx:imp_detail}
In Section~\ref{apx:rule_gen_optim}, we present two techniques that can accelerate the \sdca candidate generation and assessment process. Then, in Section~\ref{apx:detection_stage_optim}, we show how to further optimize the online prediction stage to achieve better evaluation latency.

\subsection{Optimization for \sdca Generation and Assessment Process} 
\label{apx:rule_gen_optim}

Enumerating and assessing all \sdca candidates as what we describe in Section~\ref{subsec:rule_cand_generation} and \ref{subsec:rule_quality_eval} can be inefficient. In this appendix, we present two optimization techniques that can reduce the unnecessary computations.

Given a \sdca $r$, let $\mathcal{C}^{r}_{C}$ be the set of columns in the training corpus $\mathcal{C}$ that is covered by $r$ (i.e., $\left | \mathcal{C}^{r}_{C} \right | = \left | \mathcal{C}^{r}_{C, T} \right | + \left | \mathcal{C}^{r}_{C, \overline{T}} \right |$, where $\left | \mathcal{C}^{r}_{C, T} \right |$ and $\left | \mathcal{C}^{r}_{C, \overline{T}} \right |$ are as shown in Table~\ref{tab:contingency}). 
By assuming that $\left | \mathcal{C}^{r}_{C, T} \right | = 0$, we can computed the upper bound of $r$'s confidence, which we denote as $ub(r.c)$, using the following equation.

\begin{equation}\label{eqt:conf_wilson_ub}
ub(r.c) = 1 -
\frac{z^2}{\left | \mathcal{C}^{r}_{C} \right | + z^2}
\end{equation}

Our first optimization technique is based on the following observation.

\begin{observation}\label{obs:conf_ub_monotone}
    Given two \sdcas $r$ and $r'$, if $\left | \mathcal{C}^{r}_{C} \right | \leq \left | \mathcal{C}^{r'}_{C} \right |$, then $ub(r.c) \leq ub(r'.c)$.
\end{observation}

Based on this observation, given a confidence threshold denoted as $c_{thres}$, we can find the minimum number of covered columns required for $ub(r.c) \geq c_{thres}$. All \sdca that do not cover a sufficient number of columns can be simply pruned.

For example, assume that we set $c_{thres} = 0.9$. Then, by Equation~\ref{eqt:conf_wilson_ub}, we can compute that the minimum number of covered columns required to let a \sdca's confidence upper bound exceed $c_{thres}$ is 34.All \sdcas that cover less than 34 columns can be simply pruned.

Next, we introduce our second optimization technique. Let $A(r) = \{C \in \mathcal{C} \,|\, r.P(C) =\text{true}\}$, where $\mathcal{C}$ is the entire space of possible columns, define the set of all columns to which $r$ can be applied. Given two \sdca $r$ and $r'$, we say that $r$ is \textit{in the subspace of} $r'$ if $A(r) \subseteq A(r')$. We have the following observation.

\begin{observation} \label{obs:pre_subset}
    Given two \sdca $r$ and $r'$, if $r$ is in the subspace of $r'$, then $ub(r.c) \leq ub(r'.c)$. 
\end{observation}

Assume we have a \sdca $r'$ where $ub(r'.c)$ is less than the confidence threshold, $c_{thres}$. Additionally, consider another \sdca $r$ that lies within the subspace of $r'$. In this case, $r$ can be pruned directly because, as shown in Observation~\ref{obs:pre_subset}, the confidence of $r$ cannot exceed $c_{thres}$.

It's important to note that during the \sdca candidate generation process, many candidates exist within the subspace of another. For instance, consider CTA-based \sdca. If \sdca $r_1$ has parameters $m$, $d_{in}$, $d_{out}$, and $t_i$, and another \sdca $r_2$ shares the same parameters except for $d_{in}'$, where $d_{in}' \geq d_{in}$, then $r_2$ lies within the subspace of $r_1$.
This optimization allows us to bypass all \sdca in the subspace of $r'$ once we determine that $ub(r'.c)$ falls below the confidence threshold, significantly reducing unnecessary computations within that subspace.

\subsection{Optimization for Online Prediction Stage} 
\label{apx:detection_stage_optim}

In this section, we present an optimization technique that can further improve the latency of online prediction. Observe that \sdcas with the same pre-condition can be applied on the same set of columns. Therefore, when applying a set $R$ of \sdcas on a column $C$, instead of testing each $r \in R$ on $C$ sequentially, we can adopt a two-step approach: 
(1) First, we extract the unique set of pre-conditions $\mathcal{P} = \{r \in R \,|\, r.P\}$, and determine which pre-conditions evaluate as true on $C$; 
(2) Once the applicable pre-conditions are identified, we apply the corresponding \sdcas to detect errors in $C$. 
By effectively "compressing" the evaluation of \sdcas with identical pre-conditions into a single check, the number of \sdcas needed to be tested is significantly reduced, leading to an improvement of the prediction latency.
    \section{Proofs} 
\label{apx:proofs}

\subsection{Proof of Theorem~\ref{thm:crs_np_hard}} \label{subsec:proof_crs_np_hard}

We prove Theorem~\ref{thm:crs_np_hard} by reducing from the maximum coverage (MC) problem. 
The MC problem is defined as follows: 
Given a set of elements $\mathcal{X}=\{x_1, x_2, \ldots, x_n\}$, and a collection of sets $\mathcal{S} = \{S_1, S_2, \ldots, S_m\}$ where every $S_i \subseteq X$, the goal is to find a subset $\mathcal{S}' \subseteq \mathcal{S}$ of size at most $k$, such that the total number of elements covered by $\mathcal{S}'$ is maximized.

We show that for every MC instance with parameters $(\mathcal{X}, \mathcal{S}, k)$ we can find an equivalent \cssa instance with parameters $(\mathcal{C}_{syn}, R_{all}, B_{size}, B_{FPR})$. 
Specifically, for each $x_j \in \mathcal{X}$, we create a column $C_j$ in $\mathcal{C}_{syn}$. 
For each $S_i \in \mathcal{S}$, we create a \sdca $r_i$ in $R_{all}$. 
For every $x_j \in S_i$, let $r_i$ can detect the error in $C_j$. 
We set $B_{size} = k$ and $B_{FPR} = +\infty$. 
It is easy to verify that the two instances are equivalent, and the transformation takes polynomial time. 
Therefore, \cssa is NP-hard. 
By \cite{feige98}, MC cannot be approximated with a ratio better than $(1 - 1/e)$, unless $NP \subseteq DTIME(n^{O(\log \log n)})$. The same conclusion applies to \cssa as well.

\subsection{Proof of Theorem~\ref{thm:cs}} \label{subsec:proof_cs}

We first prove the equivalence between the original \cssa and its corresponding CSS-ILP. 
Firstly, the objective of \cssa (i.e., $\max_{R \subseteq R_{all}} \left| \bigcup_{r \in R} D(r) \right|$) is equivalent to the objective of CSS-ILP (i.e., maximize $\sum_{C_j \in \mathcal{C}_{syn}} y_j$). 
The size constraint in \cssa (i.e., $\left | R \right | \leq B_{size}$) is equivalent to the first constraint in CSS-ILP (i.e., $\sum_{r_i \in R_{all}} x_i \leq B_{size}$), 
and the FPR constraint in \cssa (i.e., $\sum_{r \in R} \text{FPR}(r) \leq B_{FPR}$) is equivalent to the second constraint in CSS-ILP (i.e., $\sum_{r_i \in R_{all}} \text{FPR}(r_i) \cdot x_i \leq B_{FPR}$). 
The third constraint in CSS-ILP (i.e., $\sum_{r_i \in K_j} x_i \geq y_j,\; \forall C_j \in \mathcal{C}_{syn}$) ensures that $y_j = 1$ if and only if some $r_i$ that reports $C_j$'s ground-truth error is selected into $R$. 
Therefore, every \cssa instance is equivalent to its corresponding CSS-ILP instance, and they achieve the maximum objective simultaneously.

Next, we prove the performance guarantee of \cs. Denote the solution obtained by solving CSS-LP as 
$X = \{x'_i\}$
and $Y = \{y'_j\}$.
Since each $r_i$ is selected into $R$ with a probability of $x'_i$, we have 

\begin{equation*}\label{eqt:cs_size_guarantee}
E(\left | R \right |) = E(\sum_{r_i \in R_{all}} x'_i) = \sum_{r_i \in R_{all}} x'_i \leq B_{size}.
\end{equation*}

Similarly, we can show that 

\begin{equation*}\label{eqt:cs_fpr_guarantee}
E(\sum_{r \in R} \text{FPR}(r)) = E(\sum_{r_i \in R_{all}} \text{FPR}(r_i) \cdot x'_i) = \sum_{r_i \in R_{all}} \text{FPR}(r_i) \cdot x'_i \leq B_{FPR}.
\end{equation*}

Lastly, to prove that $E(\left | D(R) \right |)  \geq (1-1/e)OPT$ where $OPT$ is the optimal value, we first show that for each $C_j \in \mathcal{C}_{syn}$, the probability that $C_j \in D(R)$ is at least $(1-1/e) y'_j$. 
To see this, observe that $C_j \notin D(R)$ only if none of the rules in $K_j$ is selected into $R$, which happens with a probability of at most 

\begin{equation*}\label{eqt:cs_obj_guarantee_eqt1}
\prod_{r_i \in K_j} (1 - x'_i) \leq \prod_{r_i \in K_j} e^{- x'_i} = e^{ - \sum_{r_i \in K_j} x'_i} \leq e^{- y'_j},
\end{equation*}

where the last inequality comes from the third constraint of CSS-ILP. 

Since $y'_j \in [0, 1]$, the probability that $C_j \in D(R)$ is at least $1 - e^{- y'_j} \geq (1-1/e) y'_j$. 
Therefore, we have

\begin{align*}
    E(\left | D(R) \right |) & \geq E(\sum_{C_j \in \mathcal{C}_{syn}} (1-1/e) y'_j) \\
    & = (1-1/e)\sum_{C_j \in \mathcal{C}_{syn}} y'_j \\
    & \geq (1-1/e)OPT
\end{align*}

where the last inequality comes from the fact that the optimal value of CSS-LP instance is at least the optimal value of its corresponding CSS-ILP instance.

\subsection{Proof of Theorem~\ref{thm:fs}} \label{subsec:proof_fs}

We first prove that by making the two key modifications, the original \fssa instance is equivalent to its corresponding CSS-ILP instance. The remainder of the proof closely follows that of Theorem~\ref{thm:cs}.

To see this, first observe that the objective of \fssa (i.e., maximize $ _{R \subseteq R_{all}} \left | \{ C \in D(R) \,|\, \text{diff}(C, R, R_{all}) \leq \delta \} \right |$) is equivalent to the objective of CSS-ILP (i.e., maximize $\sum_{C_j \in \mathcal{C}_{syn}} y_j$). 
Note that we have modified the assignment of $y_j$ to $y_j = 1$ if $C_j \in \{ C \in D(R) \,|\, \text{diff}(C, R, R_{all}) \leq \delta \}$, and 0 otherwise.
The size constraint in \fssa (i.e., $\left | R \right | \leq B_{size}$) is equivalent to the first constraint in CSS-ILP (i.e., $\sum_{r_i \in R_{all}} x_i \leq B_{size}$), 
and the FPR constraint in \fssa (i.e., $\sum_{r \in R} \text{FPR}(r) \leq B_{FPR}$) is equivalent to the second constraint in CSS-ILP (i.e., $\sum_{r_i \in R_{all}} \text{FPR}(r_i) \cdot x_i \leq B_{FPR}$). 
The third constraint in CSS-ILP (i.e., $\sum_{r_i \in K_j} x_i \geq y_j,\; \forall C_j \in \mathcal{C}_{syn}$) ensures that a column $C_j$ belongs to $\{ C \in D(R) \,|\, diff(C, R, R_{all}) \leq \delta \}$ if and only if some $r_i \in K_j$ is selected into $R$. Note that such a $r_i$ can detect $C_j$'s ground-truth error with a confidence at least $\text{conf}(C, R_{all}) - \delta$, which guarantees that $\text{diff}(C_j, R, R_{all}) \leq \delta$ since $\text{conf}(C, R)$ cannot be smaller than the confidence of $r_i$. 
We conclude that the original \fssa instance is equivalent to its corresponding CSS-ILP instance.

% To see this, observe that the third constraint in CSS-ILP (i.e., $\sum_{r_i \in S_j} x_i \geq y_j,\; \forall C_j \in \mathcal{C}_{syn}$) ensures that $y_j = 1$ if and only if some $r_i \in S_j$ is selected into $R$. In other words, a column $C_j \in D(R)$ if and only if there is a rule $r_i \in R$ where $o(C_j, \{r_i\}) \in O(C_j)$ and $conf(r_i) \geq cs(C_j, R_{all}) - \delta$, which is equivalent to the last constraint in \fssa. 
% The equivalence between the other constraints in \fssa and CSS-ILP are already shown in the proof of Theorem~\ref{thm:cs}.
% Since the \fssa instance and the corresponding CSS-ILP instance are equivalent, 
% the proof of (1) $E(\left | R \right |) \leq B_{size}$, (2) $E(\sum_{r \in R} \text{FPR}(r)) \leq B_{FPR}$, and (3) $E(\left | D(R) \right |) \geq (1-1/e)OPT$ are very similar to that in the proof of Theorem~\ref{thm:cs}. 

% To prove the last claim in the theorem, i.e., $cs(c, R) \geq cs(c, R_{all}) - \delta$ for every $c \in D(R)$, note that $c \in D(R)$ if and only if there exists some $r \in R$ such that $o(c, \{r\}) \in O(c)$ and $conf(r) \geq cs(c, R_{all}) - \delta$. Since $cs(c, R) \geq conf(r)$ by its definition, this claim holds. 
}

% \iftoggle{fullversion}
% {
%     % removed for revision
%     \revised{}
%     \clearpage
%     \appendix
%     %\input{apx-additional-operators.tex}
% }
% {
% }

\end{sloppy}

\end{document}